\documentclass[11pt,reprint,showpacs,showkeys,nofootinbib,ssfamily]{article}

\usepackage{amsmath}
\usepackage{amsfonts,amssymb,mathtools}
\usepackage{cite} 
\usepackage{hyperref} 
\hypersetup{backref, colorlinks=true  } 
\usepackage{multirow}
\usepackage[dvips]{graphicx}
\usepackage{float}
\usepackage{appendix}
\usepackage{color}
\usepackage{url}
\usepackage{subfigure}
\usepackage{footnote}
\usepackage{authblk} 

\def\beq{\begin{equation}}
\def\eeq{\end{equation}}
\def\bea{\begin{eqnarray}}
\def\eea{\end{eqnarray}}
\def\nn{\nonumber}
\def\nl{\nonumber\\}

\def\chic1{\chi_{c1}}

\newcommand{\vep}{\varepsilon}

\def \MCDD {M_{\rm{CDD}}}

\def \Im{\text{Im}\,}
\def \Eexp{E_{\rm{exp}}}
\def \Gammas {\Gamma_{*}}
\def \cbg {\rm{cbg}}

\def \BBD {{\mathcal B}_D}
\def \BBJ {{\mathcal B}_J}

\newcommand{\vp}{\mathbf{p}}
\newcommand{\vq}{\mathbf{q}}

\newcommand{\cV}{\cal{V}}
\newcommand{\cF}{\cal{F}}
\newcommand{\cE}{\cal{E}}
\newcommand{\cf}{\mathfrak{f}}
\newcommand{\ct}{d}

\title{Different pole structures in line shapes of the $X(3872)$} 
\author[]{Xian-Wei Kang}
\author[]{J.~A. Oller}
\affil[]{Departamento de F\'\i sica,
Universidad de Murcia, E-30071 Murcia,
Spain}

\begin{document}
\maketitle
\begin{abstract}
We introduce a near-threshold parameterization that is more general than the effective-range expansion
 up to and including the effective-range because it can also handle with a near-threshold zero in the
$D^0\bar{D}^{*0}$ $S$-wave. In terms of it
we analyze the CDF data on  inclusive $p\bar{p}$ scattering to  $J/\psi \pi^+\pi^-$, and the Belle and  BaBar data on $B$ decays
to $K\, J/\psi \pi^+\pi^-$ and $K D\bar{D}^{*0}$ around the $D^0\bar{D}^{*0}$ threshold.
 It is shown that data can be reproduced with similar quality for
the $X(3872)$ being a bound {\it and/or} a virtual state. We also find that the $X(3872)$ might be a higher-order
virtual-state pole (double or triplet pole), in the limit  in which the small $D^{*0}$ width vanishes.
Once the latter is restored the corrections to the pole position are non-analytic and much bigger than the $D^{*0}$ width itself.
 The $X(3872)$  compositeness coefficient in $D^0\bar{D}^{*0}$ ranges from nearly 0 up to 1 in the different scenarios.
\end{abstract}


\section{Introduction}

The $X(3872)$ has been analyzed in great phenomenological detail
by employing $S$-wave effective-range-expansion (ERE) parameterizations in Refs.~\cite{Hanhart.141116.4,Braaten.141116.3,Braaten}.
 References~\cite{Braaten.141116.3,Braaten} includes only the $D\bar{D}^*$ scattering
length, $a$, while Ref.~\cite{Hanhart.141116.4} also includes the effective-range ($r$)
contribution.\footnote{To shorten the presentation we actually refer by $D\bar{D}^*$ to the $C=+$ combination  $(D\bar{D}^*+\bar{D}D^*)/\sqrt{2}$.}
A detailed comparison between both approaches is given in Sec.~6 of Ref.~\cite{Braaten}.
Indeed, the use of the ERE up to and including the effective range
 is more general than  employing a Flatt\'e
parameterization (also used in Refs.~\cite{ZhengHQ,Kalashnikova.141116.2,Qiang.141116.1}), because
only negative effective ranges can be generated within the latter \cite{baru.170310.1}.\footnote{As follows
from Ref.~\cite{Zbpaper}, the Flatt\'e parameterization and the ERE
 including the effective-range contribution are equivalent if the former is written in terms of
 the bare  mass and coupling squared that need to be tuned, taking a priori any sign, to reproduce
the values of the residue and pole position of the  partial wave.}

However, the ERE convergence radius might be severely limited due to the presence of
near-threshold zeroes of the partial wave, in this case the $D^0\bar{D}^{*0}$ $S$-wave.
These zeroes, also called Castillejo-Dalitz-Dyson (CDD) poles \cite{Castillejo.141116.5},  constitute
the major criticism to apply Weinberg's compositeness theorem  to evaluate the actual compositeness of a near-threshold bound
state \cite{Weinberg.141116.6}, because it is based on the ERE up
to the effective-range contribution.\footnote{This serious criticism was originally due to R.~Blankenbecler, M.~L.~Goldberger,
K.~Johnson and S.~B.~Treiman, as explicitly stated in the note added in proof in Ref.~\cite{Weinberg.141116.6},
warning about the possible presence of CDD poles for $E>-B$ in the Low equation used in this reference. The only way to skip this
problem is to ascertain the range of convergence of the ERE, typically from data.}
The same criticism is of course applicable to the papers
\cite{Hanhart.141116.4,Braaten.141116.3,Braaten,Kalashnikova.141116.2,Qiang.141116.1} referred in the
previous paragraph.

The issue about the possible presence of near-threshold zero in the $S$-wave partial wave
 and the spoil of the corresponding ERE was also discussed more recently in Ref.~\cite{baru.170310.2}.
 One of the main conclusions of this reference was that in order to end with a near-threshold zero
one needs also three shallow poles. In this way this situation was qualified as highly accidental
by the authors of Ref.~\cite{baru.170310.2}. However, this conclusion is not necessarily correct, that is,
one can have a near-threshold zero with only two nearby poles, without the need of a third one.
The reason for this misstep in the study of Ref.~\cite{baru.170310.2} was a misuse of the relation between the
position of the zero and the location of the poles in the three-momentum complex plane,
as we discuss in detail in Sec.~\ref{subsec:virtualboundI}.
 Two-coupled channels effects were included in Ref.~\cite{artoi.170310.1} along  the similar lines of mixing the exchange of a resonance with direct
interactions between the mesons, in the limit of validity of the scattering length approximation for the latter ones.
In turn, the coupled channel generalization of Ref.~\cite{baru.170310.2} was
derived in Ref.~\cite{kalas.170310.1}.
 In the energy region around the $D^0\bar{D}^{*0}$ threshold  where the $X(3872)$ sits,
the coupled channel results of Refs.~\cite{artoi.170310.1,kalas.170310.1}
reduce to a partial wave whose structure can be deduced from the elastic one-channel  $D^0\bar{D}^{* 0}$ scattering,
 because the  $D^+D^{*-}$ threshold is relatively much further  away.
 We also indicate here that Ref.~\cite{baru.170310.2}  cannot reproduce
positive values for the $D^0\bar{D}^{*0}$ $S$-wave effective range, while our approach is more general in this respect and can also give rise to
 positive values of this low-energy scattering parameter.  These two points are also shown explicitly below.

As in Refs.~\cite{Hanhart.141116.4,Braaten.141116.3,Braaten} we avoid any explicit dynamical model
for the $D\bar{D}^*$ dynamics to study the $X(3872)$ line shapes in the  BaBar \cite{BaBarJ,BaBarD}
 and Belle \cite{BelleJ,BelleD}  data on the  $B$ decays to
$K^\pm J/\psi \pi^+\pi^-$ and $K J/\psi D^0\bar{D}^{*0}$.
In addition, we also consider the higher-statistics data from the inclusive $p\bar{p}$ scattering to $J/\psi \pi^+\pi^-$ measured
by the CDF Collaboration \cite{CDF.211116.5} and that gives rise to a more precise determination of the mass of the $X(3872)$
\cite{pdg.181116.2}.
However, we employ a more general parameterization
 than the ERE expansion up to and including the effective-range contribution
by explicitly taking into account the possibility of the presence of a CDD pole very close to the
$D^0\bar{D}^{*0}$ threshold. Our formalism has as limiting cases
those of Refs.~\cite{Hanhart.141116.4,Braaten.141116.3,Braaten}, but it can also consider other cases.
In particular, while in Refs.~\cite{Hanhart.141116.4,Braaten.141116.3,Braaten}
the $X(3872)$ turns out to be  either a bound  or a virtual state pole, we also find other
qualitatively different scenarios that can reproduce data  with similar quality as well.
In two of these new situations the $X(3872)$ is simultaneously a bound and a virtual state and
 for one of them the $D^0\bar{D}^{*0}$ compositeness coefficient is
just of a few per cent. This is also an interesting counterexample for the conclusions of Ref.~\cite{baru.170310.2}, because
it has a CDD pole almost on top of threshold with only two shallow poles.
 Remarkably,  we also find other cases with two/three virtual-states poles, such that
in the limit of vanishing width of the $D^{* 0}$  these poles become degenerate and result in a second/third-order $S$-matrix pole.
Along the lines of the discussions, we also match our resulting partial-wave amplitude from $S$-matrix theory with the one
 deduced in Ref.~\cite{baru.170310.2} in
terms of the exchange of  a bare state and direct interactions between the $D^0\bar{D}^{*0}$ mesons.
Similarly, this is also done with the one-channel
reduction of Ref.~\cite{artoi.170310.1} in the $D^0\bar{D}^{*0}$ near-threshold region.

The paper is organized as follows. After this Introduction we present the formalism for the analysis of the line shapes
of the $X(3872)$ in Secs.~\ref{subsec.191116.1}, \ref{subsec.191116.2}, \ref{sec.211116.1} and \ref{formalism.141116.1}.
 The different scenarios and their characteristics are the main subject of Sec.~\ref{sec:Combined},
where we also give the numerical results of the fits in each case, the poles obtained and their properties.
 After the concluding remarks in Sec.~\ref{sec.conclusions}, we give some more technical and detailed material  in the Appendices
\ref{app.081016.1}, \ref{app.170320.1} and \ref{app.251116.1}.

\section{$J/\psi \pi^+ \pi^-$ partial-decay rate and differential cross section}
\label{subsec.191116.1}

\begin{figure}
\begin{center}
\includegraphics[width=0.2\textwidth]{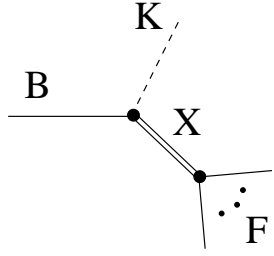}
\caption{Skeleton Feynman diagram for the $B\to KF$ decay through the $X(3872)$ resonance.}
\label{fig.181116.2}
\end{center}
\end{figure}
For the decay $B\to K F$ through the $X(3872)$ we have  the decay chain $B\to K X$ and then $X\to F$.
We can write the decay amplitude $T_F$, represented schematically by the Feynman diagram in Fig.~\ref{fig.181116.2}, as
\begin{align}
\label{18116.5}
T_F=- \frac{{\cV}_L{\cV}_X}{Q^2-P_X^2}~,
\end{align}
where ${\cV}_L$ and ${\cV}_X$ refer to the vertices from left to right in Fig.~\ref{fig.181116.2},
$Q^2$ is the invariant mass squared of the subsystem of final particles $F$ (which also coincides with the invariant mass squared of
the $X(3872)$ resonance) and the $X(3872)$ pole position is
$P_X=M_X-i\Gamma_X/2$, with $M_X$ and $\Gamma_X$ its mass and width, respectively.
 The partial-decay width for this process is
\begin{align}
\label{181116.6}
\Gamma_{B\to KF}=&\frac{1}{2M_B}\int (2\pi)^4 \delta(P-Q-p_k)\frac{d^3p_K}{(2\pi)^32E_K} d{\cF}  \frac{|{\cV}_L|^2|{\cV}_X|^2}{|Q^2-P_X^2|^2}~.
\end{align}
In this equation, $P$ is the total four-momentum of the system (or that of the $B$ meson), $p_K$ is the four-momentum of the kaon and
$Q$ is the one of the $X(3872)$ (or $F$ subsystem).
Let us denote by a subscript $i$ (with $i=1,\ldots,N_F$) the particles in $F$
and denote the four-momentum of every particle as $p_i$, so that $Q =\sum_{i=1}^{N_F} p_i$. Then, we define $d{\cF}$ as the count of
 states in the subsystem $F$,
\begin{align}
\label{181116.7}
d{\cF}& = \prod_{i=1}^{N_F}\frac{d^3p_i}{(2\pi)^3 2E_i}~,
\end{align}
being $E_i=\sqrt{m_i^2+\vp_i^2}$ the energy of the $i$th particle  with mass $m_i$ and three-momentum $\vp_i$. The phase space
factor for ${\cF}$, that we denote by $d{\cf}$, can be obtained  by extracting from $d{\cF}$ its total four-momentum contribution, so that
\begin{align}
\label{191116.1}
d\cF&= \frac{d^4Q}{(2\pi)^4} d{\cf}~.
\end{align}
We take this into the expression for $\Gamma_{B\to KF}$, Eq.~\eqref{181116.6}, and multiply and divide the integrand by
$Q_0=+\sqrt{Q^2+\mathbf{Q}^2}$, which is the energy corresponding to a particle of mass $\sqrt{Q^2}$ and three-momentum squared
$\mathbf{Q}^2$.
 Notice that $Q_0>0$ and  $Q^2>0$ because they are the total energy and invariant mass squared, in order, of the asymptotic particles in $F$.
 We then have
\begin{align}
\label{191116.2}
\Gamma_{B\to K F}&= \int \frac{dQ_0}{2\pi} d{\cf}\, 2\sqrt{Q^2+\mathbf{Q}^2} \frac{|{\cV}_X|^2}{|Q^2-P_X^2|^2}\nn \\
&\times \frac{1}{2M_B}\int (2\pi)^4 \delta(P-Q-p_K)|{\cV}_L|^2\frac{d^3p_K}{(2\pi)^32E_K} \frac{d^3Q}{(2\pi)^3 2\sqrt{Q^2+\mathbf{Q}^2}}~.
\end{align}
 The decay width of a $B$ meson into a kaon $K$ and a resonance $X$ of mass $\sqrt{Q^2}$,  $\Gamma_{B\to K X}(Q^2)$, is the term on the right-hand side of the second line in the previous equation:
\begin{align}
\label{191116.3}
\Gamma_{B\to K X}(Q^2)&=
\frac{1}{2M_B} \int (2\pi)^4 \delta(P-Q-p_K)|{\cV}_L|^2\frac{d^3p_K}{(2\pi)^32E_K} \frac{d^3Q}{(2\pi)^3 2\sqrt{Q^2+\mathbf{Q}^2}}~.
\end{align}
Similarly the decay width of $X(\sqrt{Q^2})$ into $F$, $\Gamma_{X\to F}(Q^2)$ is given by
\begin{align}
\label{191116.4}
\Gamma_{X\to F}(Q^2)&=\frac{1}{2\sqrt{Q^2}}\int d{\cf} \,|{\cV}_X|^2~.
\end{align}
We also perform the change of variables from $Q_0$ to $Q^2$, related by
\begin{align}
\label{041216.5}
Q^2=Q_0^2-\mathbf{Q}^2~.
\end{align}
Then, in terms of Eqs.~\eqref{191116.3} and \eqref{191116.4} we can rewrite Eq.~\eqref{191116.2} as
\begin{align}
\label{191116.5}
\Gamma_{B\to K F}&=\int \frac{dQ^2}{2\pi}2\sqrt{Q^2}\frac{\Gamma_{B\to K X}(Q^2)\Gamma_{X\to F}(Q^2)}{|Q^2-P_X^2|^2}~.
\end{align}
One can formulate more conveniently the previous expression by noticing that  we are interested in event distributions with invariant
mass around the nominal mass of the $X(3872)$ and $\Gamma_X\ll M_X$  ($\Gamma_X<1.2~\text{MeV}$ \cite{pdg.181116.2}). We then approximate
\begin{align}
\label{191116.8}
Q^2-P_X^2&\simeq 2\sqrt{Q^2}(\sqrt{Q^2}-P_X)~,
\end{align}
in the propagator of the $X(3872)$ in Eq.~\eqref{191116.5}.
Measuring the invariant mass of the $X(3872)$ with respect to the $D^0\bar{D}^{*0}$ threshold, we define the energy variable $E$ as
\begin{align}
\label{191116.9}
E & =\sqrt{Q^2}-M_{D^0}-M_{D^{*0}}~.
\end{align}
From Eqs.~\eqref{191116.8} and \eqref{191116.9} we rewrite the differential decay rate for Eq.~\eqref{191116.5} as
\begin{align}
\label{191116.10}
\frac{d\Gamma_{B\to K F}}{dE}&=\frac{\Gamma_{B\to K X}(Q^2)\Gamma_{X\to F}(Q^2)}{2\pi|\sqrt{Q^2}-P_X|^2}\nn\\
&=\frac{\Gamma_{B\to K X}(Q^2)\,\Gamma_{X\to F}(Q^2)}{2\pi|E-E_X+i\Gamma_X/2|^2}~,
\end{align}
with $E_X=M_X-M_{D^0}-M_{\bar{D}^{*0}}$ the mass of the resonance from the $D^0 \bar{D}^{* 0}$
threshold.\footnote{We follow a different sign convention for $E_X$ compared to Ref.~\cite{Braaten}, so that here $E_X$ is
 negative for $M_X< M_{D^0}+ M_{\bar{D}^{*0}}$.}

Next, let us assume that we describe the final state interactions of the $D^0\bar{D}^{* 0}$ system in terms of a function ${\ct}(E)$ that
 gives account of the $X(3872)$ signal properly, which is represented in Eq.~\eqref{191116.10} by the propagator factor squared
 $1/|E-E_X+i\Gamma_X/2|^2$. This is strictly the case for a bound state or for an isolated resonance such that $|\Gamma_X/E_X|\ll 1$.
For any other case (e.g. a pure virtual-state case) we make use of the analytical continuation of the expressions obtained.
 Then we can write ${\ct}(E)$ around this energy region as
\begin{align}
\label{191116.11b}
{\ct}(E)\simeq \frac{\alpha}{(E-P_X)}~,
\end{align}
with $\alpha$ the residue of ${\ct}(E)$ at the resonance pole. In this way,
\begin{align}
\label{191116.12}
\frac{1}{|E-M_X+i\Gamma_X/2|^2}\to\frac{|{\ct}(E)|^2}{|\alpha|^2}~,
\end{align}
and we express Eq.~\eqref{191116.10} as
\begin{align}
\label{191116.13}
\frac{d\Gamma_{B\to K F}}{dE}&=\Gamma_{B\to K X}(Q^2) \Gamma_{X\to F}(Q^2) \frac{|{\ct}(E)|^2}{2\pi|\alpha|^2}~.
\end{align}
As in Refs.~\cite{Hanhart.141116.4,Braaten} it is convenient to introduce in Eq.~\eqref{191116.13} the product of the branching ratios
for the decays $B\to K X$ and $X\to F$, ${\mathcal B}_F(Q^2)=\Gamma_{B\to K X}(Q^2) \Gamma_{X\to F}(Q^2)/\Gamma_{B}\Gamma_{X}$,
 with $\Gamma_B$ the total decay width of a $B$ meson. This equation then reads
\begin{align}
\label{191116.14}
\frac{d\Gamma_{B\to K F}}{dE}&= \Gamma_{B}\,{\mathcal B}_F(Q^2) \frac{\Gamma_{X}|{\ct}(E)|^2}{2\pi|\alpha|^2}~,
\end{align}
However, for a final system $F$ with a threshold relatively far away from the $D\bar{D}^{* 0}$ threshold compared to
$|E_X|$, we can neglect the $Q^2$ dependence in ${\mathcal B}_F$. This criterion can be also applied to a $B\to K J/\psi \pi^+\pi^-$ decay
because of the rather large width of the $\rho$ around 150~MeV, which washes out the sharp threshold effect
for this state if we neglected the $\rho$ width \cite{Braaten.191116.2}.
 However, this is not the case  for the
$B\to K D^0\bar{D}^{*0}$ decay measured by the BaBar \cite{BaBarD} and Belle \cite{BelleD} Collaborations,
which  is discussed in the next Section.

\begin{figure}
\begin{center}
\includegraphics[width=0.2\textwidth]{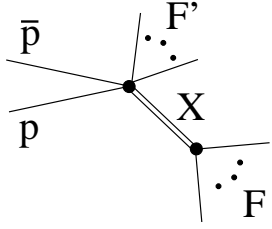}
\caption{Skeleton Feynman diagram for the $p\bar{p}\to F' F$ scattering through the $X(3872)$ resonance.}
\label{fig.041216.1}
\end{center}
\end{figure}

We also consider here the $J/\psi \pi^+\pi^-$ event distributions from the inclusive $p\bar{p}$ collisions  at $\sqrt{s}=1.65$~TeV
measured by the CDF Collaboration \cite{CDF.211116.5}.  The basic Feynman diagram now 
 shown in Fig.~\ref{fig.041216.1}. It is similar to Fig.~\ref{fig.181116.2} but changing the kaon $K$
 by a set of undetected final particles that are denoted collectively as $F'$, with the $X(3872)$ decaying into a
 set of particles denoted by $F$ as above.
 Instead of Eq.~\eqref{181116.6} we have now to calculate the cross section for $p\bar{p}$ to $F'F$ that reads
\begin{align}
\label{041216.1}
d\sigma_{p\bar{p}\to F'F}&=
\frac{1}{4\sqrt{s}|p|}\int (2\pi)^4 \delta(P-P_{F'}-Q)d{\cal F}' d{\cF}  \frac{|{\cV}_L|^2|{\cV}_X|^2}{|Q^2-P_X^2|^2}~,
\end{align}
where $|p|$ is the CM thee-momentum of the initial $p\bar{p}$ system and
\begin{align}
\label{041216.2}
d{\cal F}'&=\prod_{i=1}^{N_{F'}}\frac{d^3p_i}{(2\pi)^3 2E_i}~.
\end{align}
  Splitting $d{\cal F}$ as in Eq.~\eqref{191116.1}, we rewrite Eq.~\eqref{041216.1}
as
\begin{align}
\label{041216.3}
d\sigma_{p\bar{p}\to F'F}&=
\int  \frac{dQ_0}{2\pi} d{\cf}\, 2\sqrt{Q^2+\mathbf{Q}^2} \frac{|{\cV}_X|^2}{|Q^2-P_X^2|^2} \nn\\
&\times
\frac{1}{4\sqrt{s}|p|}   \int (2\pi)^4 \delta(P-P_{F'}-Q) |{\cV}_L|^2 d{\cal F}' \frac{d^3Q}{(2\pi)^3 2\sqrt{Q^2+\mathbf{Q}^2}}~.
\end{align}

 Next, we perform the change of integration variable from $Q_0$ to $Q^2$,
cf. Eq.~\eqref{041216.5}, and after integrating over ${\cal F}'$ and $\mathbf{Q}$
the factor on the right-hand side of the second line in the previous equation
is $\sigma_{p\bar{p}\to X \text{All}}(Q^2)$, namely,
\begin{align}
\label{041216.4}
\sigma_{p\bar{p}\to X \text{All}}(Q^2)&=
\frac{1}{4\sqrt{s}|p|}\int (2\pi)^4 \delta(P-P_{F'}-Q) |{\cV}_L|^2 d{\cal F}' \frac{d^3Q}{(2\pi)^3 2\sqrt{Q^2+\mathbf{Q}^2}}~.
\end{align}
Additionally, recalling the expression for $\Gamma_{X\to F}(Q^2)$ in Eq.~\eqref{191116.4}, the Eq.~\eqref{041216.3} becomes
\begin{align}
\label{041216.6}
d\sigma_{p\bar{p}\to F'F}&=\sigma_{p\bar{p}\to X\text{All}}(Q^2)\Gamma_{X\to F}(Q^2)\frac{2\sqrt{Q^2}}{|Q^2-P_X^2|^2}\frac{d Q^2}{2\pi}~.
\end{align}
Approaching the inverse propagator  as in Eq.~\eqref{191116.8} and
employing $d(E)$ to take into account the FSI, we finally rewrite Eq.~\eqref{041216.6} as
\begin{align}
\label{041216.7}
\frac{d\sigma_{p\bar{p}\to F'F}}{dE}&=\sigma_{p\bar{p}\to X\text{All}}(Q^2)Br(X\to F)(Q^2)\frac{\Gamma_X |d(E)|^2}{2\pi|\alpha|^2}~.
\end{align}

\section{$D^0\bar{D}^{*0}$ partial-decay rate}
\label{subsec.191116.2}

The $B\to K D^0\bar{D}^{*0}$ decay rate measured be the reconstruction of the $\bar{D}^{*0}$
 from the decay channels  $D^0\bar{D}^0\pi^0$ and $D^0\bar{D}^0\gamma$ decays  \cite{BaBarD,BelleD}
 has a strong dependence on the $D^0\bar{D}^{*0}$ invariant mass in the energy region of the $X(3872)$.
 One obvious reason is  because the $D^0\bar{D}^{*0}$ is almost at threshold.
Besides that one also has the decay chain $B\to K X(3872)$,
$X(3872)\to D^0 \bar{D}^{*0}$ and finally $\bar{D}^{*0}\to \bar{D}^0\pi^0$ or $\bar{D}^0\gamma$, so that the
$\bar{D}^{*0}$ Lorentzian has  some overlapping with the $X(3872)$ mass distribution that rapidly decreases
for increasing energy if the latter lies below threshold.
 As a result, the width of the $\bar{D}^{*0}$ has to be taken into account in the formalism  from the start
to study the decays of the $X(3872)$ through the $D^0\bar{D}^{*0}$ intermediate state, particularly if this state manifests as a
$D^0\bar{D}^{*0}$ bound state. This point was stressed originally in Ref.~\cite{Braaten.141116.3}.

A $D^0\bar{D}^0\pi^0$ event from the $B$ decays to $K X(3872)$ can be generated by either
$B\to K X(3872)$, $X(3872) \to \bar{D}^0 D^{*0}$, $D^{*0}\to D^0\pi^0$
or  $X(3872) \to D^0 \bar{D}^{*0}$, $\bar{D}^{*0}\to \bar{D}^0\pi^0$. This is an
interesting interference process for the $X(3872)$ being mostly a $D^0\bar{D}^{*0}$ molecule, as first discussed in
Ref.~\cite{voloshin.181116.1}.
 This latter reference shows that the interference effects vanish for a zero binding energy while
 Ref.~\cite{Braaten} elaborates that they can be neglected
for $|E|\ll 2 (M_{\pi^0}/M_{D^0})\delta\simeq 1$~MeV, with $\delta=M_{D^{*0}}-M_{D^0}-M_{\pi^0}\simeq 7.2~$MeV,
 the energy delivered in a $D^{*0}$ decay at rest.
 For the case of the $X(3872)$ with a nominal mass $E_X=-0.12\pm 0.20$~MeV  \cite{pdg.181116.2} (adding in quadrature the uncertainties in the masses
 of the $X(3872)$, $D^0$ and $D^{*0}$ given in the PDG  \cite{pdg.181116.2}) this inequality is operative and one might expect some suppression
of these interference effects.
 The latter were also worked out explicitly in Ref.~\cite{hanhart.170310.1} by considering the three-body $D^0\bar{D}^{0}\pi$ dynamics and
it was shown there that for  a binding energy of 0.5~MeV,
the interference effects below the $D^0\bar{D}^{*0}$ threshold  at the peak of the $X(3872)$
 are sizeable.  This result is in agreement with outcome of Ref.~\cite{voloshin.181116.1} for the
decay width of the $X(3872)$ to $D^0\bar{D}^0\pi^0$, which found that they are substantial
already for binding energies $|E_X|\gtrsim 0.1$~MeV.
However, Ref.~\cite{hanhart.170310.1} derived that above the $D^0\bar{D}^{*0}$ threshold they are very modest
and for the case of a virtual state they are so in the whole energy range (both above and below  threshold).
Additionally, these interference effects are mostly proportional to the weight of the molecular $D^0\bar{D}^{*0}$
weight of the $X(3872)$ or compositeness, as explicitly shown by Voloshin in Ref.~\cite{voloshin.181116.1}.
 In turn, the interference contributions in the decay channel $D^0\bar{D}^0\gamma$
 should be smaller because the three-momentum of the $D^0$ from the decay $D^{*0}\to D^0\gamma$
 is significantly bigger than for $D^0\pi^0$, so that the overlapping with the wave function of the $D^0$ in the
 $X(3872)$ is reduced, an argument borrowed from Ref.~\cite{voloshin.181116.1}.
  Based on these facts resulting from previous works \cite{voloshin.181116.1,hanhart.170310.1,Braaten}
and because we are mostly interested in our study in scenarios for the $X(3872)$ in which it is a double/triplet virtual state
or it has a very small molecular component,  we neglect in the following any interference effect in the $D\bar{D}^0\pi^0$
and $D\bar{D}^{0}\gamma$ decays.\footnote{Mostly for comparison with less standard
 scenarios we also discuss the case of a pure molecular $X(3872)$ generated within the scattering length  approximation~\cite{Braaten.141116.3,Braaten}.
In this case it is true that interference effects could be more important, around a 60\% of the direct term at the resonance mass
according to Refs.~\cite{voloshin.181116.1,hanhart.170310.1}.
 Nonetheless, we are interested in the $D^0\bar{D}^{*0}$ production above its threshold for which the effect at the peak of the $X(3872)$
is reduced by the remarkably narrow Lorentzian associated with the $D^{*0}$ resonance for physical energies (${\cE'}>0$), cf. Eq.~(25).
 In addition we also have the contribution from the $D^0\bar{D}^{*0}$ production above threshold,
which is of similar size as the former for the $X(3872)$ signal region  in the pure molecular bound-state case, as we have checked.
 For this case  we then expect to do an error estimated to  be smaller than a 30\%, already of  similar
size as the experimental error, which can be easily accounted for by a renormalization about the same amount
of the normalization constant multiplying the signal contribution.}
  Then, we first consider the diagonal processes and take for definiteness the chain of decays $B\to K X(3872)$, $X(3872) \to \bar{D}^0 D^{*0}$
and finally $D^{*0}\to D^0\pi^0$.
The resulting decay width is denoted by $\gamma_{X\to \bar{D}^0D^0\pi^0}(Q^2)$, which should be multiplied
 by 2 to have the corresponding partial-decay width, $\Gamma_{X\to D^0 \bar{D}^0 \pi^0}(Q^2)$, in the limit in which we can
 neglect the aforementioned interference.
 Analogous steps would apply to the decay $B\to K D^0\bar{D}^{0}\gamma$ through the $X(3872)$.

Due to the closeness of $M_X$ and $M_{D^0}+M_{D^{*0}}$ one cannot neglect the $Q^2$ dependence of $\Gamma_{X\to D^0\bar{D}^0\pi^0}(Q^2)$ in Eq.~\eqref{191116.13}
as noticed at the end of Sec.~\ref{subsec.191116.1}. All the factors on the right-hand side of Eq.~\eqref{191116.4} are Lorentz invariant and we
evaluate it in the $X(3872)$ rest frame, where one finds the expression (with $F=D^0\bar{D}^0\pi^0$)
\begin{align}
\label{191116.15}
\Gamma_{X\to \bar{D}^0D^0\pi^0}(Q^2)&=\frac{\hat{\beta}^2}{2\sqrt{Q^2}}
\int(2\pi)^4 \delta(\vp_{\bar{D}}+\vp_D+\vp_\pi)
\delta(\delta+E-\frac{\vp_{\bar{D}}^2}{2M_{\bar{D}^0}}-\frac{\vp_D^2}{2M_{D^0}}-\frac{\vp_\pi^2}{2M_{\pi^0}})
\frac{d^3p_D}{(2\pi)^32E_D}\frac{d^3p_{\bar{D}}}{(2\pi)^32E_{\bar{D}}}\nn\\
&\times \frac{d^3p_\pi}{(2\pi)^32E_\pi} \frac{p_\pi^2}{(2M_{D^{*0}})^2|E+i\frac{\Gamma_*}{2}-\frac{\vp_{\bar{D}}^2}{2\mu}|^2}~.
\end{align}
Several points need be discussed concerning this equation.
 We have explicitly indicated the potentially most rapidly varying kinematical facts in the decay $X\to F$ that comprises
the $D^{*0}$ propagator and the $P$-wave character of $D^{*0}\to D^⁰\pi^0$,
which implies  the appearance of the momentum squared of the pion.\footnote{At the end
$p_\pi\simeq \sqrt{2M_{\pi^0}\delta}$ because $\delta\gg |E_X|$, and it could be re-absorbed in $\hat{\beta}^2$ of Eq.~\eqref{191116.15}.}
In Eq.~\eqref{191116.15} we have indicated by $\hat{\beta}^2$ a coupling constant squared,
by $\mu$ the $\bar{D}^0 D^{*0}$ reduced mass ($\mu=M_{D^0} M_{D^{*0}}/(M_{D^0}+M_{D^{*0}}$) and by $\Gamma_*$ the $D^{* 0}$ width.
 We have used the non-relativistic reduction  for the energies of the
$D^0$, $\bar{D}^0$ and $\pi^0$,  as mass plus kinetic energy, in the Dirac delta function for the conservation of energy.
This is also quite valid for the pion because $\delta\ll M_{\pi^0}$.
The non-relativistic expression is used for the $D^{*0}$ propagator as well.
Let us see how it emerges from its relativistic form:
\begin{align}
\label{191116.16}
p_{D^*}^2-(M_{D^{*0}}-i\frac{\Gamma_*}{2})^2=
(M_{D^{*0}}+E-\frac{\vp_{\bar{D}}^2}{2M_{\bar {D^0}}})^2-\vp_{\bar{D}}^2-(M_{D^{*0}}-i\frac{\Gamma_*}{2})^2~,
\end{align}
where we have employed the non-relativistic expression
for the energy of the $\bar{D}^{0}$ and that in the rest frame of the $X(3872)$, $\vp_{\bar{D}}+\vp_{D^*}=0$.
Neglecting quadratic terms in $E$, kinetic energies
and $\Gamma_*$ we are lead to the expression for the $D^{*0}$ propagator used in Eq.~\eqref{191116.15}.
Next, we insert in this equation the integral identity
\begin{align}
\label{201116.1}
1&=\int (2\pi)^4 \delta(\vp_{D^*}-\vp_D-\vp_\pi)
\delta( \delta + {\cE} + \frac{\vp_{D^*}^2}{2M_{D^{*0}}} - \frac{\vp_{D}^2}{2M_{D^0}} - \frac{\vp^2_\pi}{2M_{\pi^0}} )
\frac{d^3p_{D^*}}{(2\pi)^3}\frac{d{\cE}}{2\pi}
\end{align}
that corresponds to an intermediate $D^{*0}$ with  three-momentum $\vp_{D^*}$ and energy
$E_{D^*}=M_{D^{*0}}+\vp_{D^*}^2/2M_{D^{*0}}+{\cE}$. In this way we are explicitly extracting the phase space factor corresponding
to the final $D^0\pi^0$ in the $D^{*0}$ decay, similarly as done
above in Eqs.~\eqref{191116.1} and \eqref{191116.2} for the $X(3872)$ resonance and the subsystem $F$.
We use this result to rewrite Eq.~\eqref{191116.15} as
\begin{align}
\label{201116.2}
&\gamma_{X\to \bar{D}^0D^0\pi^0}(Q^2)=\frac{\hat{\beta}_1^2}{2\sqrt{Q^2}}
\int(2\pi)^4 \delta(\vp_{\bar{D}}+\vp_{{D^*}})
\delta(E-{\cE}-\frac{\vp_{\bar{D}}^2}{2\mu})
\frac{d^3p_{\bar{D}}}{(2\pi)^32E_{\bar{D}}} \frac{d^3p_{D^*}}{(2\pi)^3 }\frac{d{\cE}}{2\pi}
\frac{1}{2M_{D^{*0}}|E+i\frac{\Gamma_*}{2}-\frac{\vp_{\bar{D}}^2}{2\mu}|^2}\nn\\
&\times
\frac{\hat{\beta}_2}{2M_{D^{*0}}}\int (2\pi)^4 \delta(\vp_{D^*}-\vp_D-\vp_\pi)
\delta( \delta + {\cE} + \frac{\vp_{\bar{D}}^2}{2M_{D^{*0}}} - \frac{\vp_D^2}{2M_{D^0}} - \frac{\vp^2_\pi}{2M_{\pi^0}})
p_\pi^2 \frac{d^3p_D}{(2\pi)^32E_D}\frac{d^3p_\pi}{(2\pi)^32E_\pi}
 ~,
\end{align}
where we have split $\hat{\beta}=\hat{\beta}_1\hat{\beta}_2$,
such that the term on the right-hand side of the last line in the previous equation
can be identified with the partial-decay width  $D^{*0}\to D^0\pi^0$ at rest, which we denote as $\Gamma_{D^{* 0}\to D^0\pi^0}$.
As in Eq.~\eqref{191116.4} we have that this partial-decay width should
be  strictly evaluated at the corresponding $D^{*0}$ invariant mass.
However,  since the $X(3872)$ is so close to the $D^0\bar{D}^{*0}$ threshold and $\sqrt{Q^2}\simeq M_X^2$ we can
simply take  the invariant mass of the $D^{*0}$ to be equal to $M_{D^{*0}}$, which furthermore has a tiny width.

Regarding the factor on the right-hand side of the first line in Eq.~\eqref{201116.2} the integration over $\vp_{D^*}$ and $\vp_{\bar{D}}$ are straightforward, and then we are left with
\begin{align}
\label{201116.3}
\frac{\hat{\beta}_1^2\mu^{\frac{3}{2}}}{4\pi\sqrt{2Q^2}M_{D^{*0}}}
\int_{-\infty}^E
\frac{d{\cE}}{2\pi}
\frac{\sqrt{E-{\cE}}}{E_{\bar{D}}({\cE}^2+\frac{\Gamma_*^2}{4})}~,
\end{align}
where $|\vp_{\bar{D}}|=\sqrt{2\mu(E-{\cE})}$. The integration in the previous equation is a convergent one
 within a range that gives rise to  tiny kinetic energies compared to $M_{\bar{D}}$  in Eq.~\eqref{201116.3}.
 Then, we can just keep the dominant $|\vp_{\bar{D}}|$ dependence that
stems from the factor $\sqrt{E-{\cE}}$ in the numerator of the integrand and replace $E_{\bar{D}}$ by $M_{D^{0}}$.
In terms of the new integration variable ${\cE}'$, defined as
\begin{align}
\label{211116.1}
{\cE}'=E-{\cE}~,
\end{align}
 the integral is now
\begin{align}
\label{201116.3b}
\frac{\hat{\beta}_1^2\mu^{\frac{3}{2}}}{4\pi\sqrt{2Q^2}M_{D^{*0}}M_{D^0}}
\int_0^{+\infty}
\frac{d{\cE}'}{2\pi}
\frac{\sqrt{ {\cE}' }}{({\cE}'-E)^2+\frac{\Gamma_*^2}{4}}
&=\frac{\hat{\beta}_1^2\mu^{\frac{3}{2}}}{8\pi\sqrt{Q^2}M_{D^{*0}}M_{D^0}\Gamma_*}
\sqrt{E+\sqrt{E^2+\Gamma_*^2/4}}\nn\\
&=\frac{\Gamma_{X\to \bar{D}^0D^{* 0}}}{\Gamma_*}\frac{ \sqrt{E+\sqrt{E^2+\Gamma_*^2/4}} }{\sqrt{E_X+\sqrt{E_X^2+\Gamma_*^2/4}} }~,
\end{align}
with
\begin{align}
\label{170312.1}
\Gamma_{X\to \bar{D}^0D^{* 0}}&=
\frac{\hat{\beta}_1^2\mu^{\frac{3}{2}}\sqrt{E_X+\sqrt{E_X^2+\Gamma_*^2/4}}}{8\pi\sqrt{Q^2}M_{D^{*0}}M_{D^0}}
~.
\end{align}
To get the total partial-decay width of the $X(3872)$ into $D^0\bar{D}^0\pi^0$ we still have to  multiply Eq.~\eqref{201116.3b} by 2
  because of the two mechanisms involved, $X(3872)\to \bar{D}^0 D^{*0}$ (the one explicitly
analyzed) and  $X(3872)\to D^0\bar{D}^{*0}$. As argued above we are neglecting interference effects.
 Then we end from Eqs.~\eqref{191116.13}, \eqref{201116.2} and \eqref{201116.3b} with the following expression
for the partial-decay rate $B\to K D^0\bar{D}^0\pi^0$
\begin{align}
\label{201116.4}
\frac{d\Gamma_{B\to KD^0\bar{D}^0\pi^0}}{dE}&={\cal B}_{D\pi}\Gamma_B\Gamma_X\frac{\sqrt{E+\sqrt{E^2+\Gamma_*^2/4}}}{\sqrt{E_X+\sqrt{E_X^2+\Gamma_*^2/4}}}
\frac{|{\ct}(E)|^2}{2\pi|\alpha|^2}~,
\end{align}
where ${\cal B}_{D\pi}=\Gamma_{B\to KX}\Gamma_{X\to D^0\bar{D}^{*0}}\Gamma_{D^{*0}\to D^0\pi^0}/\Gamma_B\Gamma_X\Gamma_*$~.
This equation was already derived in Ref.~\cite{Braaten} and its main characteristic energy dependence
 found before in Ref.~\cite{Braaten.141116.3}.

However, what is measured experimentally is the $\bar{D}^0D^{*0}$ invariant mass \cite{BaBarD,BelleD},
which is given by $|\vp_{\bar{D}}|^2/2\mu$ when measured with respect to the $D^0\bar{D}^{*0}$ threshold in the $X(3872)$ rest frame.
 As indicated above because of the energy conservation Dirac delta function in the first line of Eq.~\eqref{201116.2} this quantity
is equal to ${\cE'}$. Thus, instead of the differential rate $d\Gamma_{B\to K D^0\bar{D}^0\pi^0}/dE$ we should compare the experimental data with
$d\Gamma_{B\to K D^0\bar{D}^0\pi^0}/d{\cE}'$.
 The latter can be calculated from Eq.~\eqref{201116.3b} by removing the integration in ${\cE}'$, and replacing
$\frac{\hat{\beta}_1^2\mu^{\frac{3}{2}}}{8\pi\sqrt{Q^2}M_{D^{*0}}M_{D^0}\Gamma_*}$ in terms of $\Gamma_{X\to \bar{D}^0D^{* 0}}$, cf.~Eq.~\eqref{170312.1}.
 The result is  multiplied by $\Gamma_{D^*\to D^0\pi^0}$, which is present
in Eq.~\eqref{201116.2}, and by 2 because of the two ways of decay involved.
 This is then placed in Eq.~\eqref{191116.13} instead of $\Gamma_{X\to F}$, which is then integrated with respect to $E$.
 We end with,
\begin{align}
\label{170312.2}
\frac{d\Gamma_{B\to K D^0\bar{D}^0\pi^0}}{d{\cE}'}&=\frac{\Gamma_{B} {\cal B}_{D\pi} \sqrt{\cE'}}{\sqrt{2}\pi \sqrt{E_X+\sqrt{E^2_X+\Gamma_*^2/4}}}
\int_{-\infty}^{+\infty}dE\frac{\Gamma_*}{({\cE}'-E)^2+\frac{\Gamma_*^2}{4}} \frac{\Gamma_X|{\ct}(E)|^2}{2\pi|\alpha|^2}~.
\end{align}

This expression coincides with the one already deduced in Ref.~\cite{Braaten}.
 However, our derivation proceeds in a more straightforward manner by having split
the $D^0\bar{D}^{0}\pi^0$ phase space factor in two terms of lower dimensionality
\cite{pdg.181116.2}, attached to the decays $X(3872)\to \bar{D}^0D^{*0}$ and $D^{*0}\to D^0\pi^0$, employing Eq.~\eqref{201116.1}.
In this way the variable ${\cE'}$ enters directly into the formulae.

Analogous steps can be followed to derive the corresponding expression for
$d\Gamma_{B^+\to K^+ D^0\bar{D}^0\gamma}/d{\cE}'$ and
when summed to Eq.~\eqref{170312.2} we have
\begin{align}
\label{211116.2}
\frac{d\Gamma_{B^+\to K^+ D^0\bar{D}^{*0}}}{d{\cE}'}&=\frac{\Gamma_{B^+} \BBD \sqrt{\cE'}}{\sqrt{2}\pi \sqrt{E_X+\sqrt{E^2_X+\Gamma_*^2/4}}}
\int_{-\infty}^{+\infty}dE\frac{\Gamma_*}{({\cE}'-E)^2+\frac{\Gamma_*^2}{4}} \frac{\Gamma_X|{\ct}(E)|^2}{2\pi|\alpha|^2}~.
\end{align}
with
\begin{align}
\label{170312.3}
\BBD&=\frac{\Gamma_{B\to K X}\Gamma_{X\to D^0\bar{D}^{*0}}\left(\Gamma_{D^{*0}\to D^0\pi^0}+\Gamma_{D^{*0}\to D^0\gamma}\right)}{\Gamma_B\Gamma_X\Gamma_{*}}\nn
\\
&=\frac{\Gamma_{B\to K X}\Gamma_{X\to D^0\bar{D}^{*0}}}{\Gamma_B\Gamma_X}~.
\end{align}
The last equality follows  by taking a 100\% branching ratio for the partial decay width
of  a $D^{*0}$ into $D^0\pi^0$ and $D^0\gamma$ \cite{pdg.181116.2}.


\section{Final-state interactions}
\label{sec.211116.1}

 As discussed above in the Introduction the applicability of the ERE (and hence of a Flatt\'e parameterization as well)
to study near-threshold  resonances,  their properties and nature
 \cite{Weinberg.141116.6,Hanhart.141116.4,Braaten.141116.3,Braaten,ZhengHQ,Kalashnikova.141116.2,Qiang.141116.1},
 could be severely limited by the presence of a nearby zero in the partial-wave amplitude.

This interplay between a resonance and a close zero indeed recalls the situation with the presence of the Adler zero
required by chiral symmetry in the isoscalar scalar pion-pion ($\pi\pi$)
scattering and the associated $\sigma$ or $f_0(500)$ resonance.
 The presence of this zero distorts strongly the $f_0(500)$ resonance signal in $\pi\pi$ scattering while for several
 production processes this zero is not required by any fundamental reason and it does not show up. This is  why the
$f_0(500)$ resonance could be clearly observed experimentally with high statistics significance
in $D\to \pi\pi\pi$ decays \cite{aitala.211116.3}, where the $S$-wave $\pi\pi$
final-state interactions are mostly sensitive to the pion scalar form factor which is free of any low-energy zero,
see e.g. Refs.~\cite{oller.211116.1,bugg.211116.2,gardner.211116.4}  for related discussions.

Regarding the $X(3872)$ there are data on event distributions involving the $J/\psi$ \cite{CDF.211116.5,LHCb.211116.6,BaBarJ,BelleJ} that show
a clean event-distribution signal for this resonance without any distortion caused by a zero. However, this does not
 exclude that a zero could  be relevant for the near-threshold $D^0\bar{D}^{*0}$ scattering, as it indeed happens for the $f_0(500)$ case.
Of course, the situation is not completely analogous because here the $X(3872)$ is almost on top of the $D^0\bar{D}^{*0}$ threshold and it has a very
small width, while the $f_0(500)$ is wide and one does use the ERE to study it because it is too far away from the $\pi\pi$ threshold. This implies
 that a CDD pole in the present problem on $D^0\bar{D}^{*0}$ scattering must be really close
to its threshold so as to spoil the applicability of the ERE.

\begin{figure}
\begin{center}
\includegraphics[width=0.4\textwidth]{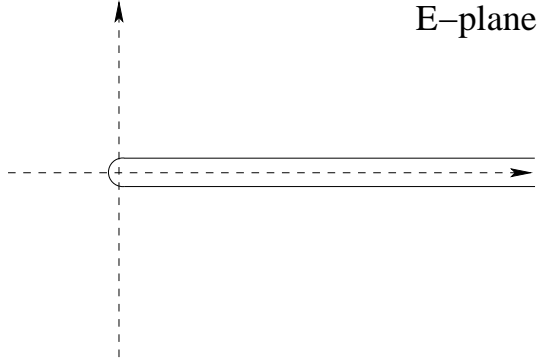}
\caption{Integration contour used for the dispersion relation of $t(E)^{-1}$ giving rise to Eq.~\eqref{211116.3}.
The contour  is closed by a circle centered at the origin and of infinite radius.}
\label{fig.170314.1}
\end{center}
\end{figure}

In this way, instead of using the ERE  as in Refs.~\cite{Hanhart.141116.4,Braaten.141116.3,Braaten,ZhengHQ,Kalashnikova.141116.2,Qiang.141116.1}
 we employ another more general parameterization that comprises the ERE up to the effective-range contribution (indeed up to the next shape parameter)
 for some limiting case but
at the same time it is also valid even in the presence of a  near-threshold CDD pole.
 This parameterization can be deduced by making use of the $N/D$ method as done in Ref.~\cite{oller.211116.5}, whose non-relativistic reduction
is given in Ref.~\cite{oller.211116.6}. The point is to perform a dispersion relation of the inverse of the $D^0\bar{D}^{*0}$ $S$-wave $t(E)$
which along the unitarity cut  fulfills the well-known unitarity relation
\begin{align}
\label{170314.1}
\text{Im}t(E)^{-1}&=-k(E)~,~E\geq 0~,
\end{align}
where $E$ is the center of mass (CM) energy of the system, cf. Eq.~\eqref{191116.9}, and $k(E)$ is the CM tree-momentum given by
its non-relativistic reduction $k(E)=\sqrt{2\mu E}$.
 Next, we neglect crossed-channel dynamics  based on the fact that the scale $\Lambda$ associated with the massless one-pion exchange
potential, as worked out in Refs.~\cite{artoi.170310.1,pavon.170314.1}, is $\Lambda=4\pi f_\pi^2/\mu g^2\sim 350$~MeV ($f_\pi=92.4$~MeV and $g\simeq 0.6$),
which is much bigger than the $D^0\bar{D}^{*0}$ three-momentum ($\lesssim 30$~MeV) in the region of the $X(3872)$.
 In this estimate one takes into account that
the denominator in the exchange of a $\pi^0$ of momentum $\vq$ between $D^{* 0}$ and $D^0$ is $\vq^2+M_{\pi^0}^2-(M_{D^{*0}}-M_{D^0})^2$ and
that   $((M_{D^{*0}}-M_{D^0})^2-M_{\pi^0}^2)/ M_{\pi^0}^2\simeq 0.1\ll 1$ because  $M_{D^{*0}}-M_{D^0}$ is larger than $M_{\pi^0}$ by only 7.2~MeV
 \cite{Suzuki.251116.1}.
 It is then appropriate to write down a dispersion relation for $t(E)^{-1}$ with at least one necessary subtraction
employing the integration contour  of Fig.~\ref{fig.170314.1}.
Then allowing for the presence of a pole of $t(E)^{-1}$ we then obtain
\begin{align}
\label{211116.3}
t(E)=&\left(\frac{\lambda}{E-M_{\rm CDD}}+\beta-ik(E)\right)^{-1}~,
\end{align}
 with  $\MCDD$ the position of the CDD pole measured with respect to the
$D^0\bar{D}^{*0}$ threshold. Notice that this is a pole in $t(E)^{-1}$ and then a zero of $t(E)$ at $E=\MCDD$.

Since the finite width effects of the $D^{*0}$ could be important as argued in Sec.~\ref{subsec.191116.2}, the
CM three-momentum $k(E)$ is finally calculated according to the expression
\begin{align}
\label{211116.4}
k(E)=\sqrt{2\mu (E+i\frac{\Gamma_*}{2})}~.
\end{align}
  For definiteness the three-momentum $k(E)$ is always defined in the 1st Riemann sheet (RS),
so that the phase of the radicand is taken between $0$ and $2\pi$.
Here  an analytical extrapolation in the mass of the $D^{* 0}$ resonance until its pole position $M_{D^{*0}}-i \Gamma_*/2$
has been performed, as also done e.g. in Refs.~\cite{Braaten,alvarez.051216.1}.
By considering explicitly the three-body channel $D^0\bar{D}^0\pi^0$  in a coupled-channel formalism, 
Ref.~\cite{hanhart.170310.1} found that  Eq.~\eqref{211116.4} is appropriate because of the smallness
of the $P$-wave $D^{* 0}$ width into $D^0\pi^0$,  which implies that $\Gamma_* /2\delta=4.5\, 10^{-3}\ll 1$.
 In Eq.~\eqref{211116.3}  the constant $\beta$ for elastic $D\bar{D}^{*0}$ scattering is real but
 it becomes complex, with negative imaginary part, when  taking  into account
inelasticities from other channels, such as $J/\psi \pi^+\pi^-$, $J/\psi \pi^+\pi^-\pi^0$, etc \cite{Hanhart.141116.4,Braaten.141116.3,Braaten}.
We finally fix this possible imaginary part in $\beta$  to zero because, as already noticed in Ref.~\cite{Braaten},
 one can reproduce data equally well, as we have also checked.

An ERE of $t(E)$ given in Eq.~\eqref{211116.3} is valid in the $k^2$ complex plane with a radius of convergence coincident with $2\mu\MCDD$.
 Notice that a zero of $t(E)$ near threshold implies that $k\cot \delta=\infty$ at this point and then
it becomes singular. As a result, its $k^2$ expansion does not converge and the ERE becomes meaningless for practical applications since
its radius of convergence is too small.
In such a case, one must consider the full expression for $t(E)$ in Eq.~\eqref{211116.3} and not its ERE, which reads
\begin{align}
\label{061015.5}
k{\rm cot}\delta&=\frac{\lambda}{k^2/2{\mu}-M_{\rm CDD}}+\beta\nn\\
&=-\frac{\lambda}{M_{\rm CDD}}+\beta -\frac{k^2
\lambda}{2{\mu}M_{\rm CDD}^2}+\ldots\nl &=-\frac{1}{a}+\frac{1}{2}r
k^2+\ldots
\end{align}
here the ellipsis indicate higher powers of $k^2$.
 This expansion can  reproduce any values of the scattering length and effective range  (as well as of the next shape parameter $v_2$) and
we obtain the expressions\footnote{Note the different sign convention for the scattering length here as compared with Ref.~\cite{oller.211116.6}.}
\begin{align}
\label{061015.6}
\frac{1}{a}=&\frac{\lambda}{M_{\rm CDD}}-\beta~,\nn\\
r=&-\frac{\lambda}{{\mu}M_{\rm CDD}^2}~.
\end{align}
It is then clear that in order to generate a large absolute value for $a$, one needs a strong cancellation between
 $\lambda/M_{\rm CDD}$ and $\beta$ unless both of them are separately small. But in order to have a small magnitude of $|a|$ and a large one for $|r|$,
one would naturally expect that $M_{\rm CDD}\to 0$, though the explicit value of $\lambda$ plays also an important role.
Equation \eqref{061015.6} clearly shows why the ERE could fail to converge even for very small values of $|k|^2$ as long as
$ M_{\rm CDD}\to 0$.

In the limit $\MCDD \to \infty$ with $\lambda/\MCDD$ fixed, the parameterization in Eq.~\eqref{211116.3} for $t(E)$ reduces to the function
$f(E)$ used in Refs.~\cite{Braaten,Braaten.141116.3}
\begin{align}
\label{211116.6}
t(E)&\underset{{\small \begin{array}{l}\MCDD\to \infty\\ \lambda/\MCDD=ct.\end{array}}}{\xrightarrow{\hspace*{1.8cm}}}
f(E)=\frac{1}{-\frac{\lambda}{\MCDD}+\beta-ik(E)} \equiv \frac{1}{-\gamma-ik(E)}
\end{align}
where $\gamma=1/a$ is the inverse of the scattering length, using the notation of Ref.~\cite{Braaten}. The function $f(E)$ has a bound(virtual) state
pole for positive(negative) $\gamma$.

While the near-threshold energy dependence of $f(E)$ is dominated by the threshold branch-point singularity  and a possible
low-energy pole associated with a bound- or virtual state,
 this is not necessarily the case for $t(E)$ as long as $\MCDD$ is small enough. In such a case one has to
explicitly remove the CDD pole from  $t(E)$ by dividing it by $E-\MCDD$. In this way, we end with the new function
${\ct}(E)$, already introduced in Sec.~\ref{subsec.191116.1} just before Eq.~\eqref{191116.11b}, which is then defined as
\begin{align}
\label{211116.7}
{\ct}(E)=&\frac{1}{1+\frac{E-\MCDD}{\lambda}(\beta-ik)}~,
\end{align}
such that its low-energy behavior is qualitatively driven by the same facts mentioned for $f(E)$. This is also the function that in general
terms drives final-state interactions (FSI) when the scattering partial-wave is given by $t(E)$ in Eq.~\eqref{211116.3}.
A detailed account on it can be found in Ref.~\cite{oller.211116.6}, although Ref.~\cite{bugg.211116.2} could be more accessible
depending on the reader's taste and education.

Next, we explicitly calculate the residue $\alpha$ for ${\ct}(E)$ needed to work out the decay rates  in Eqs.~\eqref{191116.14}
 and \eqref{211116.2} and the differential cross section of Eq.~\eqref{041216.7}.
This can be straightforwardly determined by moving to the pole position as defined in Eq.~\eqref{191116.11b}.
Thus, it results
\begin{align}
\label{211116.8}
\alpha&=\frac{\xi}{1+\xi\eta}
\end{align}
with
\begin{align}
\label{211116.8b}
\xi&=-i\frac{k_P(\beta-ik_P)}{\mu}~,\\
\eta&=\frac{\beta-ik_P}{\lambda}~.\nn
\end{align}
The three-momentum $k_P$ is evaluated at the pole position $E_P$ in the energy plane,
\begin{align}
\label{211116.9}
k_P&=\sqrt{2\mu(E_P+i\frac{\Gamma_*}{2})}~,
\end{align}
such that the phase of the radicand is between $[0,2\pi[$ for a bound-state pole in the 1st RS,
 while for a pole in the 2nd RS, a virtual-state one,  the phase is between $[2\pi,4\pi[$ and the sign of $k_P$
 is reversed compared to its value in the 1st RS.

 The constant $\alpha$, in the case of using the function $f(E)$ in Eq.~\eqref{211116.6} for the decay rates in
Eqs.~\eqref{191116.14} and \eqref{211116.2}, is defined analogously as the residue of $f(E)$ at the pole position $P_X$.
 The function $f(E)$ has a different normalization compared to ${\ct}(E)$, and $\alpha$  is  then given by
\begin{align}
\label{281116.1}
\alpha&=\frac{ik_P}{\mu}=-\frac{\gamma}{\mu}~.
\end{align}
Here we have taken into account that $k_P=i\gamma$ for the $f(E)$ parameterization.

 The limit of decoupling a bare resonance from a continuum channel, like $D^0\bar{D}^{*0}$, requires the presence of a zero
to remove the pole of the resonance from $t(E)$. This simple argument shows that CDD poles and weakly coupled bare resonances are typically related.
In this respect, we consider the resulting $t(E)$ obtained in Ref.~\cite{baru.170310.2} by considering the interplay between mesonic and quark
degrees of freedom, and that results by considering the exchange of a bare resonance together with direct scattering terms in the
mesonic channel at the level of the scattering length approximation. In the following discussions until the end of this section
the zero width limit of the $D^{* 0}$ should be understood in $k(E)$.
 The resulting $S$-wave amplitude from Ref.~\cite{baru.170310.2} is\footnote{A minus sign is included due to the different convention in
Ref.~\cite{baru.170310.2}.}
\begin{align}
 \label{170314.2}
t(E)&=-\frac{1}{4\pi^2\mu}\frac{E-E_f+\frac{1}{2}g_f \gamma_V}{(E-E_f)(\gamma_V+ik)+i\frac{i}{2}g_f\gamma_V k}
\end{align}
Here, $a_V$ is the scattering length for the direct $D^0\bar{D}^{*0}$ scattering (referred as  potential scattering in Ref.~\cite{baru.170310.2}),
$\gamma_V=1/a_V$, $g_f$ is the coupling squared between the bare resonance and the mesonic channels, while $E_f$ is the mass of the bare resonance in the
decoupling limit $g_f\to 0$. By comparing $t(E)$ in Eq.~\eqref{170314.2} with our expression above Eq.~\eqref{211116.3} one has the following
relation between parameters
\begin{align}
\label{170314.3}
\beta&=-\gamma_V~,\\
\lambda&=\frac{1}{2}g_f\gamma_V^2~,\nn\\
M_{\rm CDD}&=E_f-\frac{1}{2}g_f\gamma_V~,\nn
\end{align}
which  shows that the results of Ref.~\cite{baru.170310.2} are a particular case of ours, since it is always possible to adjust
$\lambda$, $M_{\rm CDD}$ and $\beta$ in terms of $g_f$, $E_f$ and $\gamma_V$. However, the reverse is not true because
  $g_f\geq 0$ \cite{baru.170310.2}, which implies that $\lambda$ is restricted to be positive as well,
 while the residue of the CDD pole can have a priori any sign.
This difference is also important phenomenologically because, while our parameterization for $t(E)$ can give rise to values of the
 effective range with any sign,  Ref.~\cite{baru.170310.2} generates only negative ones, cf. Eq.~\eqref{061015.6}.
  Equation~\eqref{170314.3} explicitly shows the above remark that  $E_f\to M_{\rm CDD}$
in the decoupling limit, $g_f\to 0$, with both $g_f$ and $M_{\rm CDD}-E_f$ being proportional to the residue of the CDD pole.
 It is also interesting to notice that $\beta$ corresponds to the minus the inverse of the potential scattering length $a_V$.
 The language of the exchange of a bare resonance and direct $D^0\bar{D}^{*0}$ scattering
could be more intuitive in some aspects than the direct use of $S$-matrix theory,
employed to obtain Eq.~\eqref{211116.3}, so that we will make contact with the former when discussing  our findings.
The formalism of Ref.~\cite{baru.170310.2} was extended to coupled channels in Ref.~\cite{kalas.170310.1}, and the inclusion of
inelastic channels was also addressed more recently in Ref.~\cite{guo.170316.1}.

The  scattering length approximation for the $D^0\bar{D}^{*0}$ $S$ wave  of Refs.~\cite{Braaten.141116.3,Lu.170314.1}
was further generalized in Ref.~\cite{artoi.170310.1} to include  as well
the exchange of one bare-resonance together with the explicit coupling between the channels $D^0\bar{D}^{* 0}$ and
$D^+\bar{D}^{*-}$. The expression obtained in Ref.~\cite{artoi.170310.1} for the elastic $D^0\bar{D}^{*0}$ $S$-wave amplitude is
\begin{align}
\label{170316.1}
t_{11}(E)&=\frac{(-\gamma_1-\gamma_0+2\kappa_2)(E-\nu+g^2\gamma_0)+g^2\gamma_0^2}{D(E)}\\
D(E)&=(2\gamma_1\gamma_0-(\gamma_0+\gamma_1)(\kappa_2-ik)-i2\kappa_2k)(E-\nu+g^2\gamma_0)\nn\\
&+g^2\gamma_0^2(-2\gamma_1+\kappa_2-ik)\nn
\end{align}

Here, $\kappa_2=\sqrt{2\mu({\Delta}-E-i0^+)}$, where ${\Delta}$ is the difference between the thresholds
of $D^-D^{*+}$ and $\bar{D}^0D^{*0}$. Additionally,   $\gamma_{0,1}$ are the isoscalar and isovector scattering lengths
 in the limit of decoupling the bare state with the continuum channels   and $g$ is the coupling among them.
The parameter  $\nu$ is the mass of the bare state measured with respect to the lightest threshold.
To match Eq.~\eqref{170316.1} in the near-threshold region with the expression for $t(E)$ in Eq.~\eqref{211116.3} we
rewrite the former as
\begin{align}
\label{170314.4}
t(E)&=\left(-ik+
\frac{(E-\nu+g^2\gamma_0^2)(2\gamma_0\gamma_1-(\gamma_0+\gamma_1)\kappa_2)-g^2\gamma_0^2(2\gamma_1-\kappa_2)}
{(E-\nu+g^2\gamma_0^2)(-\gamma_1-\gamma_0+2\kappa_2)+g^2\gamma_0^2}\right)^{-1}~,
\end{align}
which explicitly shows the correct form to fulfill elastic unitarity below the $D^+D^{*-}$ threshold, so that the
term involving the product  $\kappa k$ in Eq.~\eqref{170316.1} has disappeared.
 Restricting ourselves to our region of interest, $|E|\ll {\Delta}$, we can perform a Taylor expansion of
$\kappa_2$ around $E=0$ and keep only its leading term  $\kappa_2\to \hat{\kappa}_2=\sqrt{2\mu \Delta}$,
so that all the energy dependence of $t(E)^{-1}$  is dominated by the CDD pole and the right-hand cut for elastic scattering,
as in our derivation of Eq.~\eqref{211116.3}.\footnote{If higher orders are kept in the Taylor expansion of $\kappa_2$ around $E=0$
then the matching would require to include more CDD poles (with contributions suppressed by powers of $E/\Delta$),
see Ref.~\cite{oller.211116.5} for details. Another option is to
follow the coupled-channel formalism there developed as well, particularly when considering a wider energy range.}
 The explicit expressions of $\lambda$, $\beta$ and $M_{\rm CDD}$ as a function of the parameters $\gamma_{0,1}$, $\nu$, $g^2$ and
$\nu$ in Eq.~\eqref{170314.4} are:
\begin{align}
\label{170316.2}
\lambda&=\frac{2 g^2\gamma_0^2(\gamma_1-\hat\kappa_2)^2}{(\gamma_0+\gamma_1-2\hat\kappa_2)^2}\\
M_Z&=\nu-\frac{g^2\gamma_0(\gamma_1-2\hat\kappa_2)}{\gamma_0+\gamma_1-2\hat\kappa_2}\nn\\
\beta&=\frac{-2\gamma_0\gamma_1+(\gamma_0+\gamma_1)\hat\kappa_2}{\gamma_0+\gamma_1-2\hat\kappa_2}\nn
\end{align}

The same comment as performed above, concerning the  non-fully equivalence between our parameterization and the one of
Ref.~\cite{baru.170310.2}, is also in order here regarding Eq.~\eqref{170316.1}.
 The point is that the latter implies again from Eq.~\eqref{170316.2}
that $\lambda\geq 0$ while the residue of a CDD pole can have any sign.

\section{Formulae for the event distribution}
\label{formalism.141116.1}

The combination $|{\ct}(E)|^2 \Gamma_X/2\pi|\alpha|^2$ in Eqs.~\eqref{191116.14}, \eqref{041216.7} and \eqref{211116.2}
corresponds to the  normalized standard non-relativistic mass distribution for a narrow resonance or bound state
(taking in this last case $\Gamma_X\to 0$).
We then define this combination as the spectral function involved in the energy-dependent event distributions
\begin{equation}
\label{231116.1}
\frac{d\hat M}{dE}=\frac{\Gamma_X|{\ct}(E)|^2}{2\pi|\alpha|^2}~,
\end{equation}
with the same expression replacing ${\ct}(E)$ by $f(E)$ if the latter function is used \cite{Braaten}.
 The normalization integral is defined as
\begin{align}
\label{231116.2}
{\cal N}&=\int_{-\infty}^{+\infty}\!\! dE\, \frac{d\hat{M}}{dE} ~,
\end{align}
which is equal to one for the cases mentioned before. However, this is not the case when $E_P$ corresponds to a virtual state or
other situations for which the final-state interaction function ${\ct}(E)$ has a shape that
strongly departs from a non-relativistic Breit Wigner (which also includes a Dirac delta function in the limit $\Gamma_X\to 0$).
 When using  $f(E)$ the integration in Eq.~\eqref{231116.2} does not converge. Then we take as integration interval $[2E_X,0]$
as in Ref.~\cite{Braaten}, which embraces the signal region and it is enough for a semiquantitative understanding/picture
based on the near value of ${\cal N}$ to 1 in the bound-state case.

We consider data on event distributions for $J/\psi \pi^+\pi^-$ and $D^0\bar{D}^{*0}$
from $B\to K X(3872)$ decays \cite{BaBarJ,BaBarD,BelleJ,BelleD} and inclusive $p\bar{p}$ collisions \cite{CDF.211116.5}.
 In the  $B$-decay cases the number of $B\bar{B}$ pairs produced at the $\Upsilon(4S)$
 is given and we denote it by $N_{B\bar{B}}$,  with the same amount of neutral and charged $B\bar{B}$ pairs produced.
It is also the case that the experimental papers  \cite{BaBarJ,BaBarD,BelleJ,BelleD} include the charge-conjugated decay mode to the one explicitly indicated,
 a convention followed by us too.

We perform  fits to the data on the $J/\psi \pi^+\pi^-$ event distributions from charged $B^+\to K^+ J/\psi\pi^+\pi^-$ decays
measured by the Belle \cite{BelleJ} and BaBar Collaborations \cite{BaBarJ}.
 The predicted event number $N_i$ at the $i$th bin,
with the center energy $E_i$ and bin width $\Delta$, is
given by the convolution of the decay rate in Eq.~\eqref{191116.14} times $N_{B\bar{B}}/\Gamma_{B^+}$
 with the experimental energy-resolution function $R(E',E)$, and integrating over the bin width. We divide Eq.~\eqref{191116.14} by
 $\Gamma_{B^+}$  because all the charged $B^+B^-$ pairs produced, $N_{B\bar{B}}/2$, have decayed (an integration over time of the rate of decay
is implicit. The latter is given by the product of the total width times the number of $B$ mesons decaying at a given time).
In addition, one has to multiply the signal function by the experimental efficiency $\vep_J^{(+)}$.
The resulting formula is
\bea
\label{eq:eventJpsif(E)a}
N_i=\vep^{(+)}_J N_{B\bar{B}}\BBJ \int_{E_i-\Delta/2}^{E_i+\Delta/2} dE'\int_{-\infty}^\infty dE R(E', E) \frac{d\hat M}{dE} +
N_{B\bar B} \cbg_J \Delta ~.
\eea
 The constant $\BBJ$ attached to the signal contribution in Eq.~\eqref{eq:eventJpsif(E)a}
 can be interpreted as the product of the double branching ratios $Br(B^+\to K^+ X)Br(X\to J/\psi \pi^+\pi^-)$
when ${\cal N}\cong 1$, cf. Eq.~\eqref{231116.2}.  In this case the product
 $\vep^{(+)}_JN_{B\bar{B}}\BBJ $ is directly the yield $Y_J$. If this is not the case this interpretation is not possible but
we still call this product in the same way, though its meaning is just that of a normalization constant.
 In this way, we re-express Eq.~\eqref{eq:eventJpsif(E)a} as
\bea
\label{eq:eventJpsif(E)}
N_i=Y_J \int_{E_i-\Delta/2}^{E_i+\Delta/2} dE'\int_{-\infty}^\infty dE R(E', E) \frac{d\hat M}{dE} +
N_{B\bar B} \cbg_J \Delta ~.
\eea
On the other hand, the background contribution is specified by the constant
 $N_{B\bar{B}} \cbg_J \Delta$, which can be  determined by simple eye inspection from the sidebands events around the $X(3872)$ signal region.
 The energy-resolution function is the Gaussian  function
\beq\label{eq:resolution}
R(E',E)=\frac{1}{\sqrt{2\pi}\sigma}\exp\left(-\frac{(E'-E)^2}{2\sigma^2}\right).
\eeq
 Following Ref.~\cite{ZhengHQ}, as also used in
Ref.~\cite{Braaten},  we take $\sigma=3$ MeV for both BaBar \cite{BaBarJ} and Belle \cite{BelleJ}
experiments on $J/\psi\pi^+\pi^-$ event distributions. We take both $\BBJ$ and $\cbg_J$ to be the same in the fits to BaBar and Belle data
because once we take into account the different $N_{B\bar{B}}$ for both experiments ($N_{B\bar B}=4.55\cdot 10^8$ for BaBar \cite{BaBarJ},
and $N_{B\bar B}=6.57\cdot 10^8$ for Belle \cite{BelleJ}, see also Table~\ref{tab:expdata}) the yields given in the experimental papers
\cite{BelleJ,BaBarJ} coincide. This means that the ratio of the parameters $Y_J$ and $\cbg_J$ for BaBar and Belle is the same as the quotient of
their respective $N_{B\bar{B}}$. Then, after fitting data we will give only the values of the resulting parameters for the former.

We also consider the  CDF  $J/\psi \pi^+\pi^-$ event distribution from inclusive $p\bar{p}$ scattering \cite{CDF.211116.5}.
 We use  Eq.~\eqref{041216.7} times the integrated luminosity ${\cal L}$,
which for Ref.~\cite{CDF.211116.5} is 2.4~fb$^{-1}$. In addition we neglect the $Q^2$ dependence except for $d(E)$ and
 after including the bin width, experimental efficiency $\vep_{J}^{(p)}$, energy resolution and background we have
\begin{align}
\label{291116.8}
N_i= \vep_J^{(p)}{\cal L}\sigma_{p\bar{p}\to X\text{All}} Br(X\to J/\psi\pi^+\pi^-) 
\int_{E_i-\Delta/2}^{E_i+\Delta/2} dE'\int_{-\infty}^\infty dE R_{p\bar{p}}(E', E)
 \frac{d\hat M(E)}{dE} + \zeta +\varrho E_i ~.
\end{align}
Here  the bin width $\Delta$ is 1.25 MeV and the background in the $X(3872)$ region
has been parameterized as a straight line (which is easily determined from the sideband events),
 following the outcome in Fig.~1 of  the CDF Collaboration paper \cite{CDF.211116.5}.
 In this reference the experimental resolution function is expressed as the sum of two Gaussians
\begin{align}
\label{291116.9}
R_{p\bar{p}}(E',E)&=\frac{2/3}{\sqrt{2\pi}\sigma_1}
\exp\left(-\frac{(E'-E)^2}{2\sigma_1^2}\right)
+\frac{1/3}{\sqrt{2\pi}\sigma_2}\exp\left(-\frac{(E'-E)^2}{2\sigma_2^2}\right)~,\nn\\
\sigma_1&=3.2~\text{MeV},\quad \sigma_2\simeq 2 \sigma_1~.
\end{align}
 Again, when ${\cal N}\approx 1$ the product $\vep_J^{(p)}\sigma_{p\bar{p}\to X\text{All}} Br(X\to J/\psi\pi^+\pi^-)$
can be directly interpreted as the yield $Y_J^{(p)}$ but, if not, we keep this notation.
We then rewrite Eq.~\eqref{291116.8} as
\begin{align}
\label{170319.1}
N_i=Y_J^{(p)}\int_{E_i-\Delta/2}^{E_i+\Delta/2} dE'\int_{-\infty}^\infty dE R_{p\bar{p}}(E', E)
 \frac{d\hat M(E)}{dE} + \zeta +\varrho E~.
\end{align}

Concerning the $D^0\bar{D}^{*0}$ event distributions from charged and neutral $B\to K X$ decays,
the $D^{*0}$ is fully reconstructed from its decay products $D^0\pi^0$ and $D^0\gamma$ in the
data from BaBar \cite{BaBarD}. In the case of Belle data \cite{BelleD} we employ the one in which  the $D^{* 0}$ is reconstructed only
from its decay product $D^0\pi^0$, because it has a much smaller background than for $D^0\gamma$.
 To reproduce the  event distributions  we employ the decay rate of Eq.~\eqref{211116.2} and take into account
 the experimental resolution, efficiency, bin width and background contributions,
similarly as done for  Eq.~\eqref{eq:eventJpsif(E)a} above. We end with the expression:
\bea
\label{eq:eventDDbarf(E)}
N_i&=&  \int_{E_i-\Delta/2}^{E_i+\Delta/2}dE' \int_0^\infty d {\cE}' R(E',{\cE}') \sqrt{ {\cE}'} \nl
&&\times\left[\frac{Y_D}{\sqrt{2}\pi\sqrt{E_X+\sqrt{E_X^2+\Gammas^2/4}}}
\int_{-\infty}^\infty dE \frac{d\hat M}{dE}\frac{\Gamma_*}{|{\cE'}-E-i\Gamma_*/2|^2}+N_{B\bar{B}}\cbg_D\right]~,
\eea
 In this equation the background contribution
 is parameterized   as $\cbg_D\sqrt{{\cE'}}$ as in Ref.~\cite{Braaten}, giving 
rise after fitting to similar background contributions as the ones in the 
experimental papers  Refs.~\cite{BaBarD,BelleD} (though they are parameterized in somewhat 
different form).
 The constant $\cbg_D$  can be easily determined from the events above the $X(3872)$
 signal region which gives rise to a rather structureless pattern.
The constant $Y_D$ can be interpreted again as a yield for ${\cal N}\approx 1$ because, when integrating in ${E}'$ over all the energy range
the signal contribution in Eq.~\eqref{eq:eventDDbarf(E)},  the denominator below $Y_D$ is cancelled because of Eq.~\eqref{201116.3b}.
 We again follow Refs.~\cite{ZhengHQ,Braaten}, as well as the Belle experimental
analysis \cite{BelleD}, and take the Gaussian width $\sigma$
in the resolution function $R(E',\cE')$, Eq.~\eqref{eq:resolution}, to be energy dependent and given by the expression
\beq
\sigma({\cE}')=\sqrt{(0.031 \text{MeV}){\cE}'}~,
\eeq
with $\cE'$ running through the values in Eq.~\eqref{eq:eventDDbarf(E)}.
For the $D^0\bar{D}^{*0}$ event distributions the number of $B\bar{B}$ pairs produced is
  $N_{B\bar B}=3.83\cdot 10^8$ for BaBar \cite{BaBarD}, and $N_{B\bar B}=6.57\cdot 10^8$ for Belle \cite{BelleD},
as also indicated in Table~\ref{tab:expdata}.
  For this case we have to take different values for the yields and background constants for fitting the BaBar and
Belle data.

In all our formulae for the event distributions in Eqs.~\eqref{eq:eventJpsif(E)}, \eqref{170319.1} and \eqref{eq:eventDDbarf(E)} the
background contribution is added incoherently because it is mostly combinatorial. This is the same treatment as performed
in the experimental papers \cite{BaBarJ,BaBarD,BelleJ,BelleD} as well as in the phenomenological analysis
\cite{Hanhart.141116.4,Braaten.141116.3,Braaten,ZhengHQ,Kalashnikova.141116.2}. In a Laurent expansion of the signal amplitude
around the $X(3872)$  non-resonant terms appear that add coherently but they are accounted for by the function $d(E)$ in the near-threshold region,
which, as discussed in Sec.~\ref{sec.211116.1}, is assumed to have the main dynamical features. Reference~\cite{artoi.170310.1}  attempts
to unveil further dynamical information from the $B\to K D^0\bar{D}^{*0}$ event distributions by considering them in a broader energy
region beyond the $X(3872)$ signal and explicitly including the $D^+D^{*-}$  channel within the formalism.
 Nonetheless, the present experimental uncertainty   avoids extracting any definite conclusion beyond the smooth background out of the
$X(3872)$ region.

 For the fitting process it is advantageous to rewrite Eqs.~\eqref{eq:eventJpsif(E)}, \eqref{170319.1} and \eqref{eq:eventDDbarf(E)}
by using directly $|d(E)|^2$ instead of $d\hat M(E)/dE$, and re-absorbing the factor $1/|\alpha|^2$ in the normalization constant.
In this form one avoids working out the dependence of $\alpha$ on the pole position, which numerically is
very convenient since it is not known a priori where the pole lies when using ${\ct}(E)$. Once the fit is done one can actually
calculate $\alpha$, cf. Eq.~\eqref{211116.8}, and from its value and the fitted constant the corresponding yield.
This technicality is discussed in more detail in the Appendix \ref{app.081016.1}.

\begin{table}[htbp]
\begin{center}
\begin{tabular}{ccc}
\hline \hline
 $B\to K X$ & BaBar \cite{BaBarD,BaBarJ} & Belle \cite{BelleD,BelleJ} \\
\multirow{2}{*}{$D^0\bar{D}^{*0}$} & $N_{B\bar B}=3.83\cdot 10^8$, $\Delta$=2 MeV & $N_{B\bar B}=6.57\cdot 10^8$, $\Delta$=2 MeV\\
{}                    & $\sigma(\Eexp)=\sqrt{(0.031\,\text{MeV}\Eexp)}$, 50 points  & $\sigma(\Eexp)=\sqrt{(0.031\,\text{MeV}\Eexp)}$, 50 points\\[2ex]
\multirow{2}{*}{$J/\psi \pi^+ \pi^-$} & $N_{B\bar B}=4.55 \cdot 10^8$, $\Delta$=5 MeV & $N_{B\bar B}=6.57\cdot 10^8$, $\Delta$=2.5 MeV\\
{}                    & $\sigma$= 3 MeV, 20 points  & $\sigma$=3 MeV, 40 points\\
\hline
$p\bar{p}$ & CDF \cite{CDF.211116.5}& {} \\
{}   & ${\cal L}$=2.4~fb$^{-1}$, $\Delta=1.25$~MeV & {}\\
{} & $\sigma_1$=3.2 MeV, $\sigma_2=2\sigma_1$, 37 points & {}\\
\hline\hline
\end{tabular}
\caption{The $B$ decays into $KJ/\psi\pi^+\pi^-$ and $D^0\bar{D}^{*0}$ channels
are both measured by the BaBar \cite{BaBarJ,BaBarD} and Belle Collaborations \cite{BelleJ,BelleD}.
 The total number of $B\bar{B}$ pairs ($N_{B\bar B}$), bin width ($\Delta$),
and the Gaussian width ($\sigma$) used in the experimental resolution function are given.
 The number of points included in the fits are also indicated.
 For the inclusive $p\bar{p}$ collision measured by the CDF Collaboration \cite{CDF.211116.5} we
account for similar parameters, but now the luminosity (${\cal L}$) is given instead of $N_{B\bar{B}}$.
 For more details see the text.}
\label{tab:expdata}
\end{center}
\end{table}

\section{Combined fits}
\label{sec:Combined}

The data sets that we include in the fits were already introduced in Sec.~\ref{formalism.141116.1}.
 A summary of their main characteristics can be found in  in Tab.~\ref{tab:expdata}.
Apart from the data on $B\to K X$ decays we also include the high statistics $J/\psi \pi^+\pi^-$ event distribution
from $p\bar{p}$ collisions  at $\sqrt{s}=1.96$~TeV, measured by the CDF Collaboration \cite{CDF.211116.5}, which also
has the smallest bin width. In this way one can reach from these data a better determination of
$E_R$. The value given by this Collaboration for the $X(3872)$ mass is $M_X=3871.61\pm 0.22$~MeV, from which we infer
$E_R=-0.20\pm 0.22$~MeV, that has a  smaller uncertainty than the one obtained in $B$ decays from Refs.~\cite{BaBarJ,BaBarD,BelleJ,BelleD}.
 From the point of view of mutual compatibility between data sets \cite{Hanhart.141116.4}
 it is also interesting to perform a simultaneous fit to all the data on 
$B^+\to K^+ J/\psi \pi^+\pi^-$, $B^+(B^0)\to K^+(K^0)D^0\bar{D}^{*0}$
and $J/\psi \pi^+\pi^-$ from inclusive $p\bar{p}$ collisions.

 Experimental data points have typically asymmetric error bars, see e.g. the data points in Fig.~\ref{fig:resultI}.
Thus, as done in Ref.~\cite{Braaten} and also in other experimental analysis,
the best values for the free parameters are determined by using the binned maximum likelihood method,
which is also more appropriate in statistics than the $\chi^2$ method  for  bins with low statistics.
 At each bin, the number of events is assumed to obey a Poisson distribution, so that the predicted event numbers in
Eqs.~\eqref{eq:eventJpsif(E)}, \eqref{170319.1} and \eqref{eq:eventDDbarf(E)} are
the corresponding mean value at the bin ($N_i$), while the experimentally measured number is
called $Y_i$ (experimental data). The Poisson distribution at each bin reads $L_i (N_i,Y_i)=N_i^{Y_i}\exp(-N_i)/Y_i!$ and the
total  probability function for a data sample is given by their products. One wants to maximize its value so that
the function to be  minimized is defined as
\beq
\label{241116.1}
-2\log(L)=-2\sum_i^n(Y_i \log(N_i)-N_i)~.
\eeq

When including a CDD pole in the expression for $t(E)$, Eq.~\eqref{211116.3}, one has to fix 3 free parameters to characterize the
interaction, namely, $\lambda$, $\MCDD$ and $\beta$. However, this is a too numerous set of free parameters to be fitted
in terms of the data taken (additionally we have the normalization and background constants). This manifests in the fact that
there are many local minima when minimizing $-2\log L$, so that it is not clear how to extract any useful information. Instead,
we have decided to consider 5 interesting and different possible scenarios (cases 1, 2.I-II and 3.I-II),
 so that each of them gives rise to an acceptable reproduction of
 the line shapes but corresponds to quite a different picture for the $X(3872)$. In addition, we also think that 
 the  pole arrangements that result in every case are worth studying for general interest on near-threshold states.  
 For every of these cases studied the number of free parameters associated with $d(E)$ is one (only case 2.II below has 2 free parameters),
so that the interaction is well constrained by data. 

We gather together similar sets of information for each scenario, so that 
the comparison between them is more straightforward.  
 The reproduction of the data fitted for all the cases is shown in 
Fig.~\ref{fig:resultI} by the black solid (case 1), red dotted (case 2.I), 
brown dashed (case 2.II), blue dash-double-dotted (case 3.I)
 and green dash-dotted (case 3.II) lines. A detailed view of the more interesting near-threshold 
region for the $D^0\bar{D}^{*0}$ event distributions is given in the histogram of 
Fig.~\ref{fig:insetDD}. We also show  separately the reproduction of the  $J/\psi \pi^+\pi^-$ event distribution 
of the CDF Collaboration data \cite{CDF.211116.5} 
in  Fig.~\ref{fig:CDFBand}. In this figure we  include the  error bands too, in order 
to show the typical size of the uncertainty in the line shapes that stems from the systematic errors in our fits.
For all the figures we follow the same convention for the meaning of the different lines.   
  The spectroscopical information is gathered together  in Table~\ref{tab:fitsummary},  
where we give from left to right the near-threshold pole positions, 
 the compositeness for the bound-state pole (if present), 
the residues to $D^0\bar{D}^{*0}$ and the yields. 
Finally, we show in Table~\ref{tab:t.fitC} the scattering parameters characterizing 
the partial wave $t(E)$ that result from the fits. In the two rightmost columns 
 we give the scattering length and the effective range.

   All fitted parameters are given in  
Eqs.~(\ref{241116.2},\ref{para:case2a},\ref{para:case2b},\ref{para:caseiiiII})  for the cases 1, 2.I-II and 3.I-II, in order. 
 The best values of the parameters are obtained by the routine MINUIT \cite{MINUIT}.
 The error for a given parameter is defined as the change of that parameter that makes
 the function value $-2\log L$
 less than $-2\log L_{\text{min}}+1$ (one standard deviation), where $-2\log L_{\text{min}}$ is the minimum
 value.

\beq \label{241116.2}
\begin{array}{lll}
\text{Case 1:}&\gamma=18.97^{+0.49}_{-0.19}~\text{MeV}~,&
\\[1ex]
\text{10 parameters}&Y_{D1}=7.49_{-0.41}^{+0.71}~,&N_{B\bar{B};D}^{\text{BaBar}}\,\cbg_{D1}=0.18^{+0.02}_{-0.02}~\text{MeV}^{-3/2}~,\\[1ex]
&Y_{D2}=6.45_{-0.47}^{+0.32}~,&N_{B\bar{B};D}^{\text{BaBar}}\,\cbg_{D2}=0.11^{+0.01}_{-0.01}~\text{MeV}^{-3/2}~,\\[1ex]
&Y_J=79.03_{-6.11}^{+5.65}~,&N_{B\bar{B};J}^{\text{BaBar}}\,\cbg_J=3.59^{+0.13}_{-0.13}~\text{MeV}^{-1}~,\\[1ex]
&Y_J^{(p)}=5.23_{-0.11}^{+0.07}\times10^3~,&\zeta=1515.44^{+7.40}_{-7.40}~\text{MeV}^{-1}~,\quad\varrho=5.23_{-0.48}^{+0.48}~\text{MeV}^{-2}~.
\end{array}
\eeq

\bea
\label{para:case2a}
\begin{array}{lll}
\text{Case 2.I:} &E_R=-0.53_{-0.11}^{+0.09}\,\text{MeV}~, &\MCDD=-12.29_{-1.14}^{+1.11}\,\text{MeV}~, \\[1ex]
\text{11 parameters:} &Y_{D1}=83.13_{-16.15}^{+22.42}~,&N_{B\bar{B};D}^{\text{BaBar}}\,\cbg_{D1}=0.24_{-0.02}^{+0.02}~\text{MeV}^{-3/2}~,\\[1ex]
 &Y_{D2}=68.84_{-7.25}^{+11.86}~, &N_{B\bar{B};D}^{\text{BaBar}}\,\cbg_{D2}=0.14_{-0.01}^{+0.01}~\text{MeV}^{-3/2}~,\\[1ex]
 &Y_J=8.44_{-2.59}^{+3.64}\times10^3~, &N_{B\bar{B};J}^{\text{BaBar}}\,\cbg_J=3.46_{-0.14}^{+0.14}~\text{MeV}^{-1}~,\\[1ex]
 &Y_J^{(p)}=5.78_{-1.64}^{+2.29}\times10^5~, &\zeta=1498.57^{+9.73}_{-9.73}~\text{MeV}^{-1}~,\quad\varrho=5.09_{-0.49}^{+0.49}~\text{MeV}^{-2}~.
\end{array}
\eea 

\bea
\label{para:case2b}
\begin{array}{lll}
\text{Case 2.II} &E_R=-0.95_{-0.05}^{+0.07}\,\text{MeV}~, \\[1ex]
\text{10 parameters} &Y_{D1}=79.75_{-19.81}^{+22.46}~,&N_{B\bar{B};D}^{\text{BaBar}}\,\cbg_{D1}=0.25_{-0.02}^{+0.02}~\text{MeV}^{-3/2}~,\\[1ex]
 &Y_{D2}=72.39_{-8.02}^{+9.18}~, &N_{B\bar{B};D}^{\text{BaBar}}\,\cbg_{D2}=0.15_{-0.01}^{+0.01}~\text{MeV}^{-3/2}~,\\[1ex]
 &Y_J=9.23_{-1.57}^{+1.60}\times10^3~,&N_{B\bar{B};J}^{\text{BaBar}}\,\cbg_J=3.65_{-0.13}^{+0.13}~\text{MeV}^{-1}~,\\[1ex]
 &Y_J^{(p)}=6.23_{-0.84}^{+0.71}\times10^5~,&\zeta=1520.23^{+8.57}_{-8.57}~\text{MeV}^{-1}~,\quad\varrho=4.48_{-0.50}^{+0.50}~\text{MeV}^{-2}~.
\end{array}
\eea

\bea
\label{para:caseiii}
\begin{array}{lll}
\text{Case 3.I:}   &E_R=-0.49_{-0.03}^{+0.04}~\text{MeV}~, & \\[1ex]
\text{10 parameters} &Y_{D1}=25.45_{-4.15}^{+4.05}~,&N_{B\bar{B};D}^{\text{BaBar}}\,\cbg_{D1}=0.22^{+0.02}_{-0.02}~\text{MeV}^{-3/2}~,\\[1ex]
 &Y_{D2}=21.08_{-1.89}^{+1.32}~,&N_{B\bar{B};D}^{\text{BaBar}}\,\cbg_{D2}=0.13^{+0.01}_{-0.01}~\text{MeV}^{-3/2}~,\\[1ex]
 &Y_J=80.14_{-5.19}^{+5.67}~,&N_{B\bar{B};J}^{\text{BaBar}}\,\cbg_J=3.66^{+0.13}_{-0.13}~\text{MeV}^{-1}~,\\[1ex]
 &Y_J^{(p)}=5.26_{-0.08}^{+0.12} \times 10^3~,&\zeta=1524.70^{+7.91}_{-7.91}~\text{MeV}^{-1}~,\quad\varrho=5.32_{-0.49}^{+0.49}~\text{MeV}^{-2}~.
\end{array}
\eea

\bea
\label{para:caseiiiII}
\begin{array}{lll}
\text{Case 3.II:}  &E_R=-0.49_{-0.01}^{+0.02}~\text{MeV}~, & \\[1ex]
\text{10 parameters:}  &Y_{D1}=22.90_{-3.02}^{+2.94}~,&N_{B\bar{B};D}^{\text{BaBar}}\,\cbg_{D1}=0.22^{+0.02}_{-0.02}~\text{MeV}^{-3/2}~,\\[1ex]
 &Y_{D2}=18.92_{-0.77}^{+1.40}~,&N_{B\bar{B};D}^{\text{BaBar}}\,\cbg_{D2}=0.13^{+0.01}_{-0.01}~\text{MeV}^{-3/2}~,\\[1ex]
 &Y_J=80.07_{-5.36}^{+5.14}~,&N_{B\bar{B};J}^{\text{BaBar}}\,\cbg_J=3.66^{+0.13}_{-0.13}~\text{MeV}^{-1}~,\\[1ex]
 &Y_J^{(p)}=5.28_{-0.17}^{+0.05} \times 10^3~,&\zeta=1524.66^{+7.73}_{-7.73}~\text{MeV}^{-1}~,\quad\varrho=5.32_{-0.49}^{+0.49}~\text{MeV}^{-2}~.
\end{array}
\eea

\subsection{Case 1: Bound state}
\label{subsec:f(E)I}

 In this first case we fit the different data sets by  using the function $f(E)$ \cite{Braaten.141116.3,Braaten},
 in order to take into account the FSI between the $\bar{D}^0D^{*0}$, while Ref.~\cite{Braaten}
makes fits to  different data sets separately.
 We also include this standard case as a reference to compare with other less standard ones introduced below.
 As mentioned above, the inverse scattering length $\gamma$ in the expression of $f(E)$ can be taken complex (with negative imaginary part) 
to mimic inelastic channels. Indeed  complex values were  used in Ref.~\cite{Braaten}, though it was found
 that the experimental data can be equally well described by taking the imaginary part of $\gamma$ free or fixing it to 0 (as we have also
found).
 Physically, this indicates that an inelastic effect, as the transition $ D^0 \bar D^{*0} \to J/\psi\pi^+\pi^-$,  has little impact
on  FSI and we fix it always to zero in our fits, which are also checked to be stable if the imaginary part of $\gamma$ is released.

\begin{figure}[htbp]
\begin{center}
\vglue 0.5cm
\includegraphics[height=125mm,clip]{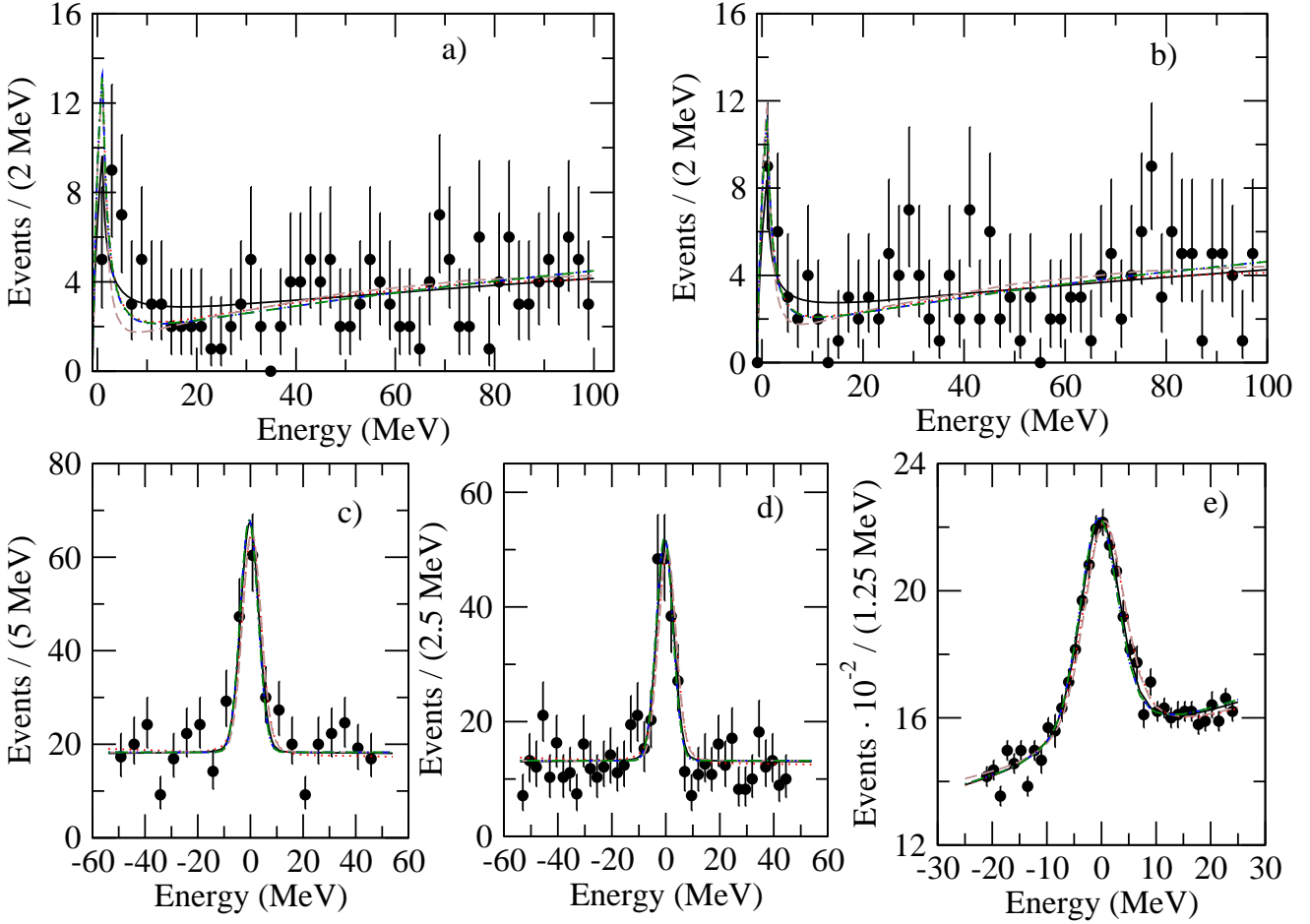}
\caption{Fit results  for the different
event distributions as a function of energy. Panels a) and b) denote
the mode $D^0\bar D^{*0}$ from BaBar \cite{BaBarD} and Belle
\cite{BelleD}, respectively. Panels c), d) and e) denote the mode
$J/\psi\pi^+\pi^-$ from BaBar \cite{BaBarJ}, Belle \cite{BelleJ},
and CDF \cite{CDF.211116.5} Collaborations, respectively. The black
solid lines correspond to case 1 that employs $f(E)$.
 The CDD pole contribution is included in the expression of ${\ct}(E)$, and the rest of lines refer to cases making use of this
parameterization. Case 2.I is shown by the red dotted lines, and
case 2.II by the brown dashed lines, while cases 3.I and 3.II are given by
the blue dash-double-dotted  and green dash-dotted lines, in
order. Part of the lines overlap so much that it is difficult to
distinguish between them in the scale of the plots.}
 \label{fig:resultI}
\end{center}
\end{figure}

\begin{figure}[htbp]
\begin{center}
\vglue 0.5cm
\includegraphics[width=0.6\textwidth]{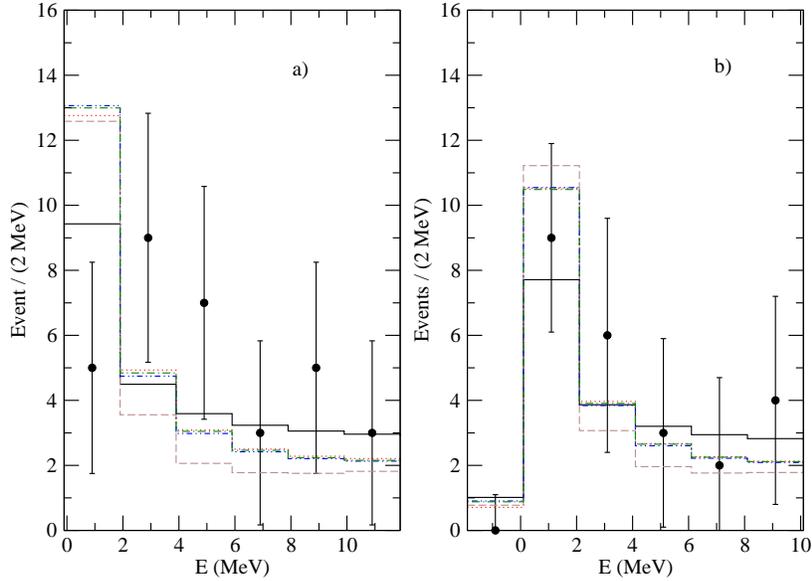}
\caption{Detail of the fit results  for the near-threshold region in the 
$D^0\bar{D}^{*0}$ event distributions. The left panel correspond to the data from the BaBar Collaboration 
\cite{BaBarD} and the right one to that of the Belle Collaboration \cite{BelleD}. 
The types of lines employed to plot the boxes for every scenario are the same as in Fig.~\ref{fig:resultI}.}
 \label{fig:insetDD}
\end{center}
\end{figure}

 The parameters corresponding to the yields and background constants are $Y_J$, $Y_J^{(p)}$, $Y_D$,  $\cbg_J$, $\xi$, $\rho$
and $\cbg_D$.  The most interesting free parameters are those fixing the interaction
$t(E)$, that for the present case just reduces to the inverse
scattering length $\gamma$. The background constants can be determined 
rather straightforwardly, because the background is very smooth in all
cases and fixed by the sidebands events around the signal region.
 The best fitted parameters that we
obtain for case 1 are given in Eq.~\eqref{241116.2}. 
 The reproduction of the event distributions is shown in Figs.~\ref{fig:resultI} and \ref{fig:insetDD} by the black solid lines.
 The different yields, that we denote globally as $Y_F$ in the following, can be interpreted properly in this way
 because ${\cal N}=0.98\simeq 1$, as expected for a bound state.

  With the values of the parameters at hand in Eq.~\eqref{241116.2},  the $X(3872)$ is a near-threshold bound-state pole 
in the function $f(E)$ located at 
 $-0.19_{-0.01}^{+0.01}- i\, \Gamma_*/2$~MeV. Here the imaginary part stems purely from
the finite width of the constituent $D^{*0}$. 
 As a result the scattering length is large and positive  with the 
value $a=10.40^{+0.10}_{-0.26}$~fm. 

The compositeness, $X$, of the resulting bound-state pole   \cite{hyodo.101016.5,aceti.101016.6,sekihara.101016.7}
 can be written as \cite{Zbpaper}
\begin{align}
\label{170319.2}
X=-ig_k^2~,
\end{align}
 with $g_k^2$ the residue of the amplitude in the momentum variable $k$.
 For $f(E)=1/(-1/a-ik)$ its residue at the pole position  in the variable $k$ is $i$, so that $X=1$. That is,
independently of which is the dynamical seed for binding (origin of $\gamma$) this is a bound state whose composition
is exhausted by the  $D^{0}\bar D^{*0}$ component \cite{Braaten.141116.3,Braaten,baru.170310.2}.
This result is in agreement with Ref.\cite{Qiang.141116.1},
 which concludes that the scattering length approximation is only valid for the bound-state case if its compositeness is 1. 
We also give in the fourth column of Table~\ref{tab:fitsummary} the residue $g^2$ of the $S$-wave scattering amplitude
 for each near-threshold pole in a more standard normalization,
 in which the partial decay width of a narrow resonance
is $\Gamma=k g^2/(8\pi M_X^2)$ \cite{pdg.181116.2}. This residue  for $f(E)$ reads
$g^2=-i16\pi k_P P_X^2/\mu$. 


\begin{table}[htbp]
\renewcommand{\arraystretch}{1.2}
\begin{center}
\begin{tabular}{rrrrr}
\hline \hline
 Case   & Pole position    & $X$   & Residue      &$Y_{D1}$ \\
        &  [MeV]           &       & [GeV$^2$]    &$Y_{D2}$ \\
        &                  &       &              &$Y_J$\\
        &                  &       &              &$Y_J^{(p)}$\\
\hline
1       &$-0.19_{-0.01}^{+0.01} - i \, 0.0325$   & $1.0$ & $14.78_{-0.14}^{+0.38}$   & $7.49_{-0.41}^{+0.71}$  \\
        &                                        &       &                           & $6.45_{-0.47}^{+0.32}$ \\
        &                                        &       &                           & $79.03_{-6.11}^{+5.65}$\\
        &                                        &       &                           & $5.23_{-0.11}^{+0.07}\times10^3$\\
\hline
2.I     &$-0.36_{-0.10}^{+0.08} - i\, 0.18_{-0.02}^{+0.01}$   &   & $-47.48_{-12.40}^{+9.75}-i\,66.06_{-13.50}^{+10.87}$  &$83.13_{-16.15}^{+22.42}$\\
        &$-0.70^{+0.11}_{-0.13} + i \, 0.17_{-0.01}^{+0.02}$  &   & $82.69_{-11.88}^{+14.84}+i\,66.03_{-10.87}^{+13.50}$ &$40.13_{-7.25}^{+11.86}$\\
        &                                                     &   &                                                      &$8.44_{-2.59}^{+3.64}\times10^3$\\
        &                                                     &   &                                                      &$5.78_{-1.65}^{+2.29}\times10^5$ \\
\hline
2.II    &$-0.33_{-0.03}^{+0.04} - i\, 0.31_{-0.01}^{+0.02}$  &   & $-6.24_{-2.20}^{+2.80}-i\,1.41_{-0.10}^{+0.14}\times10^2$  &$79.75_{-19.81}^{+22.46}$\\
        &$-0.84^{+0.07}_{-0.05} + i \, 0.77_{-0.04}^{+0.03}$  &   & $(2.32_{-0.21}^{+0.16}-i\,1.77_{-0.08}^{+0.11})\times10^2$ &$42.20_{-8.02}^{+9.18}$\\
        &$-1.67_{-0.08}^{+0.10} - i \,0.49_{-0.02}^{+0.02}$   &   & $(-3.26_{-0.16}^{+0.22}+i\,3.18_{-0.25}^{+0.18})\times10^2$&$9.23_{-1.57}^{+1.60}\times10^3$\\
        &                                                     &   &                                                            &$6.23_{-0.84}^{+0.71}\times10^5$\\
\hline
3.I     &$-0.50^{+0.04}_{-0.03}$ &$0.061_{-0.002}^{+0.003}$  &$1.52_{-0.01}^{+0.01}$   &$25.45_{-4.15}^{+4.05}$\\
        &$-0.68^{+0.05}_{-0.03}$ &                           &$2.72_{-0.04}^{+0.02}$   &$12.29_{-1.89}^{+1.32}$\\
        &                        &                           &                         &$80.14_{-5.19}^{+5.67}$\\
        &                        &                           &                         &$5.26_{-0.08}^{+0.12}\times10^3$\\
\hline
3.II    &$-0.51^{+0.03}_{-0.01}$ &$0.158_{-0.001}^{+0.001}$  &$3.96_{-0.08}^{+0.03}$   &$22.90_{-3.02}^{+2.94}$\\
        &$-1.06^{+0.05}_{-0.02}$ &                           &$7.56_{-0.20}^{+0.08}$   &$11.03_{-0.77}^{+1.40}$\\
        &                        &                           &                         &$80.07_{-5.36}^{+5.14}$\\
        &                        &                           &                         &$5.28_{-0.17}^{+0.05}\times10^3$\\
\hline\hline
\end{tabular}
\caption{Summary of the combined fits in Sec.~\ref{sec:Combined} for
all the cases. From left to right, the pole positions, compositeness
($X$), residue and yields ($Y$) are given.  $Y_{D1}$, $Y_{D2}$
denote the yields corresponding to the BaBar and Belle data on the $D^0\bar D^{*0}$ mode, 
respectively; $Y_J$ denotes the one for BaBar 
 on the $J/\psi\pi^+\pi^-$ mode and $Y_J^{(p)}$ applies to the CDF collaboration data
 with $p\bar p$ collisions. The normalization for the residues is such that 
$\Gamma=k g^2/(8\pi M_X^2)$ as in Ref.~\cite{pdg.181116.2}.} \label{tab:fitsummary}
\end{center}
\end{table}

\begin{table}[htbp]
\renewcommand{\arraystretch}{1.2}
\begin{center}
\begin{tabular}{rrrrrr}
\hline \hline
 Case & $\lambda$   & $\MCDD$    & $\beta$    & $a$ (fm)                  & $r$ (fm)   \\
      & [MeV$^2$]       &  [MeV]     & [MeV]  & $Z_a$                         & $Z_r$ \\
\hline
1    &  $\ldots$       &  $\ldots$     &  $\ldots$       & $10.40^{+0.10}_{-0.26}$   &   $0.$\\
      &             &            &            & $0.$ & $0.$ \\
\hline
2.I & $4176.35_{-462.56}^{+385.00}$   &  $-12.29^{+1.11}_{-1.14}$      & $-323.12^{+11.12}_{-6.01}$   & $-11.82^{+1.15}_{-1.21}$     & $-5.64^{+0.58}_{-0.61}$ \\
   &                            &                             &                         &    $\ldots$ & $\ldots$ \\
\hline
2.II & $324.68_{-34.45}^{+26.32}$   &  $2.84^{+0.15}_{-0.20}$      & $128.43^{+3.38}_{-4.71}$   & $-13.83^{+0.35}_{-0.53}$     & $-8.19^{+0.21}_{-0.31}$ \\
   &                            &                             &                         &    $\ldots$ & $\ldots$ \\
\hline
3.I  & $260.46_{-2.31}^{+1.39}$    & $0.25^{+0.04}_{-0.03}$ & $317.25_{-0.40}^{+0.31}$  &$0.27^{+0.07}_{-0.04}$ &$-847.84^{+212.48}_{-223.69}$ \\
        &                          &                         &                        &$\ldots$  &$\ldots$\\
\hline
3.II & $2204.29_{-76.19}^{+31.04}$  & $3.21_{-0.08}^{+0.03}$ & $561.70_{-3.41}^{+1.37}$  & $1.57_{-0.02}^{+0.05}$ &  $-43.75^{+0.24}_{-0.61}$ \\
        &                          &                         &                        &$0.86$  &$0.87$\\
\hline\hline
\end{tabular}
\caption{Parameters characterizing the $S$-wave interaction $t(E)$
for the cases 1--3.II of the combined fits in
Sec.~\ref{sec:Combined}. The ellipsis indicate that it is not
appropriate to give the corresponding magnitude in such case. The
elementariness $Z$ is calculated from the knowledge of the
bound-state mass and the Weinberg's compositeness relations of
Eq.~\eqref{271116.3} either in terms of $a$ ($Z_a$) or $r$ ($Z_r$).
The error is not given when its estimation is smaller than the
precision shown.} \label{tab:t.fitC}
\end{center}
\end{table}

\subsection{Case 2: Virtual state}
\label{subsec:virtualI}

 In the previous section, as well as in Ref.~\cite{Braaten}, only the scattering length is taken into account.
However, considering the analysis performed in Ref.~\cite{ChenGY}, the effective range should better be added into, as already
done in the pioneering analysis of Ref.~\cite{Hanhart.141116.4}, since the scattering length
approximation is only valid for pure molecular states. As discussed in Sec.~\ref{sec.211116.1},
one also has to face the problem of the possible presence of zeroes just around threshold. These two points can be better handled
 by including a CDD pole, and the  $\bar{D}^0D^{*0}$ $S$-wave scattering amplitude is given in Eq.~\eqref{211116.3}.
 In this case FSI are taken into account by the  function ${\ct}(E)$,  introduced above and given in Eq.~\eqref{211116.7}.
 We make use of this new formalism to impose the presence of a virtual state when fitting data, so as to distinguish
the virtual-state scenario from the bound-state one obtained above by using the function $f(E)$ in Sec.~\ref{subsec:f(E)I}.
We also remark that proceeding in this way drives to quite interesting situations in which the
$X(3872)$  becomes a double or triplet virtual-state  pole in the zero width limit of the $D^{* 0}$.
Reference \cite{baru.170310.2} already stressed the importance of taking care of a possible near-threshold zero in scattering and
production processes. 

The $t$-matrix for $D^0\bar D^{*0}$ scattering, in the 2nd RS sheet, is obtained from its expression in the 1st RS,
cf. Eq.~\eqref{211116.3}, but replacing
$k$ by $-k$, namely,
\beq
\label{tmatrix2nd}
t_{II}(E)=\left[\frac{\lambda}{E-\MCDD}+\beta+i k \right]^{-1},
\eeq
where $k=\sqrt{2\mu E}$ is calculated such that $\Im k>0$. Notice that here we are taking the $D^{* 0}$ without width, and then
 impose a pure virtual-state situation, that is, a pole on the real axis below threshold in the 2nd RS.
 The presence of the virtual state is granted by imposing
that $t_{II}(E)$ has a pole at $E_P=E_R-iG_R/2$, with $E_R<0$, $G_R>0$ and taking at the end the limit $G_R\to 0^+$.
For an $S$-wave resonance it is possible to have a non-zero width for a resonance mass smaller than the two-particle
threshold, see e.g. Refs.~\cite{hanhart.170319.1,albaladejo.170319.1} for particular examples
and Ref.~\cite{rios.170319.1} for general arguments.
 The vanishing of the real and imaginary parts of $t_{II}(E_P)^{-1}$ allows us to fix two parameters,
e.g. for  $\lambda$ and $\beta$  one has the expressions
\bea
\label{eq:sol}
 &&\lambda=\frac{1}{2G_R}(G_R^2+4\widetilde{\Delta}^2)v\cos u~ ,\nl
&&\beta=\frac{v}{G_R}(-2\widetilde{\Delta}\cos u +G_R\sin u)~, \nl
&&u\equiv \frac{1}{2}{\rm arg}\left(E_R+i \frac{G_R}{2}\right),\quad
v\equiv \sqrt{2\mu(E_R^2+G_R^2/4)},\quad \widetilde{\Delta}\equiv E_R-\MCDD ~.
\eea
Thus, the function ${\ct}(E)$, Eq.~\eqref{211116.7}, depends on $E_R$ and $\MCDD$, with a stable limit for $G_R\to 0^+$.
The latter can be performed algebraically from Eq.~\eqref{eq:sol} with the result
\begin{align}
\label{170319.3}
\lambda&=\frac{\mu}{\varkappa}(M_{\rm CDD}-E_R)^2~,\\
\beta&=\frac{\mu}{\varkappa}(M_{\rm CDD}-3 E_R)~,\nn
\end{align}
with $\varkappa=\sqrt{2\mu|E_R|}$. Notice that keeping $G_R$ finite and later taking the limit $G_R\to 0^+$
allows us to dispose of one more constraint (one free parameter less)
that if we had taken directly $G_R=0$ and then imposed that $t_{II}(E_R)=0$.

 Next, let us consider the secular equation for the poles of $t(E)$, Eq.~\eqref{211116.3},
in the complex $k$ plane:
\begin{align}
\label{170319.4}
\lambda+(E-M_{\rm CDD})(-ik+\beta)=0~.
\end{align}
We substitute  the expressions for $\lambda$ and $\beta$ of Eq.~\eqref{170319.3} in the previous equation and obtain:
\begin{align}
\label{170319.5}
(k+i\varkappa)^2(k+i\frac{\mu}{\varkappa}(M_{\rm CDD}+E_R))=0~,
\end{align}
where the global factor $-i/2\mu$ has been dropped. This equation explicitly shows that
$k=-i\varkappa$ is a double virtual-state pole.

It is also trivial from Eq.~\eqref{170319.5} to impose  a triplet virtual-state pole by choosing appropriately $M_{\rm CDD}$ to
\begin{align}
\label{170320.1}
M_{\rm CDD}&=-3 E_R~.
\end{align}
In the following we denote by case 2.I the one with the double virtual-state pole and by case 2.II that with the triplet pole.

 To fit data  we reinsert the finite width for the $D^{* 0}$
and use the expressions for $\lambda$ and $\beta$ in Eq.~\eqref{170319.3}.
 For the case 2.I one has two free parameters ($E_R$ and $\MCDD$) to characterize the interaction
while for the case 2.II only one free parameter remains ($E_R$)
because of the extra Eq.~\eqref{170320.1}. 
 The fitted parameters in each case 
are given in Eqs.~\eqref{para:case2a} and \eqref{para:case2b}.

 The reproduction of data for the cases 2.I and 2.II are shown by the red dotted and brown dashed lines in 
Figs.~\ref{fig:resultI} and \ref{fig:insetDD}, respectively.
 There are visible differences between case 1 and cases 2.I-II, in the peak region of the 
$D^0\bar{D}^{*0}$ and $J/\psi \pi^+\pi^-$ event distributions. For the former, the 
scenarios 2.I-II produce a signal higher in the peak that  
decreases faster with energy, while for the latter there is a displacement of the peak towards 
the threshold in the virtual-state cases. This is more visible in Fig.~\ref{fig:CDFBand} where we show 
only the reproduction of the CDF data \cite{CDF.211116.5} including error bands as well.
The reason for this displacement is  because the virtual-state poles only manifests in the physical axis above threshold,
 so that the peak of the event distribution happens almost on top of it. Nonetheless, we have to say that 
the shift in the signal peak for the cases 2.I-II and the $J/\psi \pi^+\pi^-$ data diminishes considerably if we excluded in the 
fit the $D^0\bar{D}^{*0}$ data from the BaBar Collaboration \cite{BaBarD}. Furthermore, it is clear that these data give rise to a 
line shape with a displaced peak towards higher energies as compared with the analogous data from the Belle Collaboration \cite{BelleD}, 
see Fig.\ref{fig:insetDD}. 
Thus, it is not fair just to conclude that cases 2.I-II are disfavor because of the shift of the 
 signal shape in the $J/\psi\pi^+\pi^-$ CDF data \cite{CDF.211116.5} until one also disposes of better data for the $D^0\bar{D}^{*0}$ event distributions.

 It is clear that when taking $\Gamma_*=0$ (so that standard ERE is
perfectly fine mathematically  for $|k^2|<2\mu |\MCDD|$), all the near-threshold poles are at $E_R$.
 For the case 2.I the CDD pole lies relatively far away from the $D^0\bar{D}^{*0}$ threshold. However,
if we kept only $a=-11.82~$fm in the ERE the  pole position in the 2nd RS would be $-0.14$~MeV,
if including $r=-5.64~$fm it is $-0.93$~fm, with $v_2$ then it still moves to $-0.58$~MeV and with $v_3$ one has $-0.54$~MeV.
Thus, though the CDD pole is around $-12$~MeV one needs several terms in the ERE to reproduce adequately the $S$-wave amplitude.
In particular, 
it is not enough just to keep e.g. the scattering length contribution as in case 1 or as in Ref.~\cite{Braaten.141116.3,Braaten}.
For the case 2.II the CDD pole is much closer, around $3$~MeV, so that the convergence of the ERE is much worse and
 many more terms in the ERE should be kept to properly reproduce the pole position.

\begin{figure}[h]
\begin{center}
\begin{tabular}{cc}
\includegraphics[angle=-0, width=.45\textwidth]{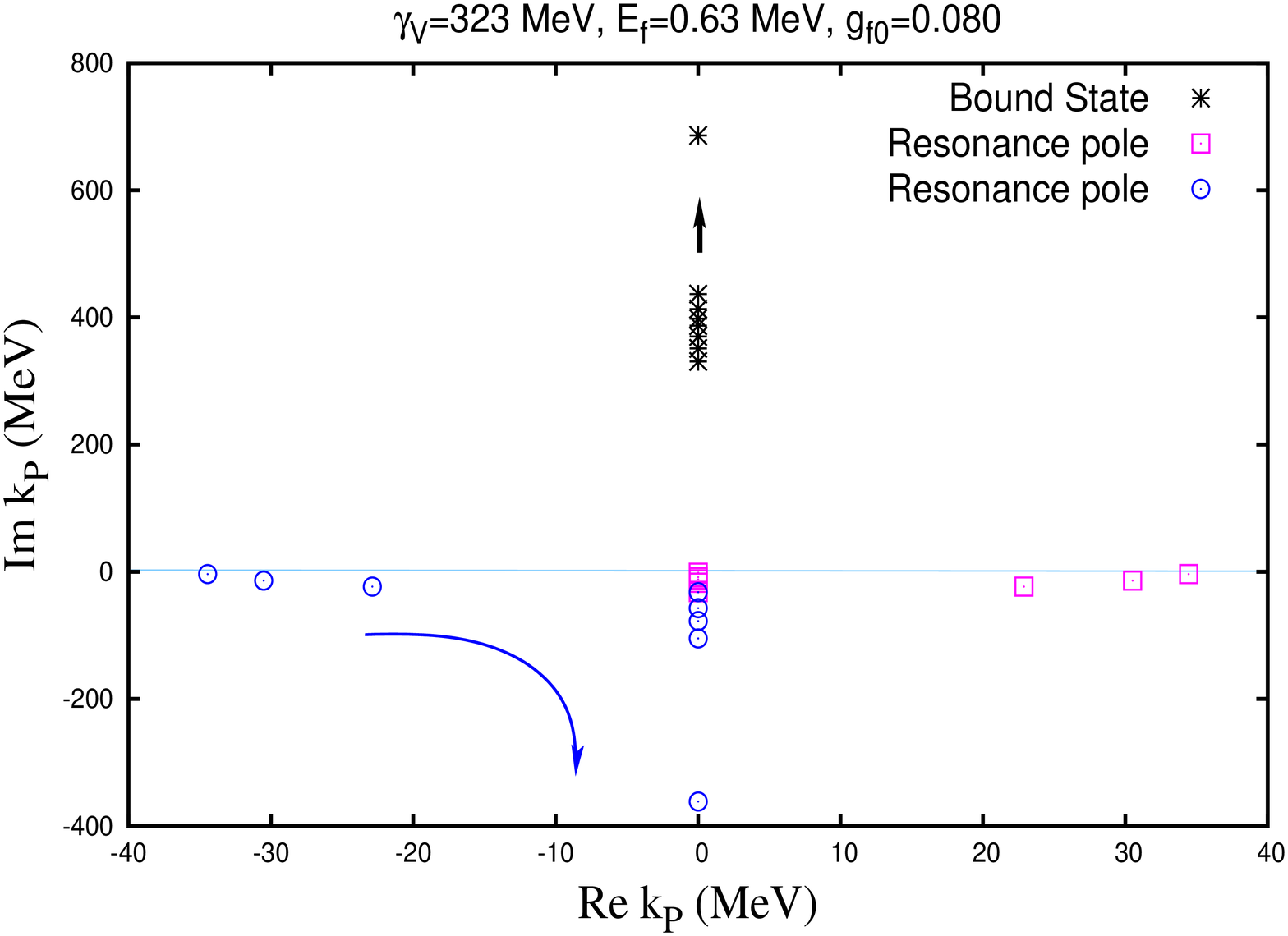} &
\includegraphics[angle=-0, width=.45\textwidth]{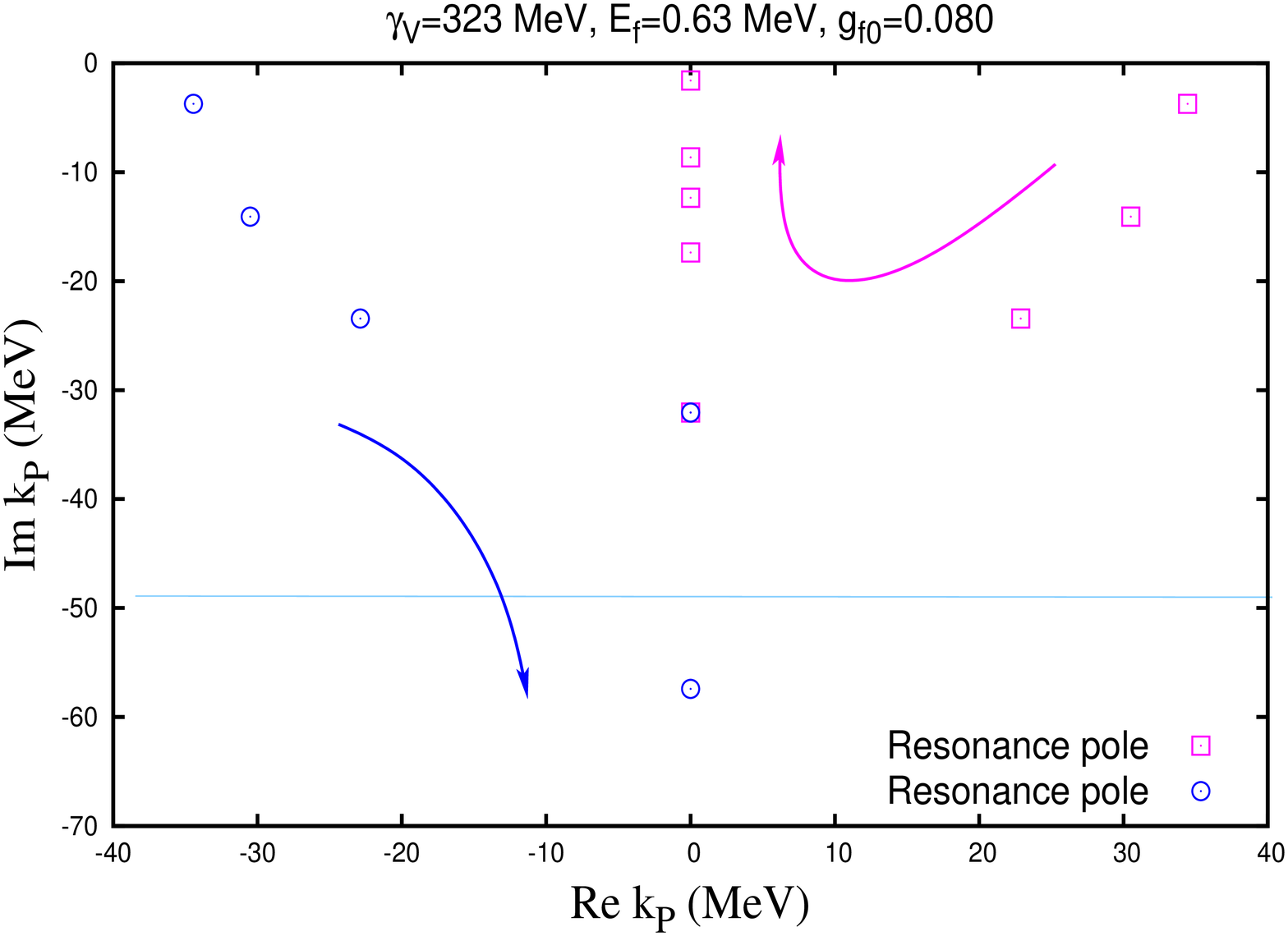}  \\
\includegraphics[angle=-0, width=.45\textwidth]{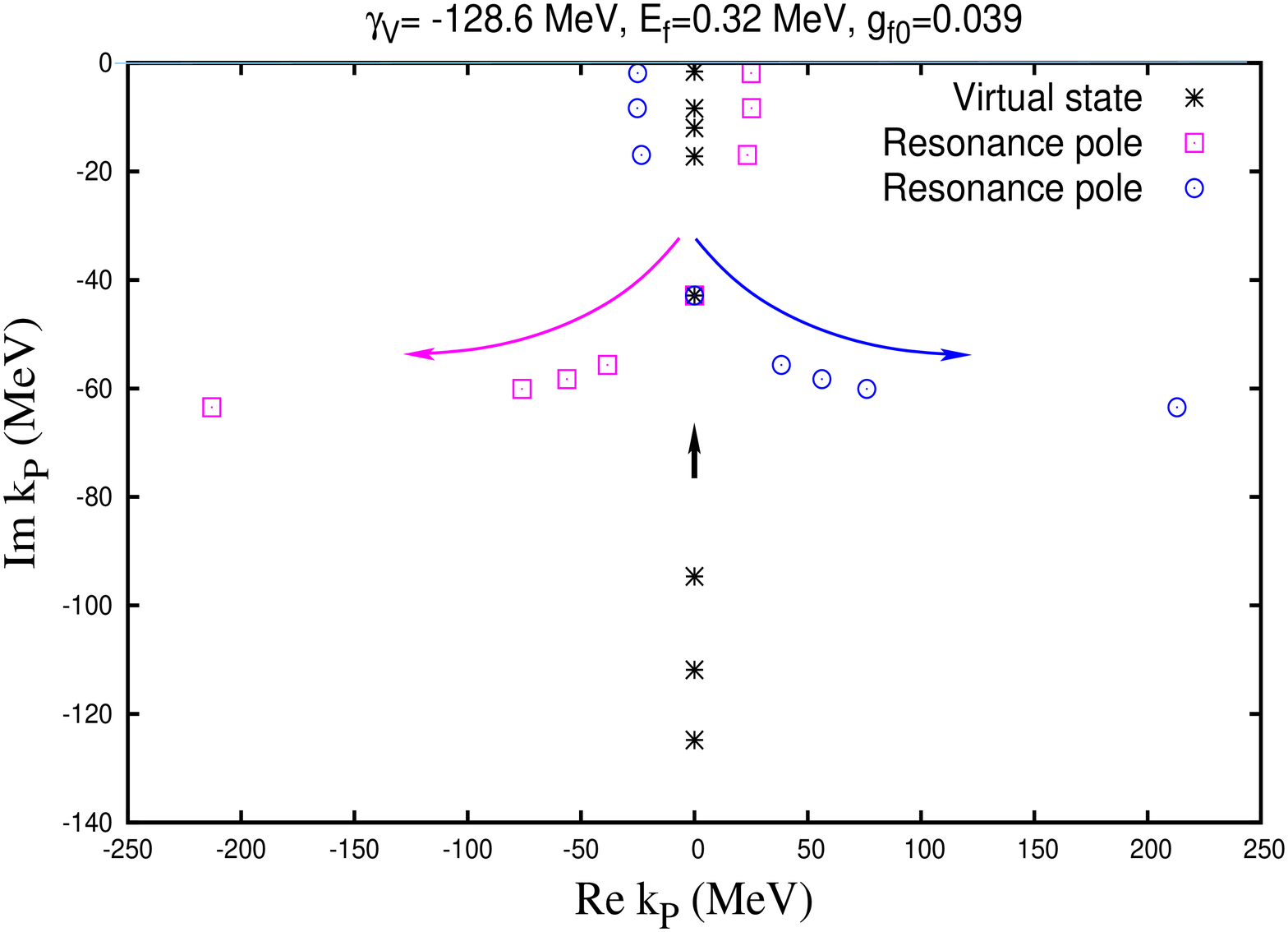}  &
\includegraphics[angle=-0, width=.45\textwidth]{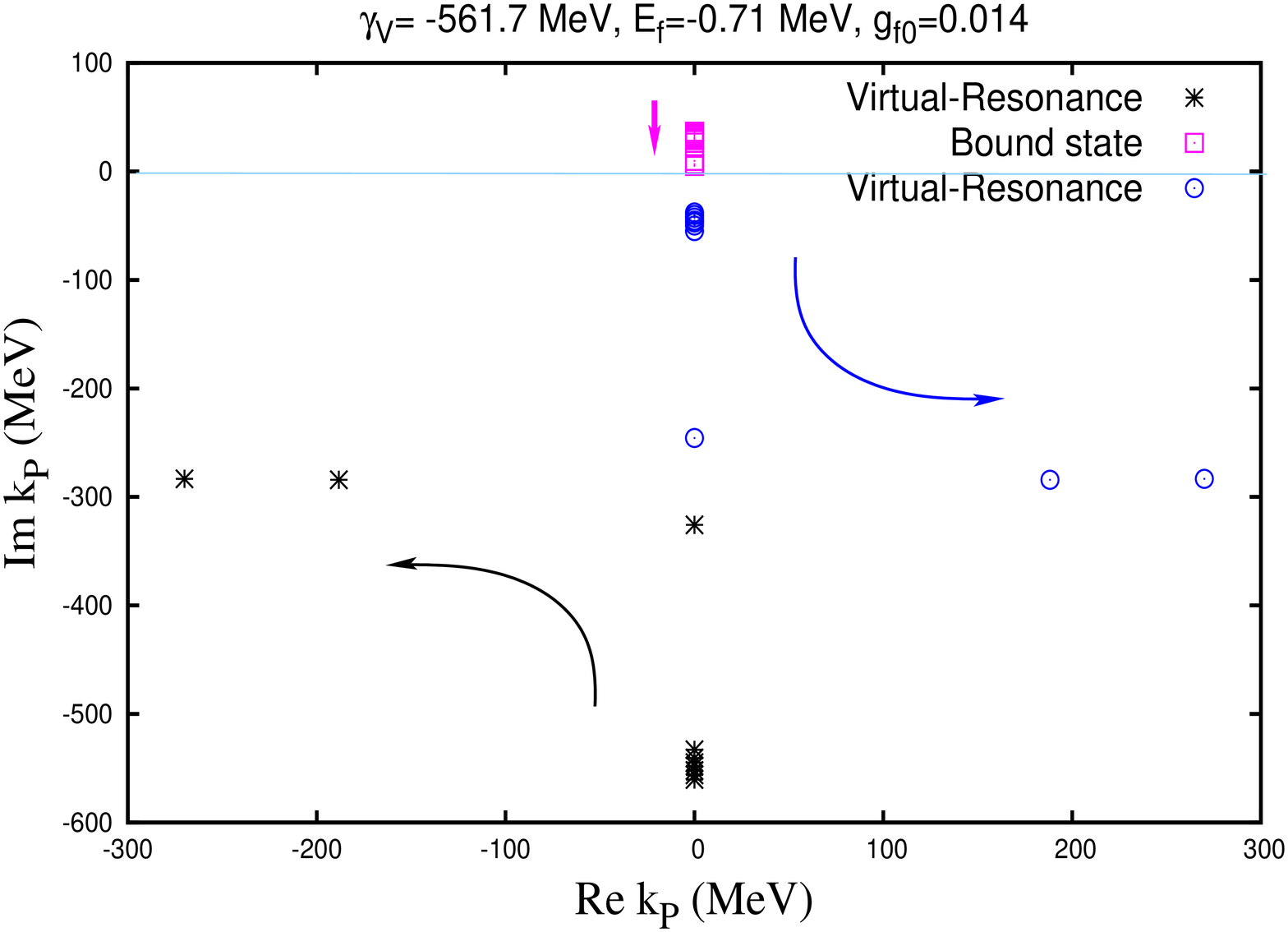}  \\
\end{tabular}
\caption{Evolution of the poles as $g_f$ increases, with $E_f$ and $\gamma_V$ fixed.
 The top panels are for case 2.I, the bottom-left panel for  case 2.II and the bottom-right one for case 3.II.
In the right top panel a detail of the near-threshold region is given for the case 2.I.}
\label{fig.170322.1}
\end{center}
\end{figure}

At this point it is interesting to display the pole trajectories as a function of $g_f$, while keeping constants $\gamma_V$
and $E_f$, cf. Eq.~\eqref{170314.3}. In this way, we have a quite intuitive decoupling limit $g_f\to 0$ in which two poles
at $\pm \sqrt{2\mu E_R}$ correspond to the bare state and an additional one at $i\gamma_V=-i\beta$ stems
 from the direct coupling between the $D^0\bar{D}^{*0}$ mesons. As $g_f$ increases an interesting interplay between the pole
  movements arises reflecting the coupling between the bare state and the continuum channel. 
 For the fit of case 2.I one has the central
values $g_{f0}=0.080$, $\gamma_V=323$~MeV and $E_f=0.63\to \pm\sqrt{2\mu E_f}=\pm 35$~MeV. Its pole trajectories, shown in
the two top panels of Fig.~\ref{fig.170322.1}, are obtained by increasing
$g_f$ from one tenth of the fitted value up to 10 times it.
 In the left panel we show the global picture, including the far away bound state, while
in the right panel we show a finer detail of the two near-threshold poles that stem from the bare state, which  for $g_f=g_{f0}$
become degenerate.
For the case 2.II we have the central  values $g_{f0}=0.039$, $\gamma_V=-128.6$~MeV and $E_f=0.32\to \pm\sqrt{2\mu E_f}=\pm 24.7$~MeV.
The three  virtual-state poles, 2 from the bare state and another from the
direct interactions between the $D^0\bar{D}^{*0}$ mesons, become degenerate for $g_f=g_{f0}$ and the triplet pole arises.
Compared with the pole trajectories explicitly shown in Ref.~\cite{baru.170310.2} ours correspond to
a much larger absolute value of $\gamma_V$ than those in Ref.~\cite{baru.170310.2} with $|\gamma_V|$ between 20-55~MeV.
There is no pole trajectory with three poles merging neither in Ref.~\cite{baru.170310.2} nor in Ref.~\cite{rios.170319.1}.

\begin{figure}
\begin{center}
\begin{tabular}{cc}
\includegraphics[angle=-90, width=.45\textwidth]{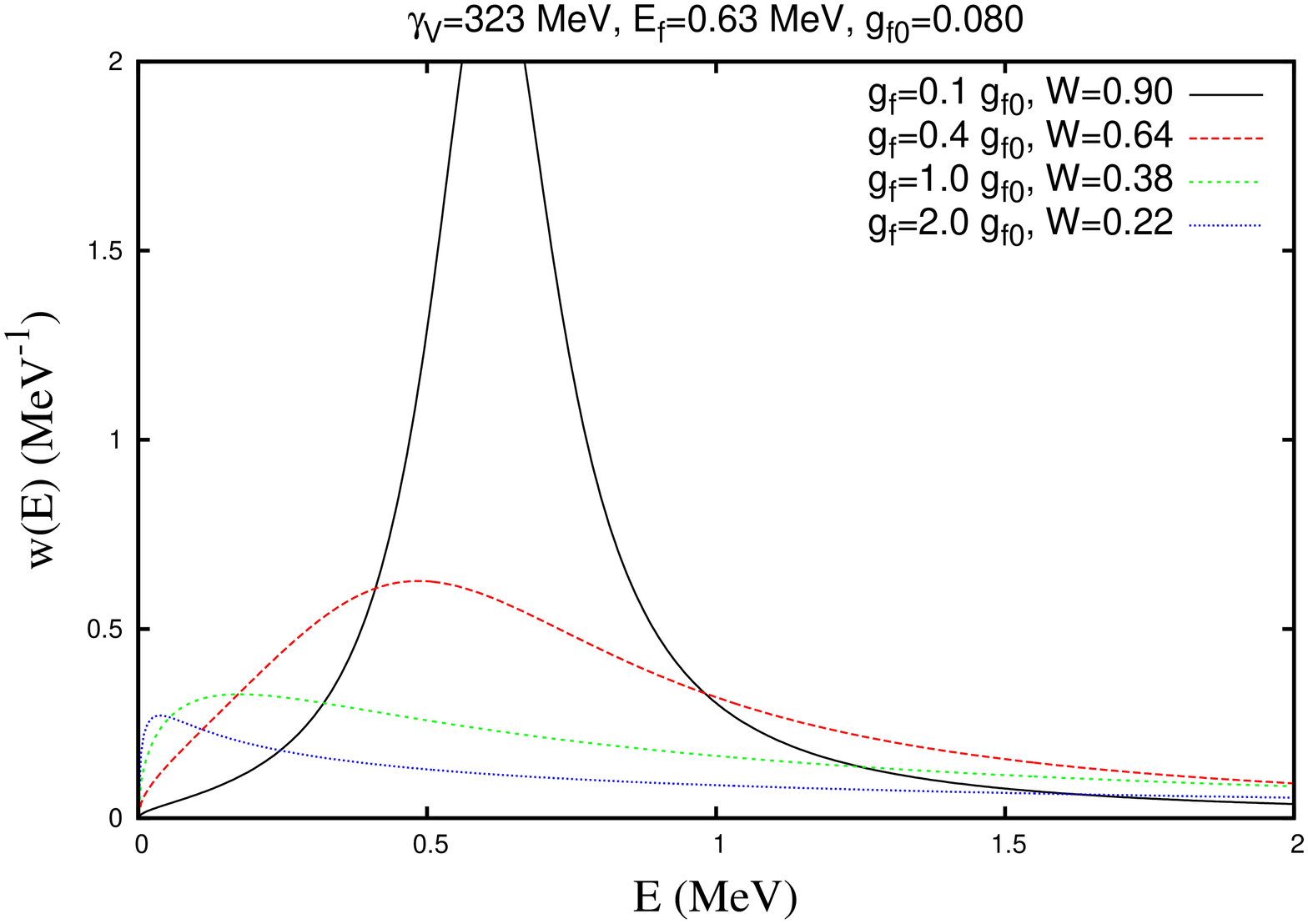} &
\includegraphics[angle=-90, width=.45\textwidth]{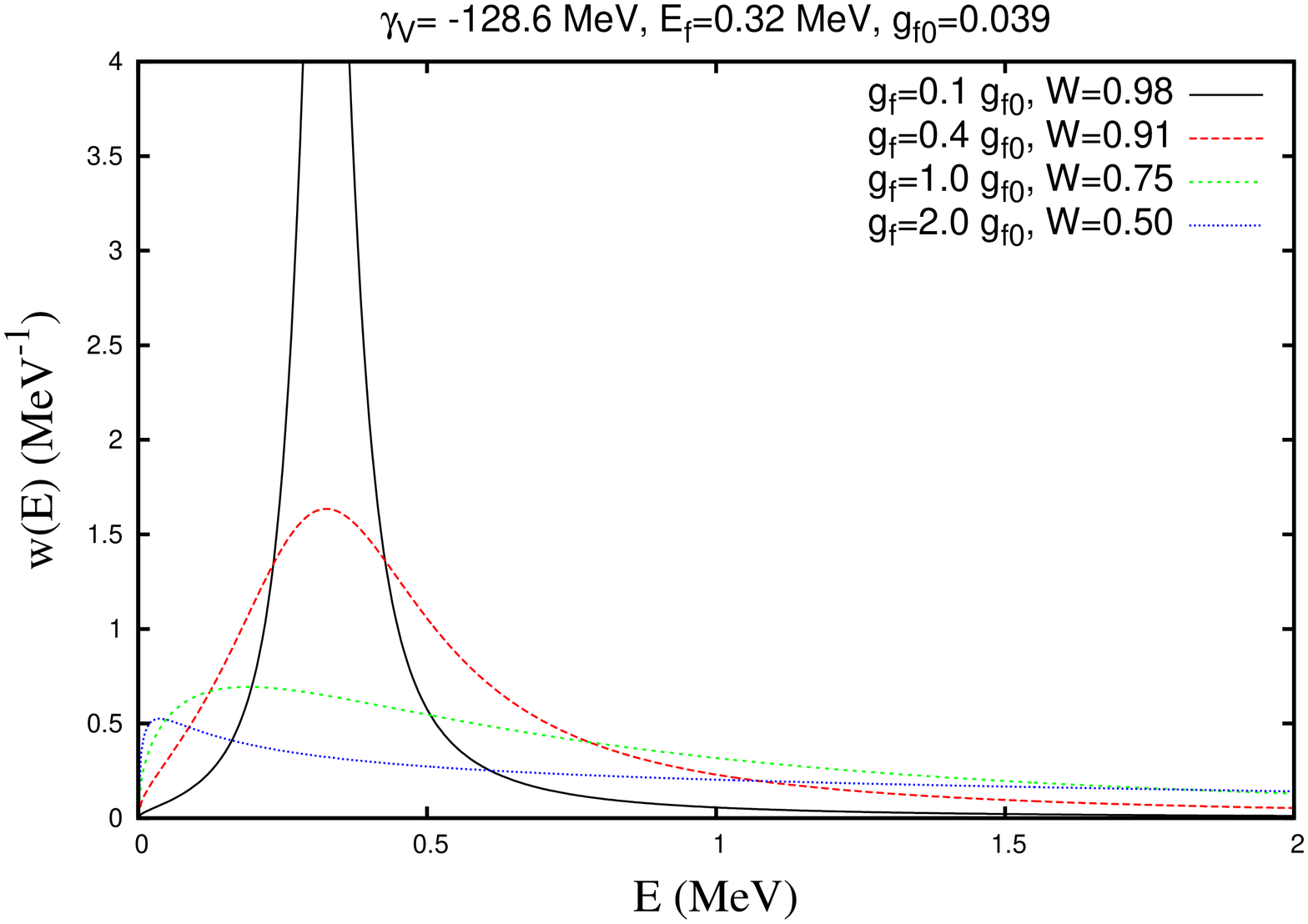}
\end{tabular}
\caption{Spectral density function for the case 2.I (left) and 2.II (right panel) for
several values of $g_f$, with $E_f$ and $\gamma_V$ fixed.
The value of $W$ is also indicated in the legends for each entry of $g_f$.}
\label{fig.170322.2}
\end{center}
\end{figure}

Due to the relationship between a near-threshold CDD pole and a bare state weakly coupled to the continuum (as exemplified explicitly in
the third expression of Eq.~\eqref{170314.3}) one expects that for the case 2.I the virtual-state pole has mostly a dynamical origin
while for the case 2.II, with a much smaller $|\MCDD|$, one anticipates an important bare component.
This expectation can be put in a more quantitative basis by using the spectral density
function $\omega(E)$ as introduced in Ref.~\cite{baru.170310.1},
which reflects the amount of the continuum spectrum in the bare state. For the dynamical model of Ref.~\cite{baru.170310.2} the
spectral function can be calculated and reads
\begin{align}
\label{170322.1}
\omega(E)&=\theta(E)\frac{\gamma_V(E_f-\MCDD)k/\pi}{|\gamma_V(E-E_f)+i(E-\MCDD)k|^2}\\
&=\theta(E)\frac{\lambda k/\pi}{|\lambda+(\beta-ik)(E-\MCDD)|^2}~,\nn
\end{align}
with $\theta(x)$ the Heaviside function. We have used the prescription argued in Ref.~\cite{baru.170310.1}, so
that the spectral density function is integrated only along the $X(3872)$ signal  region, taken as
1~MeV above threshold,
\begin{align}
\label{170322.2}
W&=\int_0^1 dE \,\omega(E)~.
\end{align}
This is a reasonable interval as explicitly shown in Fig.~\ref{fig.170322.2},
where several spectral density functions are shown
for increasing $g_f$, from 0.1$g_{f0}$ up to $2g_{f0}$, with $\gamma_V$ and $E_F$ fixed. 
 The left  panel is for case 2.I and the right one for case 2.II. In the decoupling limit the spectral
density is strongly peaked and  becomes more diluted as $g_f$ increases.
The value of the integral $W$ is interpreted as the bare component in the resonance composition, and
we obtain the values of $W=0.38$ for case 2.I and $W=0.75$ for case 2.II. 
This result is in line with our previous
conclusion based on the value of $\MCDD$, since for the former case the $D^0\bar{D}^{*0}$
component is dominant (around a 60\%) while for the latter is much smaller (around a 25\%).
For different $g_f$ we also give in the legends of the panels of Fig.~\ref{fig.170322.2}
the resulting value of $W$, that increases as $g_f$ decreases because the bare component is
 larger then. In the limit $g_f\to 1$ $W$  tends to 1, as it should.

We now discuss until the end of this section the situation used to actually fit data with $\Gamma_*\neq 0$.
 As already discussed above  for the case 2.I we find two near-threshold virtual-state poles in the 2nd RS
 and one deep bound state in the 1st RS.
 The latter is driven by the large negative value of $\beta$, so that $k\approx -i\beta$.\footnote{Of course, the deep bound state is out of the region 
of validity of our approach and it is just referred for illustrative purposes.}
   For the triplet case all poles lie close to the threshold. Let us recall that the pole positions are given in 
in the second column of Table~\ref{tab:fitsummary}.
 Contrary to case 1 their imaginary parts do not coincide with $\Gamma_*/2$ because of the energy dependence of 
the CDD pole entering in ${\ct}(E)$. One can observe that the imaginary parts of the pole positions for the case 2.I are much 
larger in absolute value than $\Gamma_*/2$, and that for the case 2.II they are even larger than for the case 2.I. 
This noticeable fact is due to   the dependence on
\begin{align}
\rho&=\frac{\Gamma_*}{E_R}~,
\end{align}
which shows a striking non-analytic behavior because of the higher order of the virtual-state
  pole in the limit $\Gamma_*\to 0$,
and corrections to the pole positions are controlled by $\rho^{1/n}$ with $n=2$ and 3 for the double and triplet virtual-state poles, respectively.
This implies that these corrections are significantly larger than expected as the order of the pole increases.
Of course, this is exemplified by the given splitting in the pole positions for the double and triplet poles (being correspondingly larger for the latter).
The dependence of the pole positions with $\rho$ is worked out explicitly in  the
Appendix \ref{app.170320.1} and we give here the final results.

For case 2.I we can simplify formulae  by taking into account that $|\MCDD|\gg |E_R|$.
The poles are located at
\begin{align}
\label{170321.1}
k_{1,2}&=-i\varkappa\left(1\pm \frac{1-i}{2}\rho^\frac{1}{2} \right)~,\\
k_3&=-i\beta~,\nn
\end{align}
with $\varkappa=\sqrt{2\mu |E_R|}$.
Their positions in the energy plane, $E=k^2/2\mu-i\Gamma_*/2$, are
\begin{align}
\label{170321.2}
E_{1,2}&=E_R \left( 1 \pm \rho^\frac{1}{2} (1-i) \right)~,\\
E_3&= -\frac{\beta^2}{2\mu}-i\frac{\Gamma_*}{2}~.\nn
\end{align}
For the triplet virtual-state pole in case 2.II, we have the pole positions 
\begin{align}
\label{170321.3}
k_1&=-i\varkappa-\varkappa \rho^\frac{1}{3}~,\\
k_2&=-i\varkappa(1+\frac{\sqrt{3}}{2}\rho^\frac{1}{3})+\frac{\varkappa}{2}\rho^\frac{1}{3}~,\nn\\
k_3&=-i\varkappa(1-\frac{\sqrt{3}}{2}\rho^\frac{1}{3})+\frac{\varkappa}{2}\rho^\frac{1}{3}~,\nn
\end{align}
which imply the energies
\begin{align}
\label{170321.3b}
E_1&=E_R(1-i2\rho^\frac{1}{3})~,\\
E_2&=E_R(1+\frac{\sqrt{3}}{2}\rho^\frac{1}{3})(1+\frac{\sqrt{3}}{2}\rho^\frac{1}{3}+i\rho^\frac{1}{3})~,\nn\\
E_3&=E_R(1-\frac{\sqrt{3}}{2}\rho^\frac{1}{3})(1-\frac{\sqrt{3}}{2}\rho^\frac{1}{3}+i\rho^\frac{1}{3})~.\nn
\end{align}
 Higher orders in $\rho^\frac{1}{3}$ have been neglected in Eqs.~\eqref{170321.3} and \eqref{170321.3b}

 As noticed above,  because of this non-analytic behavior in $\rho$,
 the imaginary parts for the pole positions in energy, except for $E_3$ in case 2.I which is just a simple pole 
in the limit $\Gamma_*=0$,  are much larger in absolute value   than a naive estimation from the 
width of the constituent $D^{* 0}$.
 In particular, one can immediately deduce from Eqs.~\eqref{170321.2}  
 that the imaginary parts have opposite signs for the  poles $E_{1,2}$ of case 2.I.
For the case 2.II it follows from Eq.~\eqref{170321.3b} that the pole at $E_1$ has 
a positive imaginary part while the latter is negative for both poles at $E_2$ and $E_3$.
 As far as we know this is the first time that it is noticed such non-analytic behavior of the pole positions
 in the width of one of its constituents for  higher degree poles.
  Of course, this might have important phenomenological implications. In particular, for our present analyses it favors
to extent the virtual-state signal to energies above the $D^0\bar{D}^{*0}$ threshold, 
 because it increases the overlapping with the
$D^{* 0}$ Lorentzian in Eq.~\eqref{eq:eventDDbarf(E)}.
 Within other context, non-analyticities of the pole positions as a function of a strength parameter near a two-body threshold around
the point where the two conjugated poles meet have been derived in Refs.~\cite{hyodo.170321.1,rios.170319.1}.
Similar behavior has also been found as a function of quark masses for chiral extrapolations
\cite{bernard.170322.1,ledwig.170322.1,guo.170322.1,guo.170321.1}.

 The residues of the poles, given in the fourth column of Table~\ref{tab:fitsummary}, are very large for the case 2.I and huge 
for the case 2.II. The point is that they are affected by the extra singularity coming from the
other coalescing poles in the limit $\Gamma_*\to 0^+$.
 For the virtual-state cases one cannot interpret the normalization constants $Y_F$
 as yields because the virtual-state pole is below threshold in the 2nd RS and then it is  blocked
 by the threshold branch-point singularity, so that
it does not directly influence the physical axis for $E<0$. 
This also manifests in that the normalization integral ${\cal N}$, Eq.~\eqref{231116.2}, is very different from 1.

\subsection{Cases 3: Simultaneous virtual and bound state}
\label{subsec:virtualboundI}

In this case we again use the more general parameterization based on ${\ct(E)}$ and move towards a scenario in which one finds
simultaneously a bound-state pole in the 1st RS and a virtual-state one in the 2nd RS. To end with such a situation
we impose that in the isospin limit there is a double virtual-state pole independently of the common  masses
taken for the isospin multiplets (either the masses of the neutral or  charged isospin $D^{(*)}$ members).
This is a way to enforce a weak coupling of the bare states with the
continuum,  have poles in different RS's  and end with a bound state with small compositeness (or large elementariness).
 At this point we adapt, as an intermediate step to end with our elastic $D^0\bar{D}^{*0}$ $S$ wave,
the main ideas developed in Ref.~\cite{guo.251116.1}. This reference takes into account the
coupled-channel structure
of $\pi^+\Sigma_c^0$, $\pi^-\Sigma_c^{++}$ and $\pi^0\Sigma_c^+$ in relation with the $\Lambda_c(2595)^+$ resonance,
where the symbol $\Sigma_c$ actually refers to the $\Sigma_c(2455)$ \cite{pdg.181116.2}.
However, the resulting expression reduces to that of Eq.~\eqref{211116.3} for single coupled-scattering
 since we focus on the $X(3872)$ signal region around the
$D^0\bar{D}^{*0}$ threshold, because of the same reasons as already discussed when matching our results with those of
Ref.~\cite{artoi.170310.1} in the last part of Sec.~\ref{sec.211116.1}. Of course, these considerations translate 
into a  different  dependence of 
$\lambda$ and $\beta$ on $E_R$  than in the cases 2.I-II analyzed in Sec.~\ref{subsec:virtualI}.

 The basic strategy is the following:
i) We take the isospin limit for $D^0 \bar{D}^{*0}$ and $D^+D^{*-}$ coupled channel scattering,
 with masses equal to either those of the neutral particles for each isospin doublet ($D^0$ and $D^{* 0}$)
or to the charged ones ($D^+$ and $D^{* +}$). At this early stage the zero width limit for the $D^{*0}$ is taken.
ii) For every isospin limit defined in i) we impose having the same virtual-state pole position located at $E_R-iG_R/2$,
 taking the limit $G_R\to 0^+$ at the end,  similarly as done in Sec.~\ref{subsec:virtualI} to end with less free parameters in a more restricted situation.
iii) The previous point provides us with four equations that are used to fix two of the
three parameters in ${\ct}(E)$, Eq.~\eqref{211116.7}, namely $\beta$ and $\lambda$.\footnote{The other two equations give the values for the
CDD pole positions in the two isospin limits considered. An information which is of not use.}
The remaining third parameter,  $\MCDD$, is also fixed by imposing
that  ${\ct}(E)^{-1}$ vanishes in the 1st RS below threshold at $E_R$.
iv) In this way all the parameters specifying ${\ct}(E)$ are given in terms of $E_R$ which is finally fitted to data once the  finite $D^{*0}$
 width is restored in the definition of the three-momentum, cf. Eq.~\eqref{211116.4}.

 The formulae derived to actually fix $\beta$, $\lambda$ and $\MCDD$ are given in the Appendix \ref{app.251116.1},
Eqs.~(\ref{291116.2},\ref{291116.3},\ref{291116.7},\ref{291116.6}).
There it is shown that indeed one has two solutions, that we indicate by case 3.I (first solution) and 3.II (second solution).
 The expressions simplify in the limit $|E_R|/\Delta\to 0$, which is relevant for the $X(3872)$  given its small energy,
 and the two solutions coalesce in just one.   In this case we show that there are two poles in different RS's,
confirming the intuitive  physical reasons given at the beginning of this Section.

 The values for the fitted parameters are given in Eq.~\eqref{para:caseiii} for the case 3.I and in Eq.~\eqref{para:caseiiiII} for 
the case 3.II.
 The resulting event distributions are shown by the blue dash-double-dotted (case 3.I) and green dash-dotted  (case 3.II) 
lines in Fig.~\ref{fig:resultI} and in the histogram of Fig.~\ref{fig:insetDD}. These lines are hardly distinguishable among them and can only 
be differentiated with respect to case 1 in the $D^0\bar{D}^{*0}$ event distribution, 
 as one can appreciate clearly from Fig.~\ref{fig:insetDD}.  With respect to
cases 2.I-II we have the already commented shift of the peak in the $J/\psi \pi^+\pi^-$ event distributions, more clearly seen in Fig.~\ref{fig:CDFBand}. 
The global  reproduction of data is of similar quality as the one
already achieved by the pure bound-state and virtual-state cases.
  The values for the pole positions and CDD parameters are given in Tables~\ref{tab:fitsummary} and \ref{tab:t.fitC}, respectively.

For the case 3.I the resulting pole position for the bound state (the pole in the 1st RS) is $-0.50^{+0.04}_{-0.03}$~MeV, 
 while the pole position for the virtual state (the pole in the 2nd RS) is $-0.68^{+0.05}_{-0.03}$~MeV.  
In both cases there is a tiny imaginary part of the order of $10^{-3}$~MeV which is beyond the precision shown.
 The compositeness of the $D^0 \bar{D}^{*0}$ state in the bound state, 
evaluated in the same way as explained at the end of Sec.~\ref{subsec:f(E)I},
is  $0.06$, i.e., the $D^0 \bar{D}^{*0}$ component only constitutes around a 6\% of the $X(3872)$ due to the extreme
proximity of the  CDD pole  to the  $D^0 \bar{D}^{*0}$ threshold. As shown in Table~\ref{tab:t.fitC} 
the CDD pole is much closer to threshold than the bound-state pole. 
 As a result, other components are dominant, e.g. one could think of the conventional $\chi_{c1}(2P)$ as $c\bar c$,
tetraquarks, hybrids, etc \cite{Nieves.251116.1,Entem.251116.1,ZhengHQ,Cillero.251116.1,Chao.251116.1,Suzuki.251116.1,ferretti.170324.1}.
These facts about the smallness of the imaginary part of the two near-threshold poles and the small compositeness for the bound state
can be understood in algebraic terms in the limit $|E_R|/\Delta\to 0$ as shown in Appendix~\ref{app.251116.1}, cf.
Eqs.~(\ref{170323.12},\ref{170323.9},\ref{170323.15}).
 Indeed, they are related because if the $X(3872)$ has such a small value for the compositeness
 then it is fairly insensitive to the width of the  $D^{*0}$.
In addition, one also has a deep virtual state located at
$E_3\approx -\beta^2/2\mu-i\Gamma_*/2$, that is quite insensitive
to the CDD pole contribution, which is strongly suppressed at those
energies as explained in more detail after Eq.~\eqref{170323.14}.

 Let us notice that  the ERE  for the present near-threshold bound state
fails because ERE is not applicable since the zero is closer to threshold than the pole. Taking $\Gamma_*=0$ and
calculating $a$ and $r$ we obtain the central values (errors are given in Table~\ref{tab:t.fitC})
\begin{align}
\label{271116.2}
a&=0.27~\text{fm}\to 0~,\quad
r=-847.8~\text{fm}\to \infty~,
\end{align}
 while the binding momentum is $\kappa=\sqrt{2\mu |E_X|}=31.0$~MeV$<< 1/|a|=723$~MeV. 
 The pole position in the 1st (2nd) RS that stems from the ERE up to the effective range  is $-0.17$ $(-0.18)$ ~MeV,
 which is indeed very different to the actual pole position of the bound (virtual) state in the full amplitude at $-0.50$ ($-0.68$)~MeV.

\begin{figure}
\begin{center}
\begin{tabular}{ll}
\includegraphics[width=0.45\textwidth]{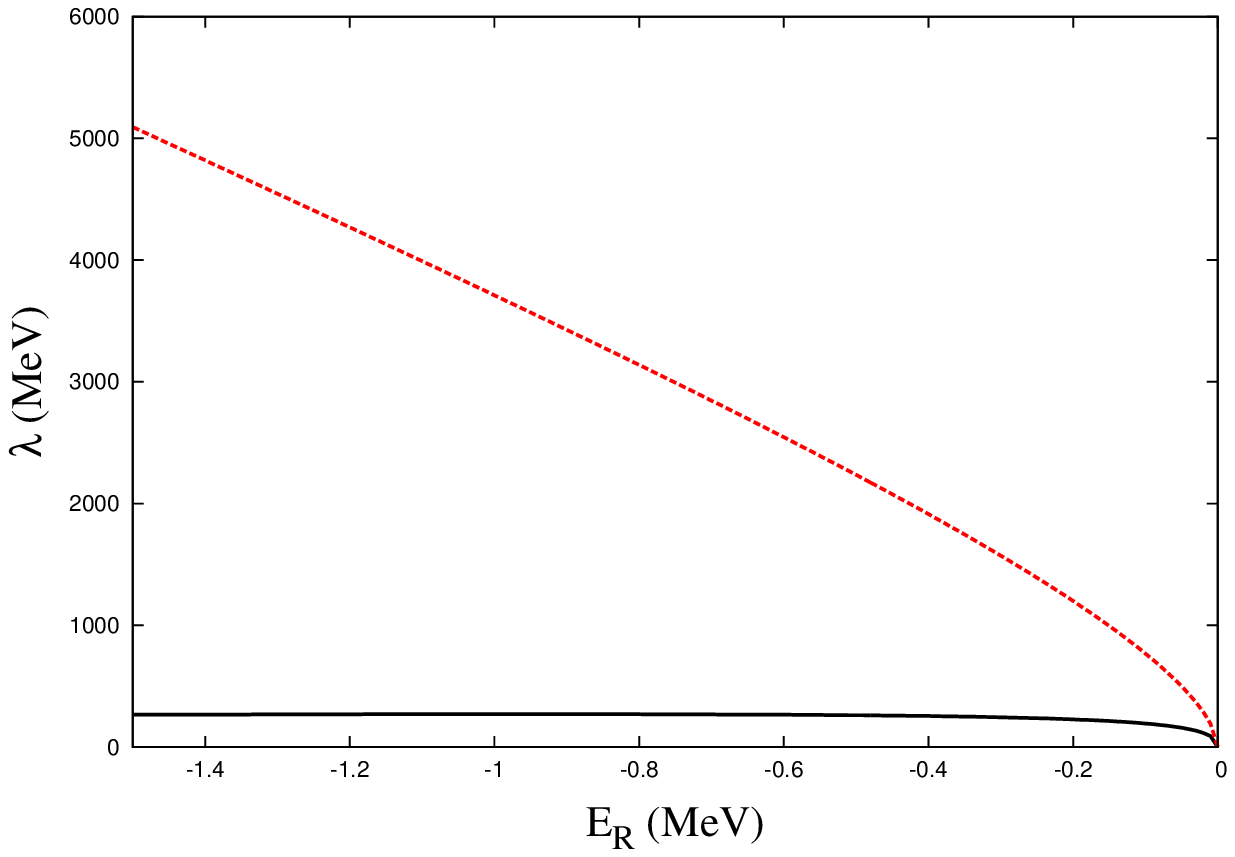} &
\includegraphics[width=0.45\textwidth]{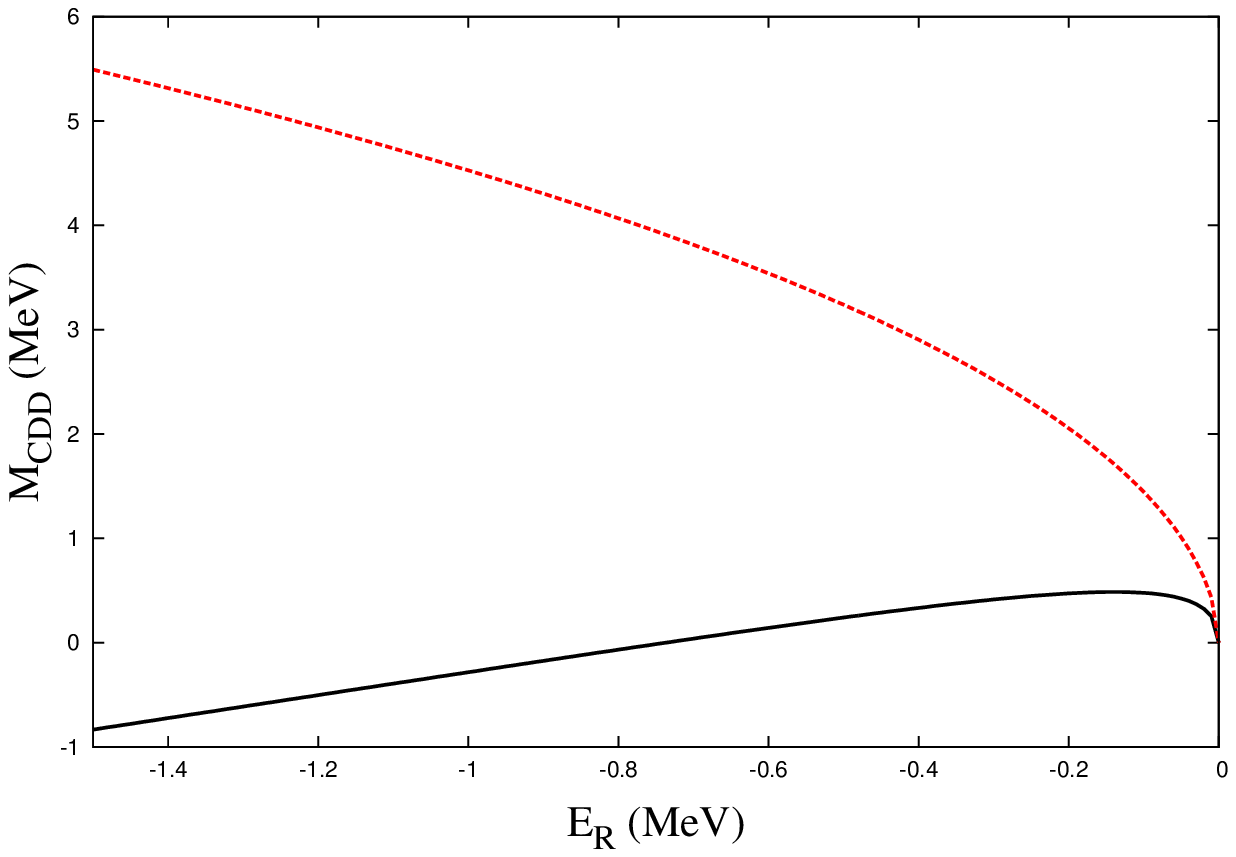}
\end{tabular}
\caption{The left panels shows $\lambda$ and the right one $\MCDD$ as a function of $E_R$ for the cases 3.I (black solid) and 3.II (red dashed lines).}
\label{fig.011216.1}
\end{center}
\end{figure}

\begin{figure}[ht]
\begin{center}
\vglue 0.5cm
\includegraphics[width=0.7\textwidth]{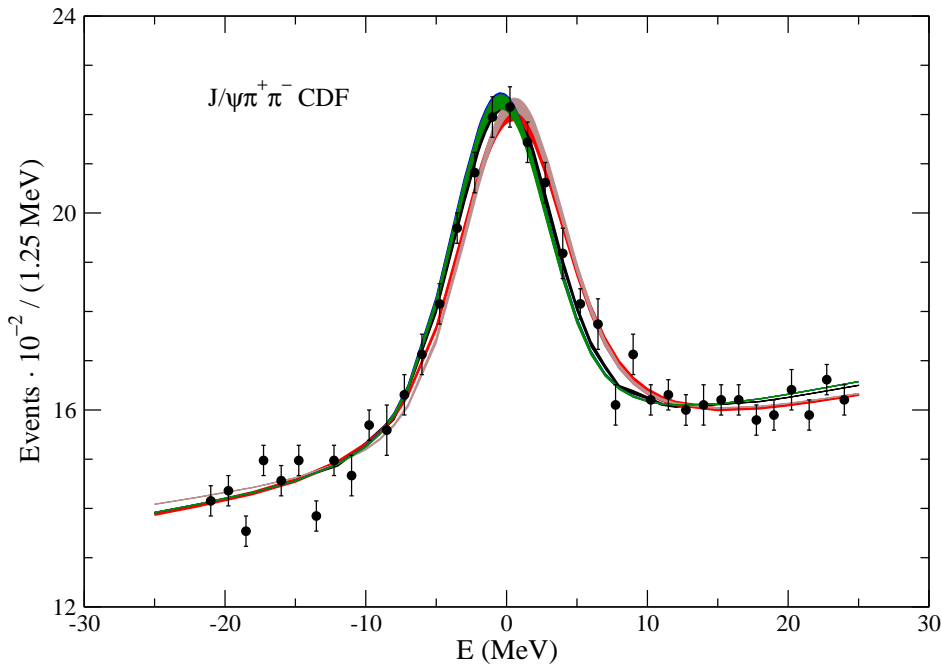}
\caption{The curves of the different cases for the $J/\psi\pi^+\pi^-$ event distribution from the inclusive $p\bar{p}$ scattering \cite{CDF.211116.5}
are given with error bands included. The black-, red-, brown-, blue-, and green-filled bands
correspond to cases 1, 2.I, 2.II,  3.I and 3.II, in order.}
\label{fig:CDFBand}
\end{center}
\end{figure}

  For the second solution, i.e. case 3.II, we have much larger values of $\lambda$ and $\MCDD$ than for the first one,
 compare between the last two rows in Table~\ref{tab:t.fitC}. 
 This is a  common characteristic to any value $E_R<0$ as shown in Fig.~\ref{fig.011216.1}, where the values
of $\lambda$ (left panel) and $\MCDD$ (right panel) are given as function of $E_R$ for the first (black solid) and second (red dashed lines) solutions.

The pole positions in the 1st and 2nd RS's are given in the second column of the last line of  Table~\ref{tab:fitsummary}.
The fact that for this second solution $\MCDD$ is further away from threshold than for the solution case 3.I is an indication that
compositeness is larger for the former than for the latter.  For the case 3.II  we obtain now that $X=0.16$,
 while before it was around 0.06. This is in agreement with our expectations, but still $X$ is small and the state is
dominantly a bare (non-molecular) one.\footnote{Taking the same value for the residue of $D^+D^{*-}$ as for $D^0\bar{D}^{*0}$,
because of isospin symmetry, we can evaluate straightforwardly the compositeness of the $D^+D^{*-}$ channel 
and is a factor 3 smaller than for $D^0\bar{D}^{*0}$.
Then, summing it to 0.15 we end with 0.20 as an estimated value for the total compositeness of the $D\bar{D}^*$ states, which is still  small.
For case 3.I it would be around only 0.05.} The
residues for this case are also given in the column four of Table~\ref{tab:fitsummary}. They are larger by around a factor 3 compared to the
 first solution, which is in line with the increase in the value of compositeness. The bound states for the cases 3.I-II have a normalization
integral ${\cal N}=1$ so that it is legitimate to interpret the $Y_F$ as yields.

The ERE expansion for the case 3.II is better behaved because $\MCDD$ is relatively  further from threshold. We now have the values ($\Gamma_*=0$
should be understood in the following discussions)
\begin{align}
\label{021216.1}
a&=+1.57_{-0.02}^{+0.05}~\text{fm}~,\\
r&=-43.75^{+0.24}_{-0.61}~\text{fm}~,
\end{align}
 where $r$ is still  much larger than a typical range of strong interactions and $a$ is much smaller than $1/\sqrt{2\mu|E_X|}$.
 These facts just reflect the dominant bare nature of the $X(3872)$ in this case.
 The ERE up to $r$ gives rise to a bound state  located
at $-0.45$~MeV, already very close to the full-solution result  at $-0.51$~MeV, which is much better than for the case 3.I.
Regarding the virtual-state pole the ERE also produces a pole in the 2nd RS at $-0.76$~MeV, while the full result is  at $-1.06$~MeV, around a 25\% of
error.  This worse behavior of the ERE to determine the location of the virtual-state pole is to be expected because the radius of convergence
of the ERE is $2\mu \MCDD$, and the virtual-state pole is closer to this limit than the bound-state one.
More contributions  are certainly needed as the ERE is applied to
 energies that are closer to the radius of convergence of the expansion.

 Related to this discussion we consider the Weinberg's compositeness theorem for
 a near-threshold bound state, which reads \cite{Weinberg.141116.6}
\begin{align}
\label{271116.3}
a&=\frac{2(1-Z)}{(2-Z)\kappa}~,\nn\\
r&=-\frac{Z}{(1-Z)\kappa}~,
\end{align}
where $Z$ is the elementariness, or $1-X$.
This criterion, as discussed in the Introduction,  cannot be applied if a CDD is closer to threshold
than the bound-state pole, as it happens for the case 3.I,  because it relies on the applicability of the ERE up to the effective range.
For the case 3.II this is not the case, but still the CDD pole is quite close so that energy dependences beyond the
effective range play a role. 
The Weinberg's compositeness relation  gives $Z_a=0.86$ and $Z_r=0.87$, when using  $a$ and $r$ to calculate it
from  the first and second expressions in Eq.~\eqref{271116.3}, respectively.
 These numbers compare very well with $Z=1-X=0.84$, where $X=0.16$ is determined above and given in Table~\ref{tab:fitsummary}.
 These values of elementariness so close to 1  for cases 3.I-II  are also in agreement with the expectation of having two poles
close to threshold in adjacent RS's (virtual- and bound-state poles simultaneously),
 which fits very well within the Morgan's criterion for a preexisting or non-molecular state \cite{Morgan.271116.1}.


Our fit for the case 3.II corresponds to the following central values for the parameters
characterizing  the scattering model of Ref.~\cite{baru.170310.2},
 in terms of the exchange of a bare state and direct scattering between the $D^0\bar{D}^{*0}$:
 $g_{f0}=0.014$, $\gamma_V=-561.7$ and $E_f=-0.71$~MeV. Compared with the cases 2.I-II one observes a
significant smaller value for $g_{f0}$ and $|E_f|$. The resulting pole trajectories as $g_f$ is increased from  $0.1g_{f0}$ up to
 $20 g_{f0}$, with $E_V$ and $\gamma_V$ held fixed,
 are shown in the bottom-right panel of Fig.~\ref{fig.170322.1}, with a similar behavior for the case 3.I which is not shown.
Notice that because $E_f<0$  one has in the decoupling limit ($g_f\to 0$) a bound- and a virtual-state pole  at $\pm i \sqrt{2\mu|E_f|}=33.2$~MeV.
 This type of pole movement as a strength parameter varies is different to those discussed in Ref.~\cite{rios.170319.1}, because
the near-threshold poles do not belong to the trajectory of two complex poles associated with the same resonance. However, this is the
case for the two virtual state poles, the shallow and deep ones, as clearly seen in the figure.

 The poles trajectories in the last panel of Fig.~\ref{fig.170322.1} do not belong either to the ones discussed explicitly in Ref.~\cite{baru.170310.2},
 where much smaller values of $|\gamma_V|$ are considered. The reason behind is the misused performed in this reference of
the relationship between the pole positions $k_i$, $i=1,2,3$ and the position of the CDD pole\footnote{Similarly, one can also derive the equalities
$\lambda=-i(k_3^2(k_1+k_2)+k_3(k_1+k_2)^2+k_1k_2(k_1+k_2))/2\mu$ and $\beta=i\sum_i k_i$.
These relations can be easily worked out from the secular equation which is a third order polynomial.}
\begin{align}
\label{170324.1}
\MCDD&=-\frac{1}{2\mu}(k_3(k_1+k_2)+k_1k_2)~.
\end{align}
The point is that Ref.~\cite{baru.170310.2} concluded from this equation that
it is necessary that the three poles be shallow ones $(|k_i|\ll \Delta)$ in order to have a near-threshold CDD pole ($|\MCDD|\ll \Delta$).
 However, this conclusion is just sufficient but not necessary.
 The other possibility is that $k_1$ and $k_2$ nearly cancel  each other (such that $|k_1+k_2|={\cal O}(|k_{1,2}/k_3|^2$),
 without being necessary that $|k_3|\ll \Delta$ (which in our case is given by $\beta\gg \Delta$).
 This is what happens particularly for the case 3.I with $k_1=i 31.0$ and
$k_2=-i 36.2$~MeV, so that the CDD pole is almost on top of threshold.
Therefore, one does not really need that three poles lie very
 close to threshold to end with a shallow CDD pole.

It is also interesting to apply the spectral density introduced in Eq.~\eqref{170322.1} to evaluate the compositeness and
elementariness of the bound states in the cases 3.I-II. For such purpose one has to integrate the spectral
density up to infinity with $W$ defined then as
\begin{align}
\label{170324.2}
W&=\int_0^\infty dE \omega(E)~,
\end{align}
and interpreted as the compositeness $X$ \cite{baru.170310.1}.
  The normalization to 1 of the bare state then guarantees that $Z=1-X$, which provides us with the elementariness.
Notice that this is the third way that we have introduced to evaluate the compositeness of a bound state. Namely, we can evaluate it
in terms of the residue of $t(E)$ at the pole position, Eq.~\eqref{170319.2}, Weinberg's relations, Eq.~\eqref{271116.3}, or
in terms of the spectral density, Eq.~\eqref{170324.2}. The latter also provides remarkably close values to the previous ones so
 that for the case 3.II one has $W=0.16$, while for the case 3.I (in which case Weinberg's result does not apply)
one obtains $W=0.06$.

We have also checked that our results are stable if the $D^+D^{*-}$ channel is explicitly included in ${\ct}(E)$ by using the same
formalism as in Ref.~\cite{guo.251116.1}. At the practical level this amounts to modifying the denominator of ${\ct}(E)$ such that
$\beta\to \beta-i[k^{(2)}(E)-\overline{k}^{(2)}(0)]$, with $k^{(2)}(E)$ the $D^+D^{*-}$ three momentum given by the expression
$k^{(2)}(E)=\sqrt{2\mu_2(E+i\widetilde{\Gamma}_*/2-{\Delta})}$, where $\mu_2$ is the
reduced mass and $\widetilde{\Gamma}_*$ the width of the $D^{*+}$ resonance \cite{pdg.181116.2}, while
$\overline{k}^{(2)}(0)$ is given by the same expression with $E+i\widetilde{\Gamma_*}/2\to 0$. For example, by redoing the fit in this case
the fitted parameters match  very well the values in Eq.~\eqref{para:caseiii} within errors.

 In order to have an extra perception on how the uncertainty in the fitted parameters influences our results, we also show
in Fig.~\ref{fig:CDFBand} the error bands of the curves obtained from the different cases considered
for the reproduction of the  CDF data on the inclusive $p\bar{p}$ scattering to $J/\psi\pi^+\pi^-$ \cite{CDF.211116.5}.
  We have not shown the error bands for the other data, and just shown the curves obtained from the central values of the fit parameters
in Figs.~\ref{fig:resultI},  because the typical width for every error band in each line is of similar size
as the one shown in Fig.~\ref{fig:CDFBand}.
 We have chosen this data because it is the one with the smallest relative errors,
having the largest statistics and smallest bin width.
 In addition, the curves are so close to each other that  in the scale of the Fig.~\ref{fig:resultI}
  it would be nearly impossible to distinguish between all the curves with additional error bands included.
Here we offer just one panel which also allows us to use a larger size for it and be able to distinguish
better between lines with error bands. But even then, one clearly sees in Fig.~\ref{fig:CDFBand}
that the bands for the cases 1 and 3.I-II mostly overlap each other so that they are hardly distinguishable.
 The cases 2.I-II can be differentiated from the rest because there is a slight shift of the peak structure to the right. 
However,  this shift becomes smaller and all the bands overlap each other if we had excluded in the fit the $D^0\bar{D}^{*0}$ 
event distribution of the BaBar Collaboration \cite{BaBarD}.

\section{Conclusions}
\label{sec.conclusions}

Since its exciting discovery \cite{Belle.051216.1} the $X(3872)$ has been extensively studied, for a recent review
see Ref.~\cite{Lebed.051216.1}. Among the many theoretical approaches \cite{Nieves.251116.1,Entem.251116.1,ZhengHQ,Cillero.251116.1,Chao.251116.1,ferretti.170324.1,Suzuki.251116.1,oset.051216.1,gamermann.051216.1,fleming.051216.1,Hanhart.141116.4,Braaten.141116.3,Braaten,Kalashnikova.141116.2,Qiang.141116.1,liu.081216.1}, we have
paid special attention to the applicability of the popular ERE approximation up to and including
the effective range contribution to study near-threshold
 states like the $X(3872)$. We have elaborated about the fact that the ERE convergence radius might be severely limited due to the presence of
near-threshold zeroes of the partial wave, the so-called Castillejo-Dalitz-Dyson poles. We have then derived a parameterization that is more general
than the ERE up to and including effective range,\footnote{Indeed up to the next shape parameter, $v_2$. Let us recall again that the ERE
is more general than a Flatt\'e parameterization.}
but it can  deal as well with the presence of a CDD pole arbitrarily close to threshold.
 We have shown too that other parameterizations based on the picture of the exchange of a bare state plus direct interactions
between the $D^0\bar{D}^{*0}$ can be also matched into our parameterization \cite{baru.170310.2,artoi.170310.1}.
 In particular, Ref.~\cite{baru.170310.2} already stressed the strong impact that a possible
near-threshold zero would have in the $D^0\bar{D}^{*0}$ $S$-wave amplitude.
 However, we have shown that the conclusion there stated about the necessity of three simultaneous shallow poles  to
 end with a near-threshold zero  is  sufficient but not necessary, because it is just enough having two such poles.
We have illustrated this conclusion with a possible scenario for the $X(3872)$ in which there are only two near-threshold poles,
 a bound state in the 1st RS and  a virtual-state one in the 2nd RS.

We have then reproduced several event distributions around the $D^0\bar{D}^{*0}$ threshold including those 
 of $D^0\bar{D}^{*0}$ and $J/\psi \pi^+\pi^-$ from charged $B$ decays
measured by the BaBar and Belle Collaborations and the higher-statistics
 CDF $J/\psi \pi^+\pi^-$ event distributions
from inclusive $p\bar{p}$ scattering at $\sqrt{s}=1.96$~TeV.  Our formalism has as limiting cases
those of Refs.~\cite{Hanhart.141116.4,Braaten}, but it can also include other cases in which the presence of a
CDD pole plays an important role.
In this respect we are able to find other interesting scenarios  beyond those
found in Refs.~\cite{Hanhart.141116.4,Braaten} that can reproduce data fairly well, and without increasing the number of free parameters.
  In two of these new situations the $X(3872)$ is simultaneously a bound and a virtual state, while in others
 the $X(3872)$ is a double or a triplet virtual-state pole.
 In the limit of vanishing  width of the $D^{* 0}$ these  poles become degenerate and produce a higher-order
  pole (of second or third order).
 Thus, our parameterization constitutes  in these latter cases a simple example for higher order  $S$-matrix poles
 that could have a clear impact on particle physics phenomenology.

In this respect, we stress that the corrections to the pole position when taking into account the finite width of the
$D^{*0}$ resonance, $\Gamma_*$, are non-analytic for the higher-order poles of order $n>1$.
 In such situations one has that the
leading corrections are proportional to $\rho^{1/n}$, with $\rho=\Gamma_*/|E_R|$, being $E_R$ the real part of the pole position
with respect to the $D^0\bar{D}^{*0}$ threshold without the $D^{*0}$ width.
This could be an important source of $D^0\bar{D}^{*0}$ partial width for the $X(3872)$. 
 Indeed, with this mechanism one that the absolute value of twice the imaginary part of the
pole positions for the triplet-pole scenario could be nearly as large as  1~MeV, despite $\Gamma_*$ is only around 0.065 keV
\cite{pdg.181116.2,Braaten.141116.3}. Thus a measurement of the total width of the $X(3872)$ might be useful to discriminate between the discussed
scenarios.

 Further, while the compositeness is equal to 1 for the bound-state case analyzed making
use of the ERE including only the scattering length \cite{Braaten},
it is nearly zero for the cases 3.I-II in which the $X(3872)$ is a simultaneous virtual and bound state.
 The case 3.I has  the closest CDD pole to threshold, even closer than the pole positions.
In this respect, we also estimate that the $X(3872)$ is mostly  $D^0\bar{D}^{*0}$
 for the double virtual-state case, because the CDD pole is relatively far away from threshold, while in the
triplet-pole case the elementariness is dominant as indicated by the closeness of the CDD to threshold.
We have verified quantitative these conclusions as well by employing the spectral density function.

From another perspective, we have shown that using a more refined treatment of $D^0\bar D^{*0}$ scattering
the $X(3872)$ can be a a bound state, a double/triplet virtual-state  pole or  two types of simultaneous virtual and bound states with poles
occurring in both the physical and unphysical sheets, respectively.
 All these scenarios can give a rather  acceptable reproduction of the experimentally measured event distributions.
 Up to some extent this situation recalls the case of the  $X(1835)$, for which
the energy-dependent $J/\psi\to \gamma p\bar{p}$ event distribution is nicely reproduced by purely final-state interactions of $p\bar p$
 \cite{Sibirtsev}.
However, this treatment  fails to describe the data  when a more elaborate model is taken.
Only the generation of a $p\bar p$ bound state in the scattering amplitude is able to reproduce
the data within this more sophisticated  model \cite{Kang1835}.

From our present results and
this experience, more efforts are still needed  to finally  unveil the nature of the acclaimed $X(3872)$,
which is the first $XYZ$ state  observed.
In this respect, we mention that there are visible differences between the different scenarios analyzed in the $D^0\bar{D}^{*0}$
 invariant mass distributions in the peak of the $X(3872)$, as shown in detail in Fig.~\ref{fig:insetDD}, in particular between 
the scenarios I, 2.II and the rest.
 We have to indicate that the present data shows a 
 clear displacement towards higher masses of the $X(3872)$ peak in the BaBar Collaboration data 
\cite{BaBarD} as compared with the Belle Collaboration one \cite{BelleD}. Indeed, if the former data is excluded in the fits 
the shift towards the right of the signal peak for the cases 2.I-II 
  in the $J/\psi \pi^+\pi^-$  CDF data \cite{CDF.211116.5}, cf. Fig.~\ref{fig:CDFBand},  diminishes considerably.  
Thus, a future high-statistic experiment on $B\to K D^0\bar{D}^{*0}$ might be very helpful to differentiate between different cases,  
if complemented with high-precision data on $J/\Psi \pi^+\pi^-$. 
   Another way to discriminate between different possibilities might be the measurement
 of the partial decay width of the $X(3872)$ to $D^0\bar{D}^{*0}$, as mentioned above.
   Lattice QCD can also provide interesting information from where one could deduce the $D^{0}\bar{D}^{0*}$ near-threshold
scattering amplitude and then determine whether there is a CDD pole or not.
 Indeed, present Lattice QCD results point towards the importance of the interplay between quark and meson degrees of freedom
to generate the $X(3872)$ \cite{lqcd.170325.1,lqcd.170325.2,lqcd.170325.3}.
 Another interesting idea was put forward by
Voloshin in Ref.~\cite{voloshin.181116.1} indicating the convenience to measure the $D^0\bar{D}^{*0}\pi^0$ Dalitz plot to
distinguish between the molecular and quarkonium picture for the $X(3872)$.

\section*{Acknowledgements}

We would like to thank  C. Hanhart for inspiring and compelling discussions.
Interesting discussions are also acknowledged to Q.~Zhao.
 This work is supported in part by
 the MINECO (Spain) and ERDF (European Commission) grant FPA2013-40483-P and
  the Spanish Excellence Network on Hadronic Physics FIS2014-57026-REDT.

\appendix

\section{Obtaining properly normalized yields}
\label{app.081016.1}
\def\theequation{\Alph{section}.\arabic{equation}}
\setcounter{equation}{0}

In order to fit the event distributions when using ${\ct}(E)$ with a CDD included, the evaluation of $\alpha$ entering in
Eq.~\eqref{231116.1} for calculating the event distributions requires to work out the pole position, cf. Eq.~\eqref{211116.8b},
 which is not easily expressed in terms of free parameters.
 This is not the case when using the function $f(E)$  because in this case $\alpha$, Eq.~\eqref{281116.1},
 is given directly in terms of $\gamma$, a fit parameter.
Thus, when ${\ct}(E)$ is employed we first perform the fits such that $|\alpha|^2$ is re-absorbed in the normalization
constants multiplying the signal contribution. In this way one avoids having to calculate the pole position for each iteration in the
fit procedure. Being specific, we use the following expression for the fits to the $D^0\bar{D}^{*0}$ event distributions
\bea
\label{eq:eventCDDfitD}
N_i(E_i)&=& \int_{E_i-\Delta/2}^{E_i+\Delta/2}dE''\int_0^\infty d E' R(E'',E') \sqrt{E'}\nl
&&\times \left\{\mathfrak{b}_D\int_{-\infty}^\infty dE |d(E)|^2 \frac{\Gamma_*}{(E'-E)^2+\Gammas^2/4}+\widetilde{\text{cbg}}_D\right\}~.
\eea
For the fits to the $J/\psi\pi^+\pi^-$ event distributions,  we employ
\beq
\label{eq:eventCDDfitJ}
N_i(E_i)=\vartheta_J\left\{\mathfrak{b}_J\int_{E_i-\Delta/2}^{E_i+\Delta/2}dE'\int_{-\infty}^{\infty}
dE R(E',E)|d(E)|^2+\widetilde{\text{cbg}}_J\Delta \right\}~,
\eeq
 where $\vartheta_{J,D}=1$ for BaBar data and for Belle it corresponds
 to the ratio of the number of $B\bar{B}$ pairs produced in Belle and BaBar for each type of $B^+\to J/\psi \pi^+\pi^-$ decays, namely,
$N_{B\bar{B}}^{\text{Belle}}/N_{B\bar{B}}^{\text{BaBar}}$. The number of $B\bar{B}$ pairs is given in Table ~\ref{tab:expdata}.

Once the fit is performed we can deduce the values of  the ``yields'' $Y_D$ and $Y_J$
 (here the quotation marks are introduced because it is required that ${\cal N}\simeq 1$, with the normalization constant ${\cal N}$
 introduced in Eq.~\eqref{231116.2}, in order to interpret meaningfully these constants as yields).
 The appropriate relations can be deduced by comparing Eqs.~\eqref{eq:eventDDbarf(E)} and \eqref{eq:eventJpsif(E)}
 with Eqs.\eqref{eq:eventCDDfitD} and \eqref{eq:eventCDDfitJ}, in order. They read
\begin{align}
\label{051216.1}
Y_D &=  \mathfrak{b}_D\frac{ \sqrt{2}\pi\sqrt{E_X+\sqrt{E_X^2+\Gammas^2/4}} }{g_{\rm{CDD}} }~,\\
Y_J &= \frac{\mathfrak{b}_{J}}{g_{\rm{CDD}} }~,\nn\\
g_{\rm CDD}&=\frac{\Gamma_X}{2\pi|\alpha|^2}~.\nn
\end{align}
 The pole position  $E_X-i\Gamma_X/2$ and associated momentum $k_P$ are determined  from the fitted values of the parameters.
We also have the trivial relations between the background parameters
\begin{align}
\label{281116.2}
\widetilde{\text{cbg}}_J& =N_{B\bar{B};J}^{\text{BaBar}} \text{cbg}_J~,\\
\widetilde{\text{cbg}}_D & =N_{B\bar{B};D}^{\text{BaBar}} \text{cbg}_D~.\nn
\end{align}

An analogous procedure is also applied when fitting the $J/\psi \pi^+\pi^-$ event distribution from the inclusive $p\bar{p}$
scattering measured by the CDF Collaboration \cite{CDF.211116.5}.  In this way, we can extract $Y_J^{(p)}$ by applying Eq.~\eqref{051216.1} too.

\section{Pole positions with finite $\Gamma_*$ in the cases 2.I,II.}
\label{app.170320.1}
\def\theequation{\Alph{section}.\arabic{equation}}
\setcounter{equation}{0}

For $\Gamma_*\neq 0$ the relation between energy and momentum is $E=k^2/2\mu-i\Gamma_*/2$, cf. Eq.~\eqref{211116.4},
and the secular equation Eq.~\eqref{170319.4} becomes
\begin{align}
\label{170321.4}
\lambda+(\frac{k^2}{2\mu}-i\frac{\Gamma_*}{2}-\MCDD)(-ik+\beta)&=0~.
\end{align}
After some simplifications it can be written as
\begin{align}
\label{170321.4}
(k+i\varkappa)^2\left(k-i\frac{\varkappa^2-2\mu\MCDD}{2\varkappa}\right)
-i\mu\Gamma_*\left(k+i\left[\frac{3\varkappa}{2}+\frac{\mu\MCDD}{\varkappa}\right]\right)&=0~,
\end{align}
with $\varkappa=\sqrt{2\mu|E_R|}$.

For the case 2.I $|\MCDD|\gg |E_R|$ and the previous equation becomes in good approximation
\begin{align}
(k+i\beta)\left((k+i\varkappa)^2-i\mu\Gamma_*\right)&=0~.
\end{align}
Its solution gives rise to the pole positions in the $k$ plane given in Eq.~\eqref{170321.1} and
in the $E$ plane are those of Eq.~\eqref{170321.2}.

For the case 2.II, $\MCDD=-3E_R$ and Eq.~\eqref{170321.4} becomes
\begin{align}
\label{170321.5}
(k+i\varkappa)^3-i\mu\Gamma_*(k+i 3\varkappa)&=0~.
\end{align}
In terms of the dimensionless variable
\begin{align}
\label{170321.6}
t&=\frac{k+i\varkappa}{\varkappa}
\end{align}
Eq.~\eqref{170321.5} reads
\begin{align}
\label{170321.7}
t\left(t^2-i\frac{\Gamma_*}{2|E_R|}\right)&=-\rho~.
\end{align}
This equation can be solved in a power expansion of $\rho$ with the leading result
\begin{align}
t_1&=-\rho^\frac{1}{3}+{\cal O}(\rho^\frac{2}{3})~,\\
t_2&=e^{-i\frac{\pi}{3}}\rho^\frac{1}{3}+{\cal O}(\rho^\frac{2}{3})~,\nn\\
t_3&=e^{i\frac{\pi}{3}}\rho^\frac{1}{3}+{\cal O}(\rho^\frac{2}{3})~.\nn
\end{align}
In the momentum and energy variables this solution gives rise to Eqs.~\eqref{170321.3} and \eqref{170321.3b}, in order.
\section{Formalism for the cases 3.I-II: Simultaneous virtual- and bound-state poles}
\label{app.251116.1}
\def\theequation{\Alph{section}.\arabic{equation}}
\setcounter{equation}{0}

 Let us connect with the $T$-matrix derived in a previous paper
by one of the authors in which the resonance $\Lambda_c(2535)^+$ was studied \cite{guo.251116.1}.
 This is a resonance that also lies very close to the $\pi\Sigma_c$ thresholds.
Namely, it has a small width of $2.6\pm 0.6$~MeV \cite{pdg.181116.2} and its mass is
 4.37~MeV above the $\pi^0\Sigma_c^+$ threshold, and 1.06 and 1.30~MeV below the nearly degenerate thresholds of $\pi^+\Sigma_c^0$ and
$\pi^-\Sigma_c^{++}$, respectively.
 Reference~\cite{guo.251116.1} explored the viability of this resonance to be a preexisting one, with $Z\simeq 1$,
so that its actual pole position is unaffected by taking as masses in the isospin limit those
 in the $\pi^0\Sigma_c^+$ or in the $\pi^-\Sigma_c^{++}$ states.\footnote{Contrarily to the more general expectations of
Ref.~\cite{guo.251116.1} the parameter $\lambda$ could depend strongly on the isospin mass taken.}

Here we have adopted a similar point of view and required the invariance of the virtual-state pole position independently of whether
one takes    in the isospin limit the masses of the neutral mesons $D^0$, $D^{*0}$ or of the charged ones $D^+$, $D^{*+}$.
 We denote the channel $D^0\bar{D}^{*0}$ by 1 and the channel made by the
charged particles $D^+D^{*-}$ by 2. In the isospin limit, we replace $\lambda$ by
$\widetilde{\lambda}$ in Eq.~\eqref{211116.3}, with
\begin{align}
\label{291116.4}
\lambda \to 2\widetilde{\lambda}
\end{align}
because of an isospin Clebsch-Gordan coefficient squared to combine $D\bar{D}^{*}$ in
isospin 0. Specifically, in the isospin limit we use the scattering amplitude
\begin{align}
\label{291116.1}
t_j(\sqrt{s})&=\left(\frac{\widetilde{\lambda}}{\sqrt{s}-\MCDD^{(j)}}+\beta_j-ik^{(j)}(\sqrt{s})\right)^{-1}~,
\end{align}
where $s$ is the usual Mandelstam variable and the subscript in $t_j(\sqrt{s})$ refers to take the isospin limit with the masses of the state $j$.
For each case the CDD pole position is indicated by $\MCDD^{(j)}$,
 and $k^{(j)}(\sqrt{s})=\sqrt{2\mu_j (\sqrt{s}-\sigma_j)}$, with $\mu_j$ the reduced mass and $\sigma_j$
the threshold mass of the $j$th $D\bar{D}^*$ state.
  In addition, following Ref.~\cite{guo.251116.1}, we have
\begin{align}
\label{291116.2}
\beta_j&=8\pi\sigma_j \alpha+\rho_j,\\
\rho_j&=\frac{1}{\pi}\left(M_{D}^{(j)}\log\frac{M_{D}^{(j)}}{M_{\pi^+}}+M_{D^*}^{(j)}\log\frac{M_{D^*}^{(j)}}{M_{\pi^{+}}}\right)~,\nn
\end{align}
The previous equation results by taking the non-relativistic reduction of the unitarity loop function for the $s$-channel intermediate state.
We have also used quite an obvious notation for the masses involved. The real and imaginary parts of the momentum at the resonance in the 2nd RS
are denoted by $k_r^{(j)}$ and $-k_i^{(j)}$ and can be calculated from
$k_r^{(j)}-ik_i^{(j)}\equiv -\sqrt{2\mu_j(M_R-i\Gamma_R/2-\sigma_j)}$, with the argument of the radicand taken between $[0,2\pi[$~.
 The expressions for $\widetilde{\lambda}$, $\alpha$ and and $\MCDD^{(j)}$
that result by imposing that each $t_j(\sqrt{s})$ has a pole in the 2nd RS at $M_R-i\Gamma_R/2$,
with $M_R=M_{D^0}+M_{D^{*0}}+E_R$ fixed can be found in Ref.~\cite{oller.211116.6}. We write them here as:
\begin{align}
\label{291116.3}
\alpha=&\frac{\Gamma_R\left(k_i^{(1)}-\rho_1\right)+2k_r^{(1)}\left(M_{\rm CDD}^{(1)}-M_R\right)}{8\pi \sigma_2\Gamma_R}~,\\
\widetilde{\lambda}=&\frac{k_r^{(1)}\left[(M_{\rm CDD}^{(1)}-M_R)^2+\Gamma_R^2/4\right]}{\Gamma_R/2}~,\nn \\
M_{\rm CDD}^{(1)}= &\frac{\chi_{11} \pm \sqrt{\chi_{12}}}{ k_r^{(1)} (k_r^{(1)}\sigma_2^2-k_r^{(2)}\sigma_1^2)}~,\nn\\
\chi_{21}=&M_R k_r^{(1)}(k_r^{(1)}\sigma_2^2 - k_r^{(2)}\sigma_1^2) +
\Gamma_R k_r^{(1)} \sigma_2(-k_i^{(1)}\sigma_2+\rho_1\sigma_2+k_i^{(2)}\sigma_1
-\rho_2\sigma_1)/2~,\nn\\
\chi_{12}=&\frac{\Gamma_R^2}{4}k_r^{(1)}k_r^{(2)}\sigma_1^2\left\{
\sigma_2^2\left[ k_r^{(1)}(k_r^{(1)}-k_r^{(2)})+(k_i^{(1)}-\rho_1)^2\right]
-2\sigma_1\sigma_2(k_i^{(2)}-\rho_2)(k_i^{(1)}-\rho_1)\right.\nn\\
+&\left.\sigma_1^2\left[k_r^{(2)}(k_r^{(2)}-k_r^{(1)})+(k_i^{(2)}-\rho_2)^2\right]
\right\}~,\nn
\end{align}
These equations provide us with two different solutions, that stem from the $\pm$ sign in the expression for $\MCDD^{(1)}$. We refer
to them as the first and second solutions.

In the limit $\Gamma_R\to 0^+$ we end with similar expressions for $\widetilde{\lambda}$ and $\beta_1$ as
Eq.~\eqref{170319.3},
\begin{align}
\label{170323.1}
\widetilde{\lambda}&=\frac{\mu_1}{\varkappa_1}(M_{\rm CDD}^{(1)}-E_R)^2~,\\
\beta_1&=\frac{\mu_1}{\varkappa_1}(M_{\rm CDD}^{(1)}-3 E_R)\to \alpha=\frac{\beta_1}{8\pi\sigma_1}-\rho_1~,\nn
\end{align}
with $\varkappa_1=\sqrt{2\mu_1|E_R|}$.

The coupled-channel $S$-wave amplitude for channels 1 and 2, using  again the correspondingly adapted expression of Ref.~\cite{guo.251116.1}, reads,
\begin{align}
\label{291116.5}
t(\sqrt{s})&=\left(
\frac{2\widetilde{\lambda}}{\sqrt{s}-\widetilde{M}_{\rm{CDD}}}+\beta_1+\beta_2-ik^{(2)}(\sqrt{s})-ik^{(1)}(\sqrt{s})\right)^{-1}~.
\end{align}
In this formula one implicitly assumes that the main isospin breaking corrections
between the different coupled channels are expected to arise
from the dependence of the three-momenta $k^{(i)}$ on their threshold because of the associated branch point singularity at
each nearby threshold \cite{guo.251116.1}.

The parameters $\widetilde{\lambda}$, $\beta_1$ and $\beta_2$, cf. Eq.~\eqref{291116.2}, are fixed here from
Eq.~\eqref{291116.3} in terms of $E_R$. We still have to determine $\widetilde{M}_{\rm{CDD}}$, which is fixed by requiring that $t(\sqrt{s})$ have a
bound state pole (in the 1st RS) at $\sqrt{s}=M_R$,
\begin{align}
\label{291116.7}
\widetilde{M}_{\rm{CDD}}&=M_R-2\widetilde{\lambda}/\left(\beta_1+\beta_2-ik^{(2)}(M_R)-ik^{(1)}(M_R)\right)~.
\end{align}
In this way, the parameters to be employed in  Eq.~\eqref{211116.3} for the case 3  introduced in Sec.~\ref{subsec:virtualboundI} are:
\begin{align}
\label{291116.6}
\lambda=&2\widetilde{\lambda}~,\\
\MCDD=&\widetilde{M}_{CDD}-\sigma_1~,\nn \\
\beta=&\beta_1+\beta_2-i k^{(2)}(\sigma_{1})~.\nn
\end{align}
Notice that the three-momentum of the channel 2 has been frozen at its value at the $D^0\bar{D}^{*0}$ threshold because the $X(3872)$
signal happens around $\sigma_1$ within an energy region $|E|\ll \Delta$ . As commented above
we have checked that our results are stable if releasing it as in Eq.~\eqref{291116.5}.

One can obtain an accurate numerical approximation to the exact expression for $\MCDD^{(1)}$
 in Eq.~\eqref{291116.3} (in the limit of $\Gamma_R\to 0^+$)
if isospin breaking corrections in $\mu_i$ and $\beta_i$ are neglected (that are set equal to $\mu_1$ and $\beta_1$).
These corrections  are of ${\cal O}(\delta M/M_D)$, with $\delta M$ an isospin splitting mass in the $D^{(*)}$ multiplets.
 The resulting simplified expression is
\begin{align}
\label{170323.2}
\MCDD^{(1)}&=\frac{E_R\pm 2\alpha\left( \sqrt{|E_R|(\Delta-E_R)}+3E_R/2\right)}{1\pm \alpha}~,\\
\alpha&=\left(\frac{\Delta-E_R}{|E_R|}\right)^\frac{1}{4}~.\nn\\
\end{align}
The $+$ applies to the first solution and the $-$ to the second.
Substituting Eq.~\eqref{170323.2} in Eq.~\eqref{170323.1} we
 have the following  expressions for $\widetilde{\lambda}$ and $\beta_1$,
\begin{align}
\label{170323.3}
\widetilde{\lambda}&=\frac{4\mu\alpha^2}{\varkappa_R(1\pm\alpha)^2}\left(E_R+\sqrt{|E_R|(\Delta-E_R)}\right)^2~,\\
\beta_1&=\frac{2\mu}{\varkappa_R(1\pm \alpha)}\left(-E_R\pm \alpha\sqrt{|E_R|(\Delta-E_R)} \right)~,\nn
\end{align}
where $\varkappa_R=\varkappa_1$ and $\varkappa_2=\sqrt{2\mu(\Delta-E_R)}$.
From Eqs.~\eqref{170323.2} and \eqref{170323.3} we also have an explicit expression for $\MCDD$,
\begin{align}
\label{170323.4}
\MCDD&=E_R+\frac{8\mu\alpha^2\left(E_R+\sqrt{|E_R|(\Delta-E_R)}\right)^2}{4\mu(1\pm \alpha)
\left(-E_R\pm \alpha \sqrt{|E_R|(\Delta-E_R)}\right)+
\varkappa_R(1\pm \alpha)^2(\varkappa_R + \varkappa_2)}~.
\end{align}

It is interesting to consider the limit $\alpha\to \infty$ because it is relevant for the $X(3872)$ given the
fact that $|E_R|\ll \Delta$ and Eqs.~\eqref{170323.2}, \eqref{170323.3} and \eqref{170323.4} largely simplify.
In this limit there is only one solution which is given by
\begin{align}
\label{170323.5}
\beta&=3\varkappa_2~,\\
\lambda&=4\varkappa_R \Delta~,\nn\\
\MCDD&=\frac{4}{3}\sqrt{\Delta|E_R|}~.\nn
\end{align}

We can also see that in this limit there is a virtual
state in the 2nd RS, with similar energy as the bound state imposed by construction.
Since $\varkappa_R$ is a root, and written the three-momentum of the new solution
as $i\varkappa_2$, we have from the secular equation the still exact relation
\begin{align}
\label{170323.6}
\varkappa_2=-\varkappa_R
-\frac{(\varkappa_2^2+2\mu\MCDD)}{\beta+\varkappa_R}~.
\end{align}
Now, implementing in this equation the values for the constants obtained
in Eq.~\eqref{170323.5} we simply have that for $\alpha\to \infty$,
\begin{align}
\label{170323.6b}
\varkappa_2= -\frac{13}{9}\varkappa_R~.
\end{align}
Equation \eqref{170323.6} considered for values of $\varkappa_2^2$ much larger than $2\mu \MCDD$
 also implies that the third solution solution in
this limit is $\varkappa_3=-\beta$.

To end this appendix let us discuss for $\alpha\to\infty$
how the poles move when including the finite width
of the $D^{*0}$, that is, with $\Gamma_*\neq 0$. First, because of the condition imposed
to guarantee the presence  of the bound state  with $\Gamma_*=0$, one can rewrite $\lambda$ as
 $\lambda=(\varkappa_R^2/2\mu+\MCDD)(\beta+\varkappa_R)$. The
secular equation to calculate its final pole position at $i\varkappa_B$ is then
\begin{align}
\label{170323.10}
(\frac{\varkappa_R^2}{2\mu}+\MCDD)(\beta+\varkappa_R)-
(\frac{\varkappa_B^2}{2\mu}+\MCDD+i\frac{\Gamma_*}{2})(\beta+\varkappa_B)&=0~.
\end{align}
In the limit $\alpha\to\infty$ we can neglect $\varkappa_{R,B}$ in front of $\beta$
and the previous equation takes us to the solution
\begin{align}
\label{170323.11}
\varkappa_B=\varkappa_R-i\mu\frac{\Gamma_*}{2\varkappa_R}~.
\end{align}
 The corresponding energy $E_B$ is
\begin{align}
\label{170323.12}
E_B&=-\frac{\varkappa_B^2}{2\mu}-i\frac{\Gamma_*}{2}=-\frac{\varkappa_R^2}{2\mu}=E_R~,
\end{align}
with quadratic terms in $\Gamma_*$ neglected both in Eqs.~\eqref{170323.11} and
\eqref{170323.12}.

Let us move to calculate the pole position of the near-threshold virtual state.
Instead of Eq.~\eqref{170323.6} we now have the exact relation,
\begin{align}
\label{170323.7}
\varkappa=-\varkappa_R+i\mu\frac{\Gamma_*}{-\varkappa+\varkappa_R}
-\frac{\varkappa^2+i\mu\Gamma_*+2\mu\MCDD}{\beta+\varkappa_R}~.
\end{align}
The dominant contribution to the imaginary part stems from the
second term on the right-hand side of the previous equation since
$\beta\gg \varkappa_R$. We then have
\begin{align}
\label{170323.8}
\varkappa_2=-\frac{13}{9}\varkappa_R-i\frac{9}{22}\frac{\mu\Gamma_*}{\varkappa_R}~.
\end{align}
The associated energy $E_2$ is
\begin{align}
\label{170323.9}
E_2&=-\frac{\varkappa_2^2}{2\mu}-i\frac{\Gamma_*}{2}
=\left(\frac{13}{9}\right)^2E_R+i\frac{\Gamma_*}{11}~,
\end{align}
where quadratic term in $\Gamma_*$ have been neglected.
Notice how Eqs.~\eqref{170323.12} and \eqref{170323.9} imply a much smaller
imaginary part in absolute value for $E_B$ and $E_2$ than the half of the width of the
constituent $D^{*0}$, in agreement with the numerical results reported
in Sec.~\ref{subsec:virtualboundI}.

For the deep virtual-state pole we consider again Eq.~\eqref{170323.7}
and neglect $\varkappa_R$ and $\sqrt{2\mu|\MCDD|}$  in front of $\varkappa$ and $\beta$ (as $\varkappa_3\approx -\beta$).
We then have the following equation for the solution $\varkappa_3$ that gives rise to the leading contribution to its imaginary part,
\begin{align}
\label{170323.13}
\frac{\varkappa_3^2}{\beta}+\varkappa_3+i\mu\Gamma_*
\left(\frac{1}{\varkappa_3}+\frac{1}{\beta}\right)&=0~.
\end{align}
Neglecting quadratic terms in $\Gamma_*$ the imaginary part in
$\varkappa_3$ cancels and we obtain again the same result as above with $\Gamma_*=0$,
$\varkappa_3=-\beta$. Its energy $E_3$ is
\begin{align}
\label{170323.14}
E_3&=-\frac{\beta^2}{2\mu}-i\frac{\Gamma_*}{2}~,
\end{align}
and its width is just determined by that of its constituent $D^{*0}$. This
result is as expected because this pole is a deep one that
stems from the direct $D^0\bar{D}^{*0}$ scattering, since at those energies
the CDD pole contributions is negligible compared to $\beta$ as
$\lambda/E\beta\simeq  2\mu \lambda/\beta^3\propto \sqrt{|E_R|/\Delta}$ and tends to zero.
This is not the case for the lighter poles because they are associated with the bare
state with a small component of $D^0\bar{D}^{*0}$.

Indeed one can easily calculate
the residue of $t(E)$ in the variable $k$ at the bound-state pole position for
$\Gamma_*=0$ and $\alpha\to \infty$ and apply Eq.~\eqref{170319.2}. The following limit result is obtained
\begin{align}
\label{170323.15}
X=\frac{2}{11}~,
\end{align}
of similar size as those reported in Sec.~\ref{subsec:virtualboundI}.
This finite small number is related to the fact that for $|E_R|\to 0$ also $\MCDD\to 0$
but the quotient $|E_R|/\MCDD=3/4\,\sqrt{E_R/\Delta}\to 0$, so that in relative terms
 the binding energy is much closer to zero than $\MCDD$ for the limit $\alpha\to\infty$.

It is also worth remarking that for all the three poles the corrections in their
pole positions as a function of $\Gamma_*$ are analytic because they are simple poles (isolated
singularities).


\end{document}